\documentclass[twocolumn]{aastex6}
\usepackage[]{units}
\usepackage[update,prepend]{epstopdf}
\usepackage{graphicx}
\usepackage{amssymb}
\usepackage{amsmath}
\usepackage{threeparttable}
\usepackage{hyperref}
\usepackage{color}

\usepackage{mathtools}

\newcommand{\Ha}{H$\alpha$}
\newcommand{\Hb}{H$\beta$}
\newcommand{\Hii}{H\,{\scriptsize II}}

\newcommand{\oiii}{\hbox{[O\,{\scriptsize III}]}}
\newcommand{\oii}{\hbox{[O\,{\scriptsize II}]}}
\newcommand{\sii}{\hbox{[S\,{\scriptsize II}]}}

\newcommand{\nii}{\hbox{[N\,{\scriptsize II}]}}

\newcommand{\Msun}{\,M$_\odot$}
\newcommand{\Msunyr}{\,M$_\odot$\,yr$^{-1}$}
\newcommand{\ang}{\,\AA}

\newcommand{\Ometallicity}{$12+\log(\rm{O/H})$}

\newcommand{\Usigma}{\Msun\ $\rm{kpc^{-2}}$}

\mathchardef\mhyphen="2D
\shorttitle{Low-metallicity regions in star-forming MaNGA galaxies}
\shortauthors{Hwang et al.}

\graphicspath {{figs/} {figs/merger/} {figs/ring/} {figs/remnant/} {figs/close_pair/} {figs/local_properties/} {figs/MZrelation/} {figs/global_properties/} {figs/inter/} {figs/example/} {figs/calibrator_test/}}

\begin{document}

\title{Anomalously low metallicity regions in MaNGA star-forming galaxies: Accretion Caught in Action?}

%\title{Accretion caught in action: anomalously low metallicity regions in MaNGA galaxies}

\author{
Hsiang-Chih Hwang\altaffilmark{1}, 
Jorge K. Barrera-Ballesteros\altaffilmark{1}, 
Timothy M. Heckman\altaffilmark{1}, 
Kate Rowlands\altaffilmark{1},  
Lihwai Lin\altaffilmark{2}, 
Vicente Rodriguez-Gomez\altaffilmark{1}, 
Hsi-An Pan\altaffilmark{2}, 
Bau-Ching Hsieh\altaffilmark{2}, 
Sebastian S\'anchez\altaffilmark{3}, 
Dmitry Bizyaev\altaffilmark{4,5}, 
Jorge S\'anchez Almeida\altaffilmark{6,7}, 
David A. Thilker\altaffilmark{1}, 
Jennifer M. Lotz\altaffilmark{1,8}, 
Amy Jones\altaffilmark{9}, 
Preethi Nair\altaffilmark{9},
Brett H. Andrews\altaffilmark{10,11,12}, 
Niv Drory\altaffilmark{13}
}
\altaffiltext{1}{Department of Physics \& Astronomy, Johns Hopkins University, Baltimore, MD 21218, USA}

\altaffiltext{2}{Institute of Astronomy \& Astrophysics, Academia Sinica, Taipei 10617, Taiwan}

\altaffiltext{3}{Instituto de Astronom\'ia, Universidad Nacional Aut\'onoma de M\'exico, A. P. 70-264, C.P. 04510, M\'exico, D.F., Mexico}

\altaffiltext{4}{Apache Point Observatory, P.O. Box 59, Sunspot, NM 88349, USA}

\altaffiltext{5}{Sternberg Astronomical Institute, Moscow State
University, Moscow, Russia}

\altaffiltext{6}{Instituto de Astrof\'isica de Canarias, 38205 La Laguna, Tenerife, Spain}

\altaffiltext{7}{Departamento de Astrof\'isica, Universidad de la Laguna, E-38205 La Laguna, Tenerife, Spain}

\altaffiltext{8}{Space Telescope Science Institute, 3700 San Martin Drive, Baltimore, MD 21218, USA}

\altaffiltext{9}{ Department of Physics and Astronomy, The University of Alabama, Tuscaloosa, AL 35487, USA}

\altaffiltext{10}{Department of Physics and Astronomy, University of Pittsburgh, Pittsburgh, PA 15260, USA}

\altaffiltext{11}{Department of Astronomy, The Ohio State University, Columbus, OH 43210, USA}

\altaffiltext{12}{Center for Cosmology and Astro-Particle Physics, The Ohio State University, Columbus, OH 43210, USA}

\altaffiltext{13}{McDonald Observatory, The University of Texas at Austin, Austin, TX 78712, USA}

\begin{abstract}
We use data from 1222 late-type star-forming galaxies in the SDSS IV MaNGA survey to identify regions in which the gas-phase metallicity is anomalously-low compared to expectations from the tight empirical relation between metallicity and stellar surface mass-density at a given stellar mass. We find anomalously low metallicity (ALM) gas in 10\% of the star-forming spaxels, and in 25\% of the galaxies in the sample. The incidence rate of ALM gas increases strongly with both global and local measures of the specific star-formation rate, and is higher in lower mass galaxies and in the outer regions of galaxies. The incidence rate is also significantly higher in morphologically disturbed galaxies. We estimate that the lifetimes of the ALM regions are a few hundred Myr. We argue that the ALM gas has been delivered to its present location by a combination of interactions, mergers, and accretion from the halo, and that this infusion of gas stimulates star-formation. Given the estimated lifetime and duty cycle of such events, we estimate that the time-averaged accretion rate of ALM gas is similar to the star-formation rate in late type galaxies over the mass-range M$_* \sim10^9$ to 10$^{10}$ M$_{\odot}$.

\end{abstract}
\keywords{galaxies: abundances -- galaxies: evolution -- galaxies: statistics -- surveys -- techniques: imaging spectroscopy}

\section{Introduction}

Gas accretion has long been known to play a crucial role in galaxy formation and evolution \citep{Rees1977, White1978}. Indeed, the growth in the stellar masses of galaxies over cosmic time occurs mainly through the accretion of gas from the environment, rather than through galaxy mergers \citep{Dekel2009, Papovich2011, Behroozi2013, Rodriguez-Gomez2016}. Moreover, if there were no gas replenishment from the outside, the observed rate of star formation is so high that most star-forming galaxies in the present-day Universe would soon run out of gas and cease to form stars \citep{Kennicutt2012}. Our own Milky Way is a good example. It currently has a star formation rate of about $2$\Msunyr\ \citep{Robitaille2010, Chomiuk2011, Licquia2015} with a roughly constant star formation history over the past 8 Gyr \citep{Snaith2014}. This star formation rate (SFR) would consume all the available gas within a few Gyr if there were no gas inflow. As the star formation activity in galaxies is strongly related to their gas content \citep{Kennicutt1998}, gas accretion is therefore at the center of our understanding of galaxy growth and quenching.

However, direct observation of accreting gas is challenging, and it is even more difficult to directly determine the accretion rates. The immediate source of such gas in star-forming galaxies is the Circum-Galactic Medium (CGM) or the Inter-Galactic Medium (IGM; \citealt{SanchezAlmeida2014}), which consists of ionized gas and neutral hydrogen. For our Galaxy, the accretion rate is dominated by the Magellanic stream ($4$\Msunyr) with only $0.4$\Msunyr\ from the other high-velocity clouds \citep{Putman2012}. Such estimates are very hard to derive for extragalactic samples. Also, the accreted gas does not necessarily become the fuel for star formation because gas clouds may get heated up by shocks when they fall onto the disks, preventing them from cooling and forming stars.

The chemical abundances (metallicities) in the interstellar medium can serve as an indirect tool to study the gas circulation between a galaxy and its surroundings. Inflows from the CGM, IGM, or satellite galaxies can provide low-metallicity gas for galaxies. After the gas is processed by nucleosynthesis in stars, metals are released to the Interstellar Medium (ISM) via stellar winds and supernovae. If the star formation activity is strong enough, it is able to launch a galactic outflow \citep{Heckman1990, Heckman2002}. The outflow takes metals away from galaxies to enrich the CGM. This process and its consequences can be studied using absorption line spectroscopy of the CGM using background quasars \citep{Chen2010,Rubin2014,Lehner2013,Lan2017,Heckman2017}. The ejected metals may re-accrete onto the disks on a dynamical timescale \citep{Oppenheimer2010}. Consequently, the metallicity in the ISM is produced by the interplay between star formation, inflow, and outflow.

Over the past decade, we have learned a great deal about the systematic relationships between ISM metallicity and galaxy properties. Single-fiber spectroscopic surveys have shown that central oxygen metallicity increases with galaxy stellar mass: the so-called global M-Z relation \citep{Lequeux1979, Garnett1987, Vila-Costas1992, Tremonti2004}. The relation can be explained in part by mass-dependent metal loss because of the shallower potential wells in low-mass galaxies, and/or by the inflow of metal-poor gas. The global M-Z relation may have an additional dependence on global SFR such that galaxies with higher SFRs have lower metallicity at a fixed stellar mass M$_*$\ \citep{Ellison2008}. The form of the relation between galaxy stellar mass, metallicity, and SFR may be independent of redshift, and therefore it is called the fundamental metallicity relation \citep{Mannucci2010, Lara-Lopez2010}. It has been debated whether or not this additional dependence on SFR is due to observational systematics like single-aperture bias \citep{Sanchez2013, Sanchez2017, Barrera-Ballesteros2017}, the nature of the metallicity calibrators used \citep{Kashino2016}, and whether or not its form is really independent of redshift \citep{Steidel2014}. Also, rather than SFR, the global M-Z relation may have a stronger dependence on other properties like gas fraction \citep{Hughes2013, Bothwell2013} and stellar age \citep{Lian2015}.

Single-aperture spectroscopy limits the metallicity study to the galactic centers, while metallicity is known to have significant spatial variation. Long-slit spectroscopy and IFU observations have shown the radial dependence of gas-phase metallicity in late-type galaxies with metallicity decreasing outwards, i.e. negative metallicity gradient \citep{Searle1971, Vila-Costas1992, Zaritsky1994, Moran2012, Sanchez2012a, Sanchez2014, Belfiore2017, Poetrodjojo2018}. Furthermore, the azimuthal variations imprinted on the radial gradients suggest that the local chemical evolution may be influenced by local enrichment and dilution caused by spiral density waves \citep{Ho2017}. 

Indeed, evidence is now growing that the global relations between mass, metallicity, and arguably SFR may emerge from relations at local scales. On $\sim$\,kpc scales, local metallicity tightly correlates with local stellar surface mass density ($\Sigma_*$-Z relation, \citealt{Moran2012, Rosales-Ortega2012}). This local relation can reproduce both the global M-Z relation and radial metallicity gradients \citep{Barrera-Ballesteros2016}. The local metallicity also correlates with both the local gas-mass fraction and local escape velocity \citep{Barrera-Ballesteros2018}, as expected in simple models for chemical evolution that incorporate accretion and outflows. Metallicity inhomogeneities associated with enhanced SF activity have been observed in some galaxies. Extremely metal poor galaxies tend to have a large lopsided star-forming region of low metallicity (e.g. \citealt{SanchezAlmeida2013, SanchezAlmeida2015}). There also seems to be tendency for the star-forming regions of larger SFR to have lower metallicity \citep{Cresci2010, Richards2014, SanchezAlmeida2018}. These metallicity drops are usually attributed to recent gas accretion events triggering star formation.

While the existence of these excellent empirical correlations between local galaxy properties and the local ISM metallicity provide important insights into galaxy evolution, in this paper we will take a very different approach. In particular, we use these empirical correlations to identify anomalous regions whose metallicity is significantly lower than expected. We will then compare the global properties of the galaxies that contain such regions to those that do not. We will also compare the local properties of the regions with anomalously low metallicity to regions with normal metallicity. Based on this analysis, we will argue that these regions likely trace the sites of the recent accretion of gas that has stimulated on-going star-formation. By constraining the lifetimes of the low-metallicity regions, we will try to estimate the frequency of such accretion events and the implied accretion rates, and thereby assess their implications for galaxy evolution.

The paper is structured as follows. In Sec.~\ref{sec:observation} we will describe the observational dataset, our sample selection, and our measurements of metallicity and other local and global properties. In Sec.~\ref{sec:result} we will present our results linking the low-metallicity regions with both global and local galaxy properties. In Sec.~\ref{sec:discussion} we will discuss the possible sources for the low-metallicity gas and quantify the properties of the accretion. We summarize our conclusions in Sec.~\ref{sec:conclusion}. Throughout, we use an $h=0.7$, $\Omega_m = 0.3$, $\Omega_\Lambda = 0.7$ cosmology. We use lower-case `$z$' for redshifts, and capital `$\rm Z$' for metallicity.

\section{Observations and Data Reduction}
\label{sec:observation}

\subsection{MaNGA overview}

\begin{figure}[thbp] %  figure placement: here, top, bottom, or page
   \centering
   \includegraphics[width=3.4in]{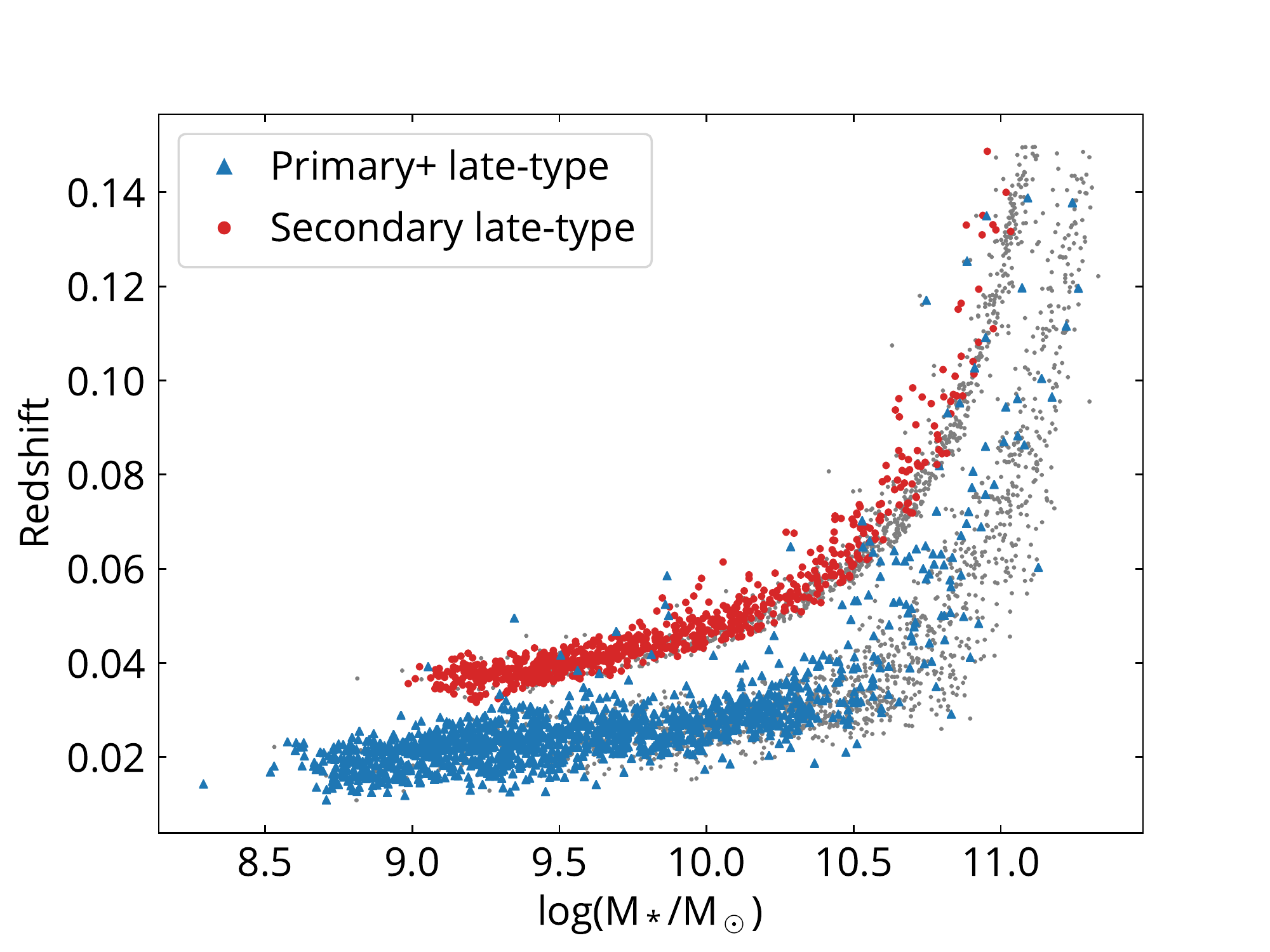} 
   \includegraphics[width=3.4in]{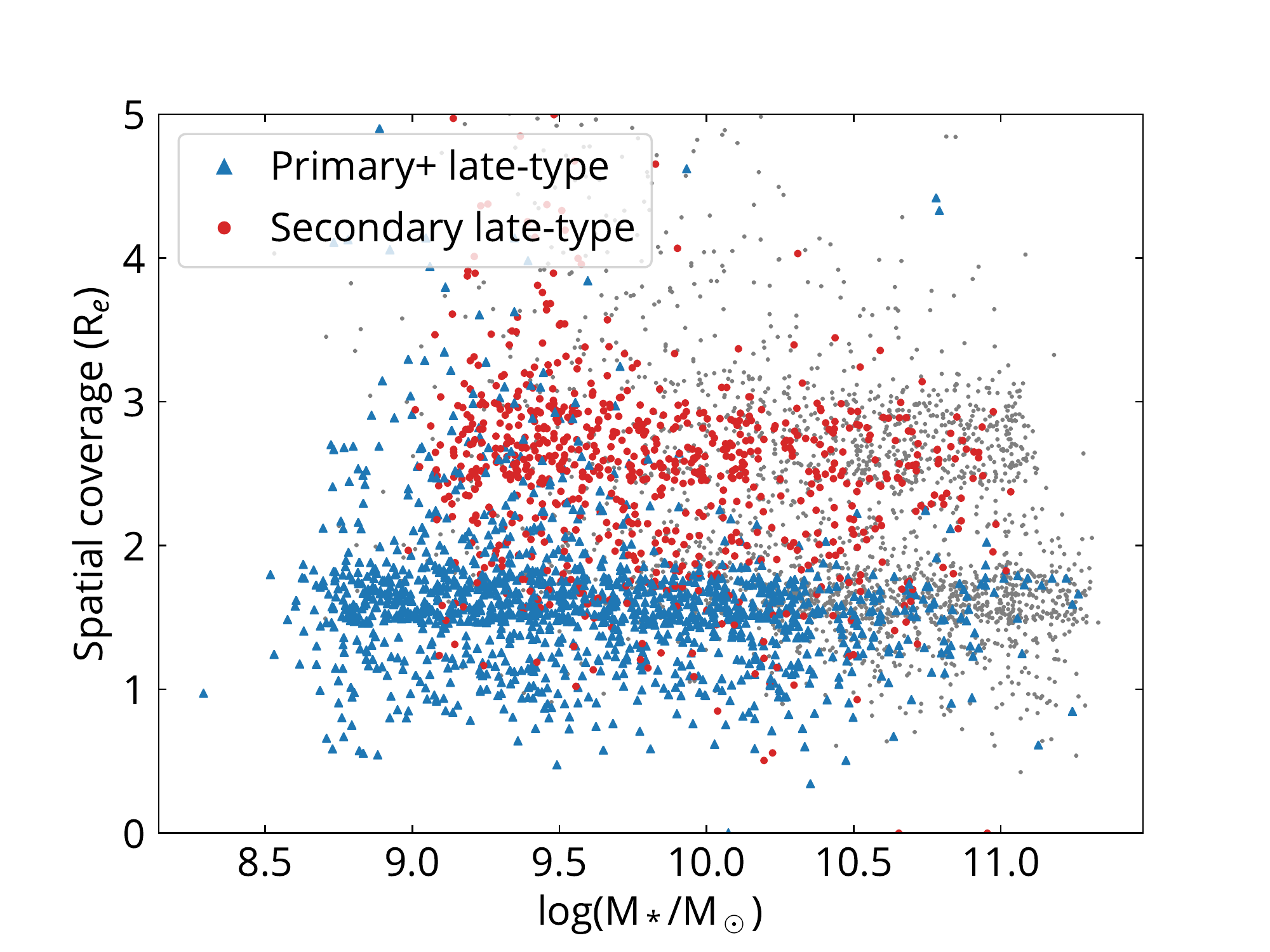} 
   \caption{The distributions of redshift and spatial coverage with respect to stellar mass. Blue triangles and red circles represent Primary$+$ and Secondary late-type galaxies respectively. The small grey dots are early-type galaxies. The Primary$+$ sample is covered by MaNGA over regions interior to $\sim1.5R_e$ while the secondary sample extends to $\sim 2.5 R_e$. }
   \label{fig:pri_sec_sample}
\end{figure}

Mapping Nearby Galaxies at Apache Point Observatory (MaNGA; \citealt{Bundy2015}) observes galaxies in the local Universe ($z\lesssim 0.15$) using integral field units (IFUs) as part of Sloan Digital Sky Survey (SDSS)-IV \citep{Blanton2017}. MaNGA provides wavelength coverage from 3600 to 10,300\ang\ with a spectral resolution of $R\sim2000$ \citep{Smee2013} and a reconstructed PSF of $\sim2.5$\,arcsec in FWHM \citep{Law2016}. The MaNGA IFUs consist of bundles of 19, 37, 61, 91, and 127 fibers, providing effective coverage over radii from 5.45 to 14.54\,arcsec \citep{Drory2015, Yan2016}. Started in 2014, the MaNGA survey will observe $\sim$ 10,400 galaxies in 6\,years using the Sloan 2.5\,m telescope \citep{Gunn2006}. For more details about MaNGA observing strategy and spectrophotometry calibration, we refer to \cite{Law2015} and \cite{Yan2016a}.

The MaNGA galaxy sample constitutes the Primary sample, Secondary sample, Color-Enhanced sample, and objects for ancillary programs. As shown in Fig.~\ref{fig:pri_sec_sample}, the Primary and Secondary samples are targeted to have IFU areal coverage out to $\sim1.5$ and $\sim2.5 R_e$ respectively with roughly uniform stellar mass distribution \citep{Wake2017}. The Color-Enhanced sample is designed to select galaxies with IFU coverage similar to the Primary sample ($\sim1.5R_e$), but which are underrepresented because of their colors (especially high-mass blue galaxies, low-mass red galaxies, and green-valley galaxies). As the Primary and Color-Enhanced samples have the same targeted IFU coverage of $1.5R_e$, their combination is referred to as the Primary+ sample. To cover all galaxies out to similar effective radii, massive (large) galaxies are observed at higher redshifts than are low-mass (small) galaxies (Fig.~\ref{fig:pri_sec_sample}).

In this study, the galaxy sample comes from the MaNGA Product Launches 6 (MPL-6), an internal collaboration release comprising 4682 galaxies. The raw observed data are calibrated and sky-subtracted by the Data Reduction Pipeline v2.3.1 \citep{Law2016}, providing processed data cubes with spaxel sizes of 0.5\,arcsec. We use the measurements of emission lines (Gaussian fluxes and equivalent widths) and relevant stellar continuum indices (D$_n4000$) from the MPL-6 Data Analysis Pipeline v2.1.3 (Westfall et al. in preparation). Galaxy global properties such as redshift, total stellar mass, axis ratio from elliptical Petrosian fitting ($b/a$), broad-band photometry ($NUV$ and $r$ in this paper), and effective radius are drawn from the NASA-Sloan catalog v1\_0\_1 \citep{Blanton2011}.

\subsection{Sample selection}

Our goal is to identify and investigate regions of anomalously low gas-phase metallicities. These metallicities are measured using emission-lines arising in gas photo-ionized by massive stars. We therefore focus on late-type star-forming galaxies (in which the gas phase-metallicities can be measured over much of the galaxy). The SDSS pipeline fits galaxies with a de Vaucouleurs profile and an exponential profile, and provides the quantity \texttt{fracdeV} which is the fraction of fluxes contributed from the de Vaucouleurs profile. Therefore, pure de Vaucouleurs early-type galaxies have \texttt{fracdeV}$=1$, and pure exponential late-type galaxies have \texttt{fracdeV}$=0$. We follow \cite{Zheng2015} and select late-type galaxies as those having \texttt{fracdeV}$< 0.7$ in r-band, where \texttt{fracdeV}$=0.7$ roughly corresponds to Sa spiral galaxies \citep{Masters2010}. In the rest of the paper, late-type galaxies are referred to galaxies selected using \texttt{fracdeV}$<0.7$. High galaxy inclination makes deprojected quantities more uncertain, so we exclude galaxies having $b/a<0.3$ from our sample, corresponding to an inclination of 72.5\,deg. In case that some $b/a$ ratios are not reliable due to foreground stars or other image artifacts, we also use the MaNGA MPL-7 catalog of highly inclined galaxies, which are visually identified by P. Nair following the procedure described in \cite{Nair2010}. For the full highly inclined sample in this catalog, the median $b/a$ is 0.3 and the mean is 0.31 with a standard deviation of 0.12. The distributions of redshift and spatial coverage for the resulting late-type galaxy sample are shown in Fig.~\ref{fig:pri_sec_sample}.

%\blue{So my final sample is...? And they depend on local properties (pure star-forming spaxels?...)}

%\blue{Gas-phase metallicity of star-forming spaxels is measured using emission lines.} 

%\blue{We refer star-forming galaxies to galaxies .} 

Emission-lines can be produced by different ionization sources, like short-lived massive stars born from recent star formation, active galactic nuclei (AGN), and evolved stars. To distinguish between these ionization sources, spaxels are classified based on their line ratios of \nii/\Ha\ and \oiii/\Hb, the so-called Baldwin, Phillips \& Terlevich (BPT; \citealt{Baldwin1981}) diagram. Specifically, based on the location in the BPT diagram, spaxels are classified as either AGN/LINER (above \citealt{Kewley2001} demarcation line), pure star-forming (below \citealt{Kauffmann2003b} line), or composite spaxels (between the aforementioned two lines) where multiple ionization sources are present. Only pure star-forming spaxels enter our sample because the metallicity calibrators we use do not apply to other ionization sources. We will describe this in detail in Section~\ref{sec:Z-calibrator}. We apply a S/N ratio cut of 10 on the each of the emission lines used for the BPT diagram and metallicity calibrators, including \Ha, \Hb, \oiii$\lambda5007$, \oii$\lambda3926,9$, and \nii$\lambda6584$. As detailed in Sec.~\ref{sec:def-lowZ}, because we use the local stellar surface mass density to estimate expected metallicities, we only consider spaxels with deprojected stellar surface mass density $>10^7$\Usigma.

We will refer to those late-type galaxies having star-forming regions as star forming galaxies. Motivated by the size of the PSF, we require that star-forming galaxies have at least 20 pure star-forming spaxels because the angular area of a typical MaNGA PSF ($\sim2.5$\,arcsec, or $\sim1.5$\,kpc at $z=0.03$) is $\sim19.6$ spaxels. With these selection criteria, out of a total of 4784 MaNGA galaxies, there are 1685 late-type galaxies with stellar masses M$_*>10^9$\Msun, and our final star-forming galaxy sample constitutes 1222 sources with 526613 star-forming spaxels. As detailed later in Sec.~\ref{sec:def-lowZ}, we identify anomalously low-metallicity regions in 307 of these 1222 star-forming galaxies.

\subsection{Galaxy physical quantities}

%The local stellar surface mass density is measured by the PIPE3D pipeline v2.3.1 \citep{Sanchez2016b, Sanchez2016a}. In short, spaxels are spatially binned so that the continua reach a target signal-to-noise (S/N) ratio of 50. To derive the underlying stellar properties like surface mass density, the continuum is fit by the simple stellar population library MIUSCAT \citep{Vazdekis2012}. MIUSCAT covers the wavelength range 3465$-$9469\,\AA\ with four stellar ages (0.06, 0.20, 2.00, and 17.78\,Gyr) and three metallicities ($Z$=0.0004, 0.019, and 0.03), where $Z=0.019$ is the solar value. 

The local stellar surface mass density is measured by the PIPE3D pipeline v2.3.1 \citep{Sanchez2016b, Sanchez2016a}. In short, spaxels are spatially binned so that the continua reach a target S/N ratio of 50. PIPE3D performs two analysis iterations to derive the underlying stellar properties like surface mass density. In the first one, PIPE3D uses a MUISCAT simple stellar population (SSP) library \citep{Vazdekis2012} with a limited number of templates (covering 4 ages and 3 metallicities), and it is used just to analyze the kinematic properties of the stellar populations, the dust attenuation, and to remove the contamination by the emission lines. After removing the contamination from the emission lines, PIPE3D performs a second more detailed modeling of the stellar population, adopting the GSD156 SSP library \citep{CidFernandes2013} that comprises 156 templates covering 39 stellar ages (from 1Myr to 14.1Gyr), and 4 metallicities (Z/Z$_\odot$=0.2, 0.4, 1, and 1.5). These templates have been extensively used within the CALIFA collaboration \citep[e.g.][]{Perez2013,GonzalezDelgado2014}, and for other surveys \citep[e.g.][]{Ibarra-Medel2016, Sanchez-Menguiano2017}. Its current implementation for the MaNGA dataset is described in \cite{Sanchez2018}.

The projected surface mass density is further derived assuming the Salpeter \citep{Salpeter1955} initial mass function (IMF). Following \cite{Barrera-Ballesteros2016}, we deproject the observed surface mass densities using $b/a$ ratios to take inclination into account :
\begin{equation}
\Sigma_* = \Sigma_{*,obs} \times (b/a),
\end{equation}
where $\Sigma_{*,obs}$ is the observed, projected surface mass density derived by PIPE3D and $\Sigma_*$ is the deprojected surface mass density. The SFR  per unit area ($\Sigma_{\rm SFR}$) is also deprojected in the same way. In the text hereafter, stellar surface $\Sigma_*$ and $\Sigma_{\rm SFR}$ always refer to the deprojected quantities. 

SFRs are measured from extinction-corrected \Ha\ luminosity. Emission lines, including \Ha\ and those used for metallicity calibrators, are corrected for dust extinction using the Balmer decrement, assuming the star-forming extinction law of \cite{Calzetti2000} with $R_V=4.05$. For the Balmer decrement, we assume a theoretical flux ratio of \Ha/\Hb $=2.86$ \citep{Osterbrock2006}. For spaxels with \Ha/\Hb\,$<2.86$, we do not correct for extinction. After extinction correction, star formation rates (SFRs) are computed from \Ha\ luminosity \citep{Kennicutt1998}. Using the \cite{Cardelli1989} extinction law with $R_V=3.1$ only results in 3\% difference in \Ha\ luminosity \citep{Catalan-Torrecilla2015}.

In this paper, both SFRs and surface mass densities are derived using the Salpeter IMF. Compared to a Chabrier IMF \citep{Chabrier2003}, the Salpeter IMF produces SFRs that are higher by 0.2\,dex and stellar masses that are higher by 0.22\,dex \citep{Belfiore2018}. The specific star formation rate (sSFR) is the ratio between SFR and stellar mass, and so the difference between using either the Salpeter or Chabrier IMF to derive sSFR is negligible.
 
\subsection{Galaxy morphology indicators}

Several quantitative morphology indicators have been developed for galaxies. To search for ongoing interactions, the asymmetry index ($A$, \citealt{Schade1995, Abraham1996, Bershady2000, Conselice2000}) is designed to look for disturbed morphology. For a similar purpose, the Gini coefficient together with the second-moment of a galaxy's brightest regions (M$_{20}$) is used to search for mergers, especially for high-redshift sample \citep{Lotz2004}. The shape asymmetry index \citep{Pawlik2016} and outer asymmetry index \citep{Wen2014} are revisions of the asymmetry index which are more sensitive to low-surface-brightness structures. The clumpiness index \citep{Conselice2003} is used to look for clumpy structures like compact star-forming clusters \citep{Isserstedt1986, Takamiya1999}. The prominence of bulges (e.g. bulge-to-disk ratios) can be inferred by either the model-dependent Sersic index, or the non-parametric concentration index ($C$, \citealt{Abraham1994, Abraham1996, Bershady2000}). 

In this paper, we use the model-independent morphology indicators of concentration ($C$) and asymmetry ($A$). They are measured on the $i$-band images from Panoramic Survey Telescope and Rapid Response System (PanSTARRS) Data Release 1 \citep{Chambers2016, Flewelling2016}. With a mean point source $5\sigma$ depth of 23.1 (AB\,mag) in stacked images and a median PSF of 1.11\,arcsec in $i$-band, PanSTARRS provides deeper images and higher angular resolution than SDSS, resulting in more reliable morphology measurements.

The asymmetry index $A$ is measured by comparing the original image with one rotated by $180\,^{\circ}$:

\begin{equation}
A = \frac{\sum_{i,j}  |I(i,j) - I_{180}(i,j)|}{\sum_{i,j} |I(i,j)|} - A_{bgr},
\end{equation}
where $I$ is the original image and $I_{180}$ is the one rotated by $180\,^{\circ}$ about the galaxy's center, which is chosen to minimize $A$. $A_{bgr}$ is the asymmetry measured in the $32\times32$-pixel background where no structure is detected. The asymmetry index is computed over all pixels of a galaxy within 1.5$R_p$, where $R_p$ is the circular Petrosian radius \citep{Petrosian1976}. For a completely symmetric case, $A=0$, or it could be slightly negative due to the subtraction of positive $A_{bgr}$. Typically, galaxies with $A\gtrsim0.2$ are considered morphologically disturbed.

The concentration index is derived using the radii that enclose 80 ($R_{80}$) and 20 ($R_{20}$) per\,cent of the galaxy's total flux. Specifically, \begin{equation}
C = 5 \log \left(\frac{R_{80}}{R_{20}}\right).
\end{equation}
We use the definition of the total flux as the flux enclosed in 1.5$R_p$ from the galaxy's center. Higher concentration index means the light is more centrally concentrated. This correlates with the bulge-to-total-light ratios in late-type galaxies \citep{Conselice2003}. 

%For reference, a de Vaucouleurs profile has \blue{$C=?$}, while an exponential profile has \blue{$C=?$}. FIX THIS??

The morphology measurements were performed with the \texttt{statmorph} v0.2.0 code \citep{Rodriguez-Gomez2018} \footnote{https://github.com/vrodgom/statmorph}. When the morphology measurements are used, we only include galaxies where S/N per pixel $>4$ in the $i$-band image. We also have removed galaxies with  \texttt{flag==0}, which indicates that there was a problem with the basic measurements, for example the presence of foreground stars and image artifacts.

%and with S/N per pixel $>4$ in the i-band image, resulting in 1778 late-type galaxies with stellar masses M$_*>10^9$\Msun and 1248 star-forming galaxies.

\subsection{Classification of interaction stages}
\label{sec:inter-class}

The asymmetry index is mainly sensitive to the disturbed morphology at the first passage and the final coalescence during the interaction \citep{Lotz2008}. However, galaxies in interaction spend most of the time in close pairs without strong morphological disturbance, while inflows may have played a role in shaping the metallicity distribution since this stage \citep{Ellison2008}. Therefore, taking different interaction stages into account is crucial to understanding how interactions drive the inflows. 

We classify galaxies into three interaction categories: mergers, close pairs, and isolated galaxies. Mergers here refer to those interacting systems at the final coalescence stage. In the literature, mergers are often identified using morphology indicators, such as CAS (concentration, asymmetry, clumpiness; \citealt{Conselice2003}), and Gini -- M$_{20}$ coefficients \citep{Lotz2004}. To avoid irregular galaxies and to be more sensitive to low-surface-brightness signature, we adopt the classification of interaction via visual inspection. 

The visual classification of interaction stages is performed by Pan et al. (in prep.) for interacting galaxies selected from the MaNGA MPL6 sample. They selected the interacting candidates by meeting one of two criteria: (1) spectroscopic close pairs with a projected separation of 50\,kpc ($h=70$) and line-of-sight velocity difference of $< 500$\,km\,s$^{-1}$ \citep{Lin2004,Keenan2014} selected from the NASA-Sloan Atlas (NSA) \footnote{http://nsatlas.org/} matched to the MaNGA MPL6 sample; (2) the $f_m >0.4$, where $f_m$ is the `weighted-merger-vote fraction' based on the citizen classifications in the Galaxy Zoo project \citep{Darg2010,Lintott2011}; (3) galaxies having a companion within the same MaNGA bundle. This selection results in $\sim1300$ interacting candidates, which are further visually classified into 5 classes:

\begin{itemize}
\item Class 0: Non-interacting galaxies or a potential paired galaxy without a spec-$z$ confirmation for its companion.
\item Class 1: well-separated pairs not showing any morphology distortion, including very close pairs without distortions (i.e., just before the 1st passage).
\item Class 2: well-separated pairs showing strong signs of interaction, such as tails, bridges, etc. (after the 1st passage).
\item Class 3: well-separated pairs showing a weak morphology distortion.
\item Class 4 (final coalescence): two components strongly overlap with each other and show a strong morphology distortion.
\end{itemize}

We consider Class 4 as merging galaxies, Class 1-3 as close pairs, and other galaxies except Class 0 as isolated galaxies because Class 0 still has potential paired galaxies. Out of the sample of 1222 star-forming galaxies, 55 (4.5\%) are mergers, 161 (13.2\%) are in close pairs, and 929 (76.0\%) are classified as isolated galaxies.

%Mergers are defined as cases in which the morphology is highly disturbed and two galaxy components are strongly overlapped. Galaxies in close pairs are defined using SDSS spectroscopy such that they have a projected separation of $<50$\,kpc ($h=70$) and a line-of-sight velocity difference of $< 500$\kms\ \citep{Keenan2014}. Galaxies that are neither mergers nor close pairs are considered as isolated galaxies. 

%Using these classifications, galaxies with dwarf satellite galaxies are considered to be isolated.

%The visual classification of interaction stages is performed by Lin et al. (in prep.), resulting in a complete interacting catalog for all 4682 MPL-6 MaNGA galaxies. With the classifications, we will investigate the properties of anomalously low-metallicity regions in different interacting environments.

%Out of xxxx (xxxx) late-type (star-forming) galaxies, xxxx (xxxx) are mergers, xxxx (xxxx) are in close pairs, and xxxx (xxxx) are isolated galaxies.

\subsection{Metallicity}

\subsubsection{Metallicity indicators}
\label{sec:Z-calibrator}

\begin{figure*}[htbp] %  figure placement: here, top, bottom, or page
   \centering
   \includegraphics[width=3.5in]{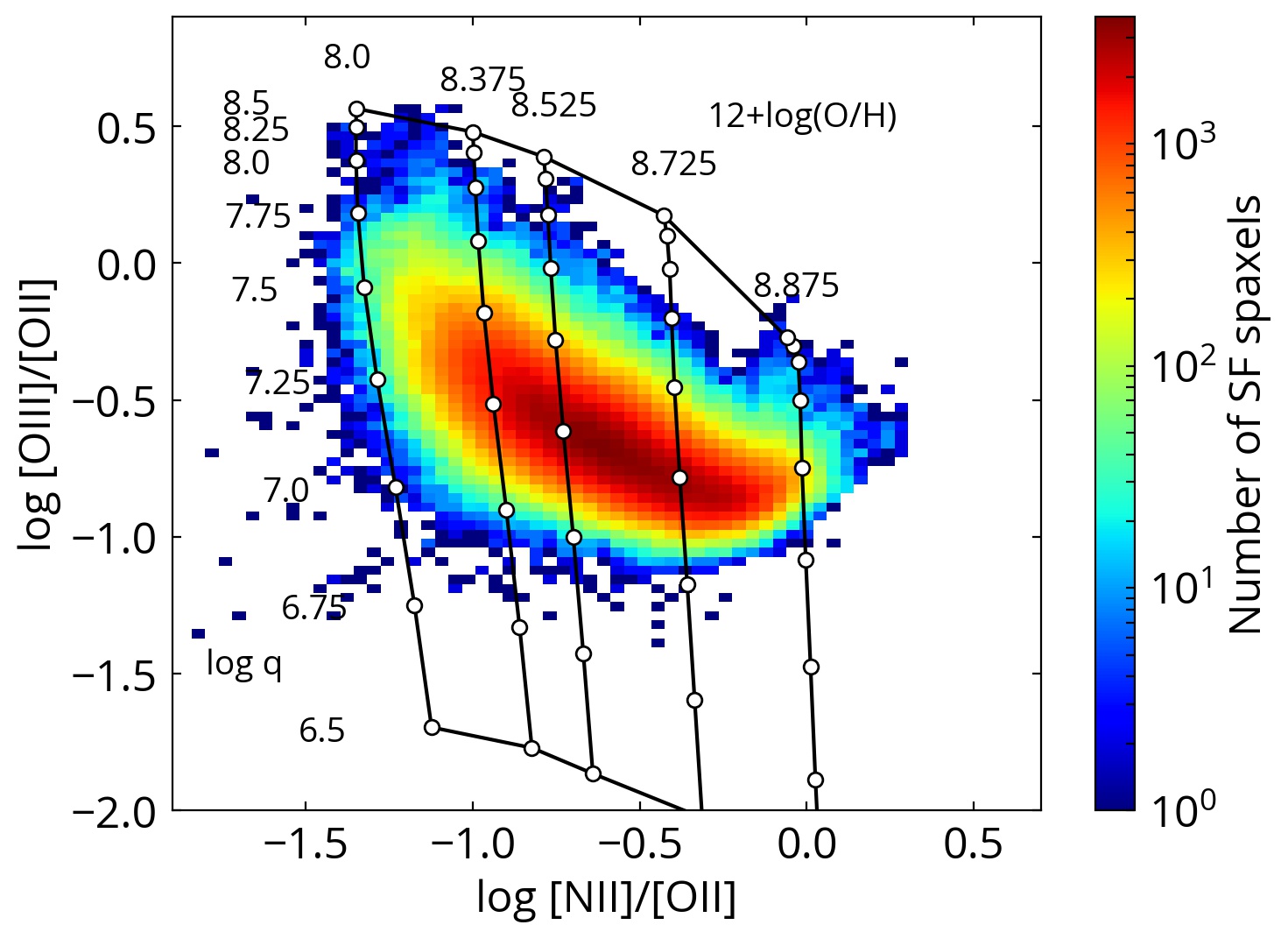} 
   \includegraphics[width=3.5in]{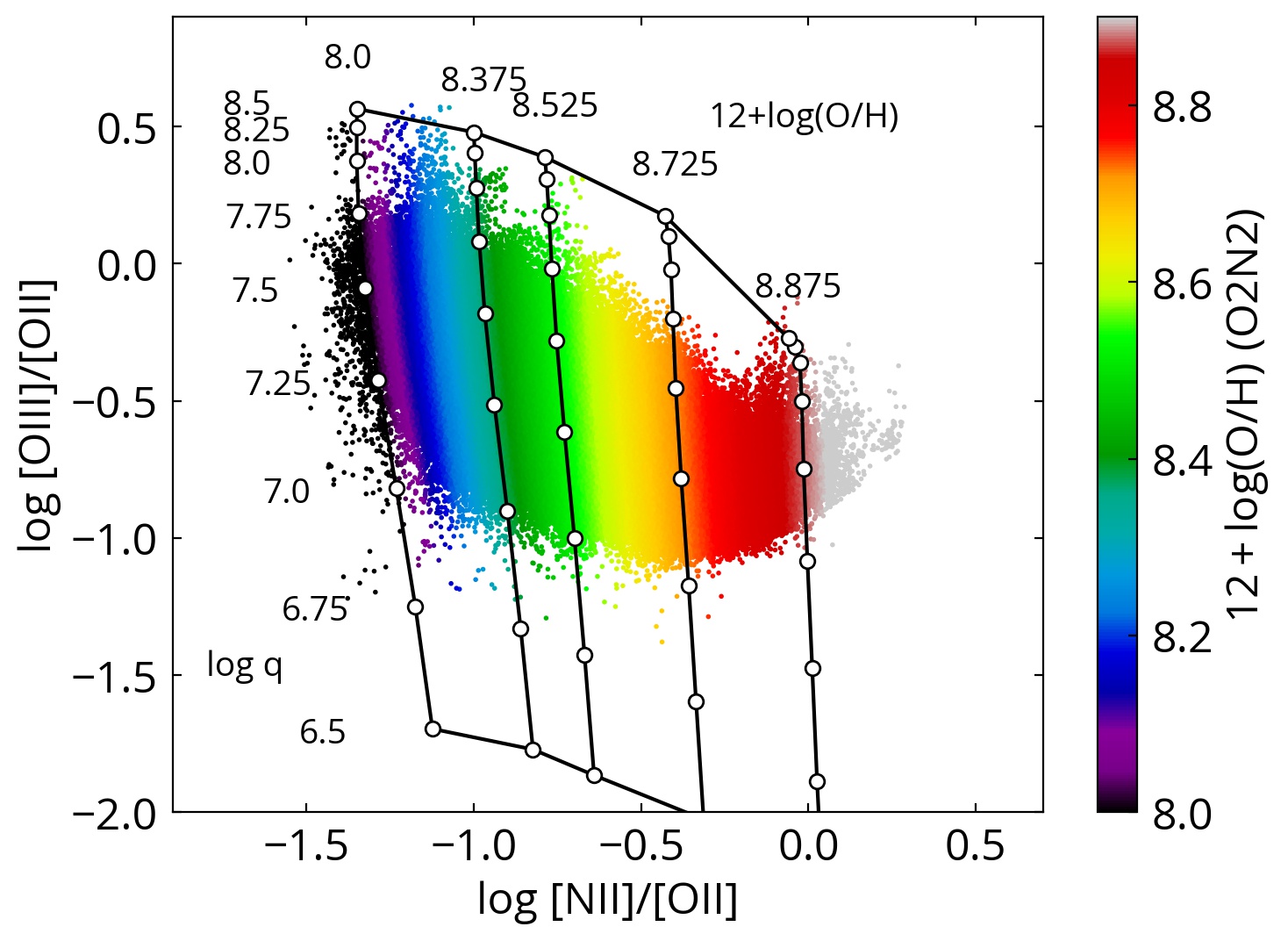} 
   \caption{
    Left: distribution of pure star-forming spaxels in late-type galaxies with M$_*>10^{8.5}$\Msun\ in the extinction-corrected \oiii/\oii-\oii/\nii\ plane. The overlaid MAPPINGS V model grid (Sutherland el al. in prep.) shows the line ratios in \Hii\ regions with different metallicities (from left to right along the x axis: \Ometallicity$=$8, 8.375, 8.525, 8.725, and 8.875) and ionization parameters (from bottom to top along the y axis: $\log q=$6.5 to 8.5 with a step of 0.25). \oii/\nii\ is mainly sensitive to metallicity while \oiii/\oii\ is mainly sensitive to the ionization parameter. Right: the same dataset color-coded by measured metallicity using the O2N2-based calibrator from the overlaid grid.
   %Upper left: distribution of pure star-forming spaxels in late-type galaxies with M$_*>10^{8.5}$\Msun\ in the extinction-corrected \oiii/\oii-\oii/\nii\ plane. The overlaid MAPPINGS V model grid (Sutherland el al. in prep.) shows the line ratios in \Hii\ regions with different metallicities (from left to right along the x axis: \Ometallicity$=$8, 8.375, 8.525, 8.725, and 8.875) and ionization parameters (from bottom to top along the y axis: $\log q=$6.5 to 8.5 with a step of 0.25). \oii/\nii\ is mainly sensitive to metallicity while \oiii/\oii\ is mainly sensitive to the ionization parameter. Upper right: the same dataset but color-coded by measured metallicity using the O3N2 calibrator \citep{Pettini2004}. The tilt in color distribution shows that O3N2 metallicity correlates strongly with ionization parameter, such that the estimated metallicity decreases with increasing ionization parameter for a fixed model metallicity. Bottom left and right: metallicity from the R23 calibrator (bottom left; \citealt{Tremonti2004}) and the O2N2-based calibrator using the overlaid grid (bottom right; Sutherland et al. in prep.). The N2S2\Ha\ and O2N2 calibrators are consistent in the sense that the measured metallicity changes strongly along the x axis (\nii/\oii) but not the y axis (\oiii/\oii).
   }
   \label{fig:Z-calis}
\end{figure*}

The oxygen abundance in the ISM can be measured using electron temperature-sensitive line ratios between nebular and auroral lines (e.g. \oiii$\lambda5007$ and \oiii$\lambda4363$), known as the direct method (e.g. \citealt{Izotov2006}). However, auroral lines are $\sim100$ times fainter than nebular lines, making the direct method challenging in typical S/N spectra. Alternatively, various strong-line methods have been developed on the basis that the flux ratios between strong lines also contain information on metallicity. Commonly used strong-line combinations include: (\oii3927,9 + \oiii4959 + \oiii5007)/\Hb\ (R23, \citealt{Pagel1979, Tremonti2004}), (\oiii5007/\Hb)/(\nii6584/\Ha) (O3N2, \citealt{Pettini2004, Marino2013}), \oii3927,9/\nii6584 (O2N2, \citealt{Kewley2002a, Dopita2013}), and a Bayesian-based method which takes multiple lines into account \citep{Blanc2015}. Some strong-line methods are calibrated by the direct method, while others are calibrated through photoionization models. The absolute oxygen abundances from different metallicity calibrators can differ by $\sim0.4$\,dex \citep{Kewley2008}. Since our analysis depends only on the relative values of metallicities, the accuracy of the absolute values of the metallicity calibrators is not the main concern.

However, strong-line flux ratios also depend on other physical conditions in the ISM and not just the metallicity. Specifically, one needs to distinguish amongst the effects of metallicity, ionization parameter, and the hardness of ionizing spectra \citep{Kewley2013}. These properties may also be physically correlated, e.g. lower metallicity regions tend to have higher ionization parameters \citep{Dopita2013, Morisset2016}. Using line ratios that depend strongly on the ionization parameter or the hardness of the ionizing radiation can complicate the interpretation of the derived metallicities. The photoionization models show that \oiii/\Hb\ is also sensitive to ionization parameter \citep{Dopita2013}. Also, the ionization potential to produce doubly ionized oxygen (35.1\,eV) is much higher than for singly ionized hydrogen, nitrogen, and oxygen (13.6, 14.5, and 13.6\,eV respectively). For these reasons, neither the R23 nor the O3N2 metallicity calibrators are ideal for the purpose of this study. 

In this paper, metallicity is estimated using the \oii/\nii\ ratio. We use the \texttt{pyqz} code v0.8.4 (\citealt{Dopita2013}; Vogt et al., in prep.) with the MAPPINGS V model grid of \oiii/\oii\ versus \oii/\nii\ (\citealt{Sutherland1993, Allen2008}; Sutherland et al., in prep.). The main idea is similar to what is described in \cite{Dopita2013} where they used MAPPINGS IV model grid, but MAPPINGS V has upgraded some input atomic physics, the depletion of heavy elements onto dust, and the methods of solution \citep{Dopita2016}. We use models with the abundance pattern and depletion fractions as described in \cite{Dopita2016} and \cite{Nicholls2017}. As shown in Fig.~\ref{fig:Z-calis}, \oii/\nii\ is mainly sensitive to metallicity while \oiii/\oii\ is sensitive to ionization parameter. The effect of the non-Maxwellian $\kappa$-distributed electrons \citep{Nicholls2012,Pierrard2010} is small on the model grid in the \oiii/\oii-\oii/\nii\ space \citep{Dopita2013}. We note that a recent study shows that the electron energy distributions in \Hii\ regions closely follow the Maxwell-Boltzmann distribution \citep{Draine2018}. Also, the ionization potentials are very similar for \nii\ and \oii, so the hardness of ionizing spectra plays a minor role in the resulting flux ratios.

We adopt the model grid assuming a Maxwellian electron distribution and an ISM pressure $\log(P/k)=5.5$ in spherical \Hii\ regions. Using different ISM pressures mainly affects the measured ionization parameters but leaves the estimated metallicities nearly unchanged. For spaxels located inside the grid, we derive their metallicities and ionization parameters from the model grid using cubic interpolation. For spaxels outside the grid, we extrapolate these quantities for them but discard those located more than $0.3$\,dex away from the grid boundary.

Fig.~\ref{fig:Z-calis} shows the distributions of pure star-forming spaxels in MaNGA late-type galaxies in the extinction-corrected \oiii/\oii-\oii/\nii\ space, overlaid with the model grid described above. The majority of pure star-forming spaxels have ionization parameters of $\log q\sim 7.25$ \footnote{Following \cite{Dopita2013}, ionization parameter $q$ is defined as the number of ionizing photons per unit area per second divided by the gas particle number density. Therefore, $q$ is in units of cm\,s$^{-1}$.}, slightly tilting to higher ionization parameters at the low-metallicity end (\Ometallicity $<8.5$). It is consistent with previous work that \Hii\ regions with lower gas-phase metallicities on average have higher ionization parameters \citep{Dopita2013, Morisset2016}. 

%In Fig.~\ref{fig:Z-calis}, we also present the measured metallicities using the O3N2 calibrator \citep{Pettini2004}, the R23 calibrator \citep{Tremonti2004}, and the overlaid model grid from MAPPINGS V (\citealt{Dopita2013}; Sutherland et al., in prep.). Compared with the model grid, the O3N2 calibrator has stronger dependence on ionization parameter. At a fixed model-grid metallicity, the O3N2 metallicities can change up to $\sim0.4$\,dex. On the other hand, the R23 and O2N2 calibrators are consistent in a way that the measured metallicities change mainly along the \oii/\nii\ axis. 

Although we attempt to select a calibrator most suitable for our purpose, the reality is that every metallicity calibrator has its pros and cons. We choose O2N2-based metallicity calibrator here because it has a weak dependence on ionization parameter. However, its main weakness is that the wider wavelength range of the emission lines used makes it more sensitive to dust extinction. Also, the photoionization models for this indicator assume a relation between O/H and N/O abundance ratios \citep{Dopita2016}. Therefore, while we select O2N2-based calibrator carefully for the aforementioned reasons, we also reproduce our key results using many other metallicity calibrators in the Appendix. The conclusions of this paper do not change with different calibrators, and the O2N2-based calibrator we choose turns out to be representative among all calibrators we test (see the Appendix). Therefore, in the rest of the paper, we present the results using the O2N2-based metallicity calibrator.

\subsubsection{Definition of anomalously-low-metallicity regions}
\label{sec:def-lowZ}

\begin{figure}[htbp] %  figure placement: here, top, bottom, or page
   \centering
   \includegraphics[width=3.5in]{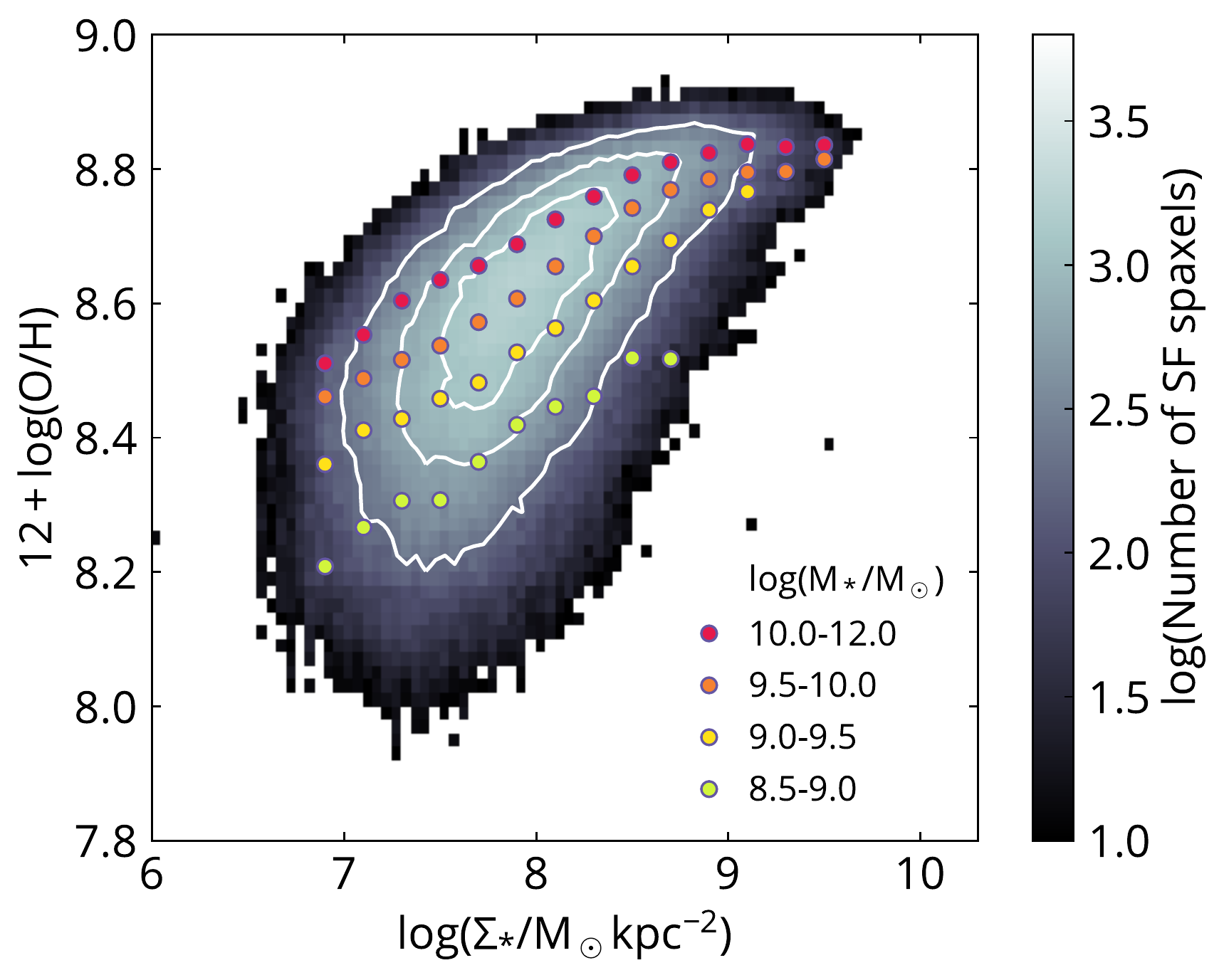} 
   \caption{The local relation between metallicity and stellar surface mass density. The symbols represent the peaks of metallicity distributions in each stellar mass bin. The stellar mass bin $10^{8.5-9}$\Msun\ is used for interpolation only and these galaxies do not enter the sample for further analysis. The background distribution and contours show the distribution of all pure star-forming spaxels in late-type galaxies with stellar masses $>10^{8.5}$\Msun.}
   \label{fig:new-MZrelation}
\end{figure}

\begin{figure}[htbp] %  figure placement: here, top, bottom, or page
   \centering
   \plotone{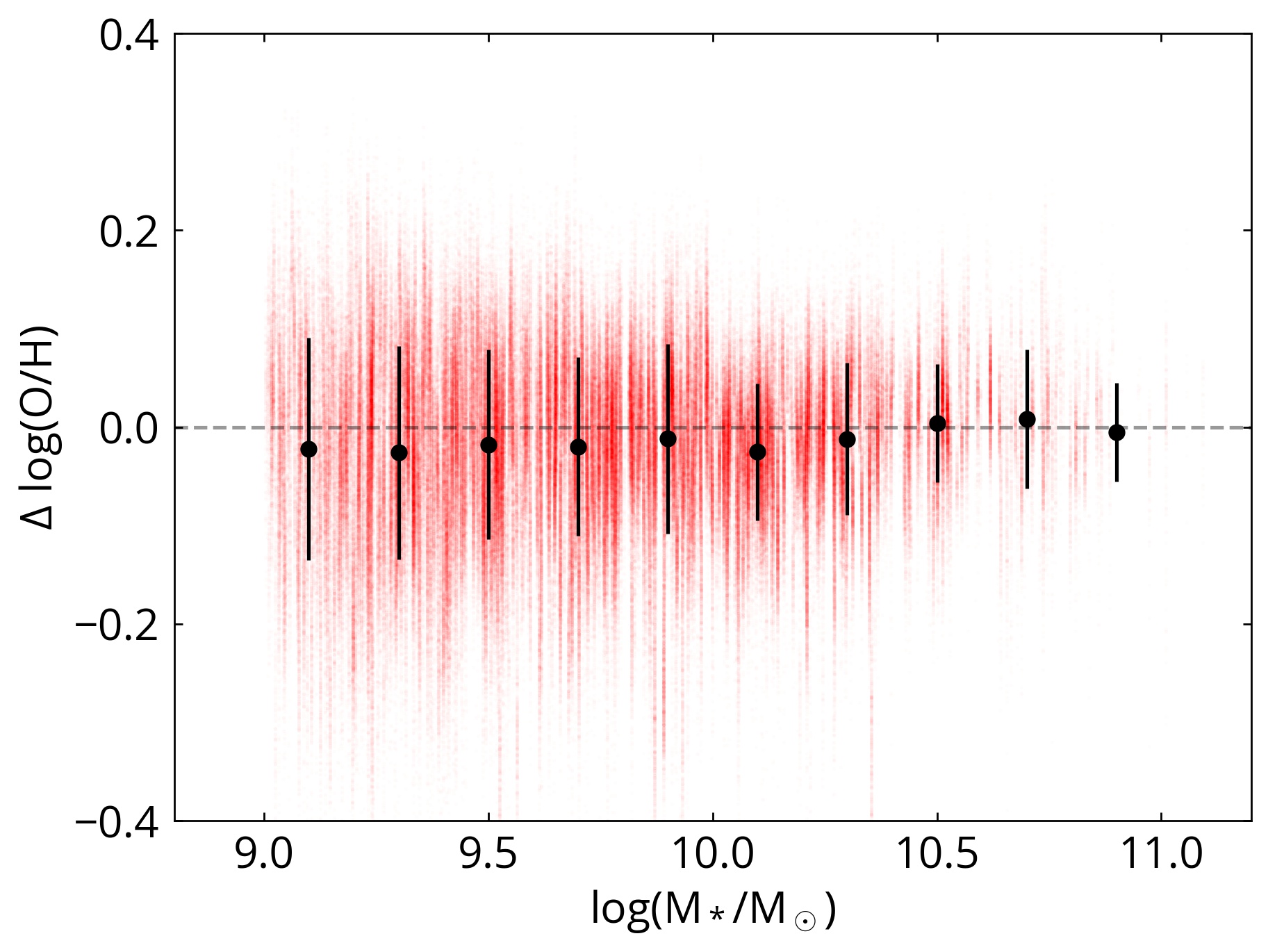}
   \caption{The metallicity deviation $\Delta \log(\rm{O/H})$ with respect to stellar mass. The red points are all star-forming spaxels in the sample, the black dots show the medians, and the bars represent the standard deviations. The median values slightly offset from $\Delta \log(\rm{O/H})=0$ because of the low-metallicity tail in the metallicity distribution (Fig.~\ref{fig:new-hist_dZ}). Metallicity deviation has nearly no dependence on stellar mass. }
   \label{fig:mass-dep}
\end{figure}

\begin{figure}[htbp] %  figure placement: here, top, bottom, or page
   \centering
   \plotone{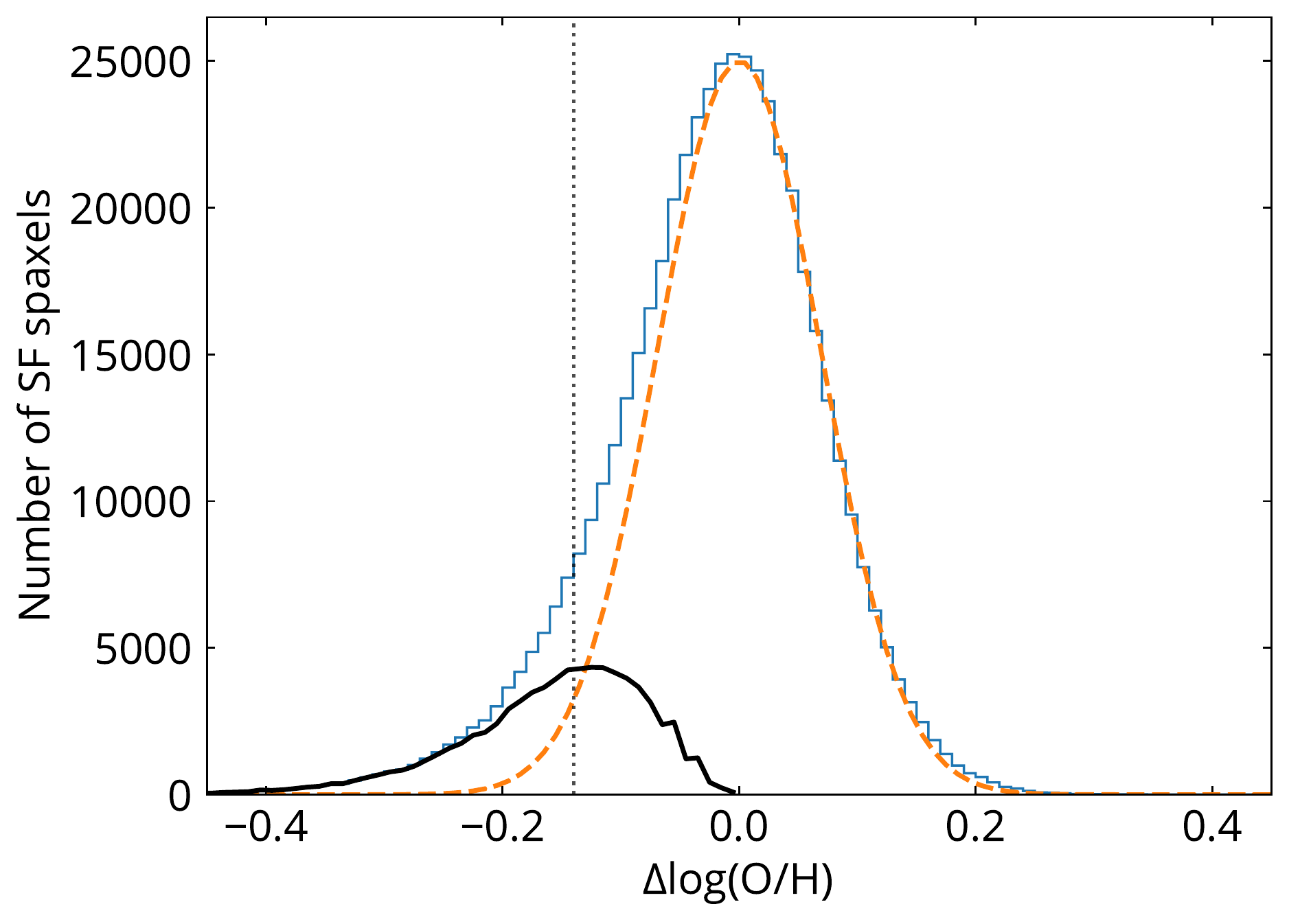}
   \caption{The distribution of metallicity deviation for star-forming spaxels in late-type galaxies with stellar mass $>10^{9}$\Msun\ (blue histogram). Clearly, there is a low-metallicity tail. The orange dashed line is a Gaussian profile for `normal' spaxels, with a standard deviation of $\sigma_{\rm Z} = 0.07$. The black solid line is the residual by subtracting the negative side of the histogram from the fitted Gaussian profile. The residual intersects with the fitted Gaussian at $\Delta\ \log({\rm O/H})=-0.13$. We refer those spaxels deviating to lower metallicity by more than $0.14$\,dex (the vertical dotted line) as anomalously-low-metallicity (ALM) spaxels.  } 
   \label{fig:new-hist_dZ}
\end{figure}

\begin{figure}[htbp] %  figure placement: here, top, bottom, or page
   \centering
   \includegraphics[width=3.5in]{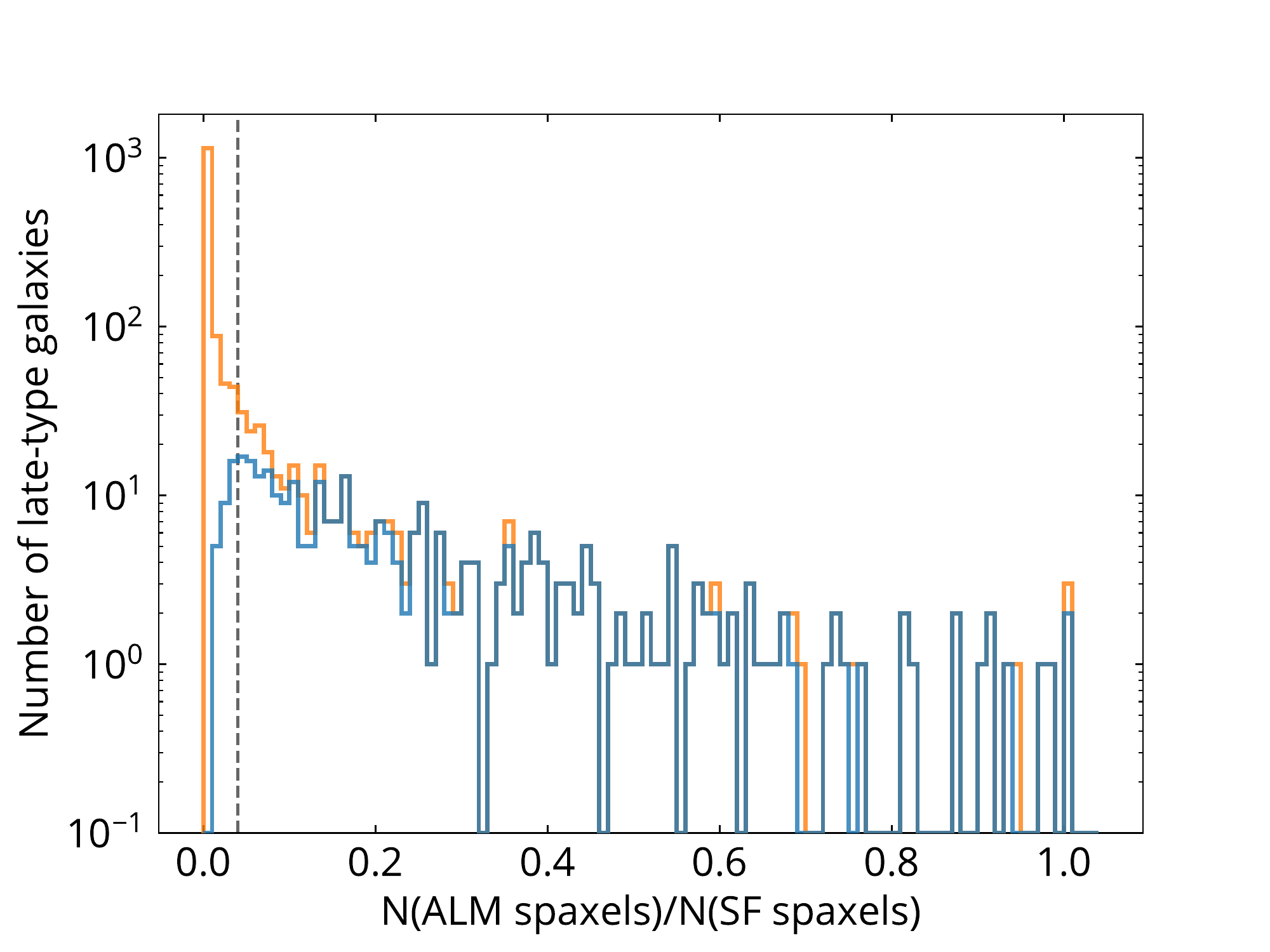} 
   \caption{Histograms of the fraction of star-forming spaxels being classified as anomalously-low-metallicity (ALM) in each galaxy. The orange (higher) histogram represents all late-type galaxies, while the blue (lower) histogram refers to those late-type galaxies having more than 20 ALM spaxels. The dashed line located at $0.04$ is the fraction used as the criterion in this paper to define galaxies having ALM regions.} 
   \label{fig:new-hist_lowZfrac}
\end{figure}

We identify anomalously low-metallicity regions by comparing the observed metallicity to the expected metallicity based on existing empirical correlations. Central oxygen abundance increases with galaxy stellar mass and gradually saturates at the high-mass end \citep{Tremonti2004}. A similar correlation is found at $\sim$\,kpc scales: the local metallicity is also correlated with $\Sigma_*$, the local stellar surface mass density \citep{Moran2012, Rosales-Ortega2012}. The local $\Sigma_*$-Z relation may explain the global M-Z relation, but the local $\Sigma_*$-Z relation in low-mass ($\lesssim10^{9.5}$\Msun) galaxies systematically deviates to lower metallicities \citep{Barrera-Ballesteros2016}. Therefore, to predict the local metallicity, at least two parameters are required: total stellar mass and the local stellar surface mass density. 

%\blue{Alternatively, rather than surface mass density, one can use metallicity radial profiles or metallicity gradients. In principle, they should give similar results.} 

For every star-forming spaxel, the expected metallicity is computed based on the correlations with total stellar mass and local surface mass density. As shown in Fig.~\ref{fig:new-MZrelation}, for each stellar mass bin, we correlate local metallicity with local stellar surface mass density. Specifically, for late-type galaxies in a certain stellar mass bin, we select the pure star-forming spaxels that have measured metallicities. Then for each surface mass density bin of these spaxels, we compute the `mode' of the metallicity distribution. We use the mode because the distribution has a low-metallicity tail (like Fig.~\ref{fig:new-hist_dZ}) such that the median values are offset from the peaks. The `mode' of the metallicity distribution is derived using the peak of a fitted skew Gaussian profile which takes the low-metallicity tail into account. The `modes' in each stellar mass bin and surface mass bin are present as circles in Fig.~\ref{fig:new-MZrelation}. The step of stellar mass bin is $0.5$\,dex, except for the bin of $10^{10-12}$\Msun\ because fewer galaxies have masses $>10^{10.5}$\Msun\ and because the relation between metallicity and surface mass density does not vary significantly in this mass range. 

The shape of the $\Sigma_*$-Z relation depends on the metallicity calibrator used. For example, using the O3N2 calibrator from \cite{Marino2013} gives a flatter $\Sigma_*$-Z relation on the high surface mass density end \citep{Barrera-Ballesteros2016}. But, as detailed later, because we are computing the metallicity difference at a fixed surface mass density, the overall shape of the $\Sigma_*$-Z relation does not affect our result. Some key results using different calibrators are shown in the Appendix. 

We do not restrict our sample to non-interacting galaxies when we derive the $\Sigma_*$-Z relation in Fig.~\ref{fig:new-MZrelation}. In fact, the result is nearly the same if we exclude them. This is because (1) interacting galaxies (mergers and close pairs) only constitutes 18\% of the star-forming galaxies, and (2) we use the `modes' to construct the $\Sigma_*$-Z relation. While the mean values are easily affected by outliers, the `mode' of the distribution is determined by the entire population so it remains almost unchanged. 
 
The expected metallicity $(12+\log(\rm{O/H}))_{exp}$ is computed via the interpolation of both stellar mass and surface mass density from the symbols in Fig.~\ref{fig:new-MZrelation}. For a given a galaxy, we first interpolate the expected local $\Sigma_*$-Z relation for its total stellar mass. During the interpolation, each stellar mass bin represents its central value of the range. For example, the local $\Sigma_*$-Z relation for the stellar mass bin $10^{9-9.5}$\Msun\ represents the relation for a $10^{9.25}$\Msun\ galaxy. The exception is the most massive bin of $10^{10-12}$\Msun\ that represents a $10^{10.25}$\Msun\ galaxy (since most galaxies in this bin are in the range of $10^{10-10.5}$\Msun). The bin of $10^{8.5-9}$\Msun\ is only used for interpolation, and we do not further analyze the galaxies in this bin. In the stellar mass bin $10^{8.5-9}$ and $10^{9-9.5}$\Msun, galaxies do not cover the high surface density end ($\gtrsim10^{8.8}$\Usigma), and we assume the same metallicities as their highest surface density bins for the uncovered bins during the interpolation. Therefore, the local $\Sigma_*$-Z relation for a galaxy of a specific total stellar mass is interpolated from these points. With the local $\Sigma_*$-Z relation for this galaxy, the metallicity expected for every spaxel is derived based on its surface mass density using interpolation. 

The metallicity deviation is computed as $\Delta \log(\rm{O/H}) = (12+\log(\rm{O/H}))_{obs} - (12+\log(\rm{O/H}))_{exp}$. Fig.~\ref{fig:mass-dep} presents the metallicity deviation as a function of stellar mass. It shows that there is no dependence on stellar mass over the range $10^{9-11}$\Msun, which is the stellar mass range of interest in this paper.

The blue histogram in Fig.~\ref{fig:new-hist_dZ} shows the distribution of metallicity deviation for the entire sample of spaxels. Because of the use of `modes' instead of medians in individual bins, the peak of the distribution is well located at $\sim0$ as expected. The distribution is notably asymmetric, with a tail extending to low-metallicity and no corresponding feature on the high-metallicity side. 

We fit the metallicity deviation distribution in Fig.~\ref{fig:new-hist_dZ} with two components: anomalously-low-metallicity spaxels whose metallicities are lowered by some mechanisms, and `normal' spaxels whose metallicities follow the local $\Sigma_*$-Z relation very well. Therefore, the positive side of the metallicity deviation distribution is dominated by `normal' spaxels. We determine $\sigma_{\rm Z}$, the standard deviation for `normal' spaxels, by fitting a Gaussian profile to the positive side and an assumed symmetric negative side. The best fit is shown as the dashed orange profile in Fig.~\ref{fig:new-hist_dZ}, giving $\sigma_{\rm Z} = 0.07$. The positive side of the metallicity deviation is well described by the best-fit Gaussian profile. 

The term $\sigma_{\rm Z}$ contains several effects: possible mechanisms that affect the metallicities, the intrinsic scatter of the metallicity calibrator, uncertainty in the interpolations, all measurement errors including $b/a$ ratios, total stellar mass, surface mass density, flux ratios, and extinction correction. Therefore, by using $\sigma_{\rm Z}$ as a criterion, we consider that this effectively take these uncertainties into account.

%The same procedure is used for the O3N2 and N2S2\Ha\ metallicity calibrators, and the resulting $\sigma_{\rm Z}$ are 0.055 and 0.10, respectively \red{(Sebastian: why O3N2 scatter smaller?)}. The O3N2 metallicity calibrator is not used because of its dependence on ionization parameter, as detailed in Sec.~\ref{sec:Z-calibrator}. Although the N2S2\Ha\ calibrator has a weak dependence on ionization parameter, its higher $\sigma_{\rm Z}$ implies a higher intrinsic scatter in the metallicity measurement compared to the O2N2 calibrator. For this reason, we use the O2N2 metallicity calibrator rather than N2S2\Ha, and adopt $\sigma_{\rm Z} = 0.07$ throughout the paper to characterize the spaxels with normal metallicities. \red{(This comparison may not be fair because the scale from different calibrators differ...) (do I need to move this paragraph downward?)}

To quantify the anomalously-low-metallicity spaxels in Fig.~\ref{fig:new-hist_dZ}, we derive the residual (the black solid line) by subtracting the fitted Gaussian profile of the normal spaxels from the total. The residual intersects with the fitted Gaussian profile at $\Delta({\rm O/H})=-0.13$. Motivated by this, we define the low-metallicity spaxels as those in which the observed metallicities are lower than the expected ones by more than $2\sigma_{\rm Z} = 0.14$, or $\Delta \log(\rm{O/H})/\sigma_{\rm Z} < -2$. {\it Hereafter, we will use the term `anomalously-low-metallicity (ALM)' to refer to those spaxels where $\Delta \log(\rm{O/H}) < -2\sigma_{\rm Z}$.}

With this criterion for ALM spaxels, here we further define the galaxies having ALM regions. The definition is motivated by the size of the PSF, and by the requirement that these low-metallicity spaxels need to constitute at least a certain fraction of the pure star-forming spaxels in a galaxy. Specifically, with a typical PSF of $2.5$\,arcsec ($\sim1.5$\,kpc at $z=0.03$) in FWHM and 0.5\,arcsec for the size of a spaxel, the angular area of the PSF corresponds to $\sim19.6$ spaxels. Therefore, we require that galaxies defined as having ALM regions have at least 20 ALM spaxels. To decide how to define the subset of galaxies with ALM spaxels, we use two other arguments. First, if the metallicity deviation of normal spaxels is a perfect Gaussian profile with a standard deviation of $\sigma_{\rm Z}$, statistically there is $\sim2.1$ per cent of normal spaxels satisfying the criterion for ALM spaxels. Second, in Fig.~\ref{fig:new-hist_lowZfrac} we show the distribution for the fraction of star-forming spaxels in a galaxy that are classified as ALM, which we call the ` ALM covering fraction'. The peak at 0 in late-type galaxies means that the majority of galaxies do not have ALM spaxels. The distribution then drops steeply from a fraction of 0 to 4 percent, and then levels off, indicating an additional component. 

Based on these two factors, we introduce a cut of 4 per cent as our criterion. Afterwards, we visually inspect the candidate galaxies having ALM regions, and exclude the galaxies where their ALM spaxels are randomly distributed and therefore no convincing ALM regions are identified. {\it In conclusion, galaxies having ALM regions are defined such that they have at least 20 ALM spaxels and that these ALM spaxels constitute more than 4 per cent of the star-forming spaxels in the galaxy.}

\section{Results}
\label{sec:result}

\begin{figure*}[htbp] %  figure placement: here, top, bottom, or page
   \centering
   \includegraphics[height=1.8in]{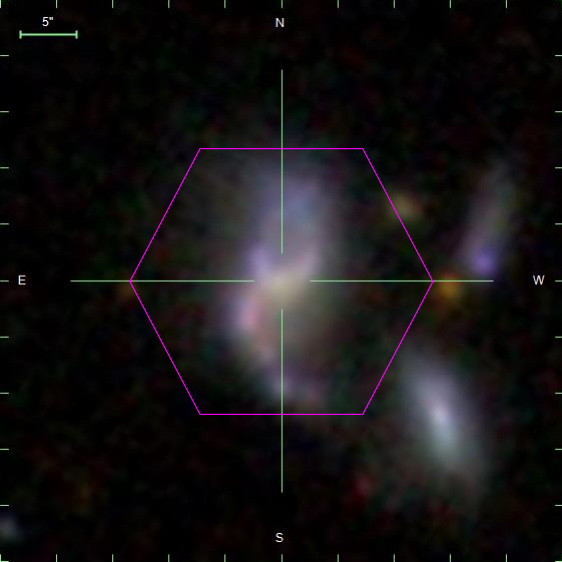}
   \includegraphics[height=1.8in]{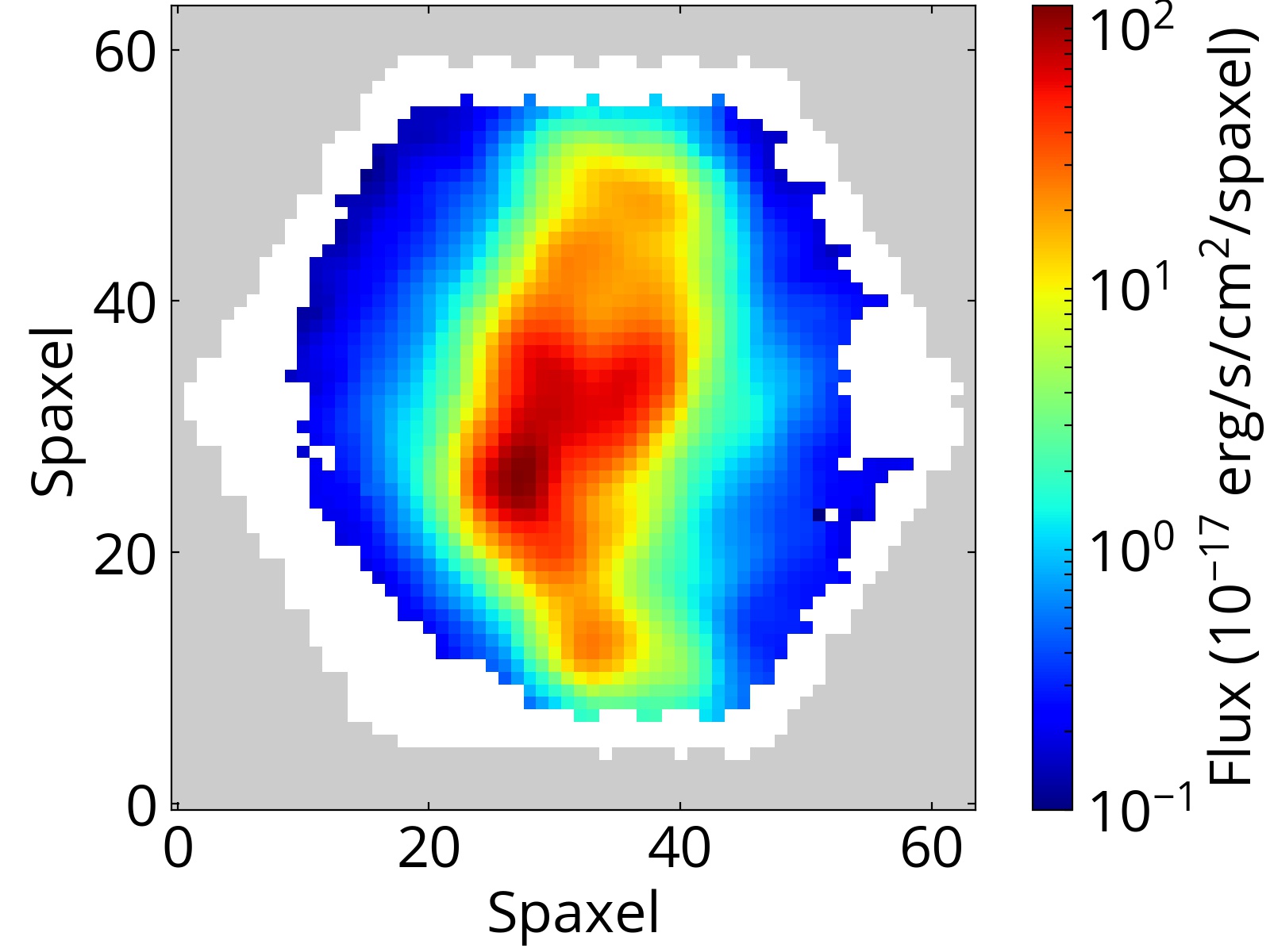}
   \includegraphics[height=1.8in]{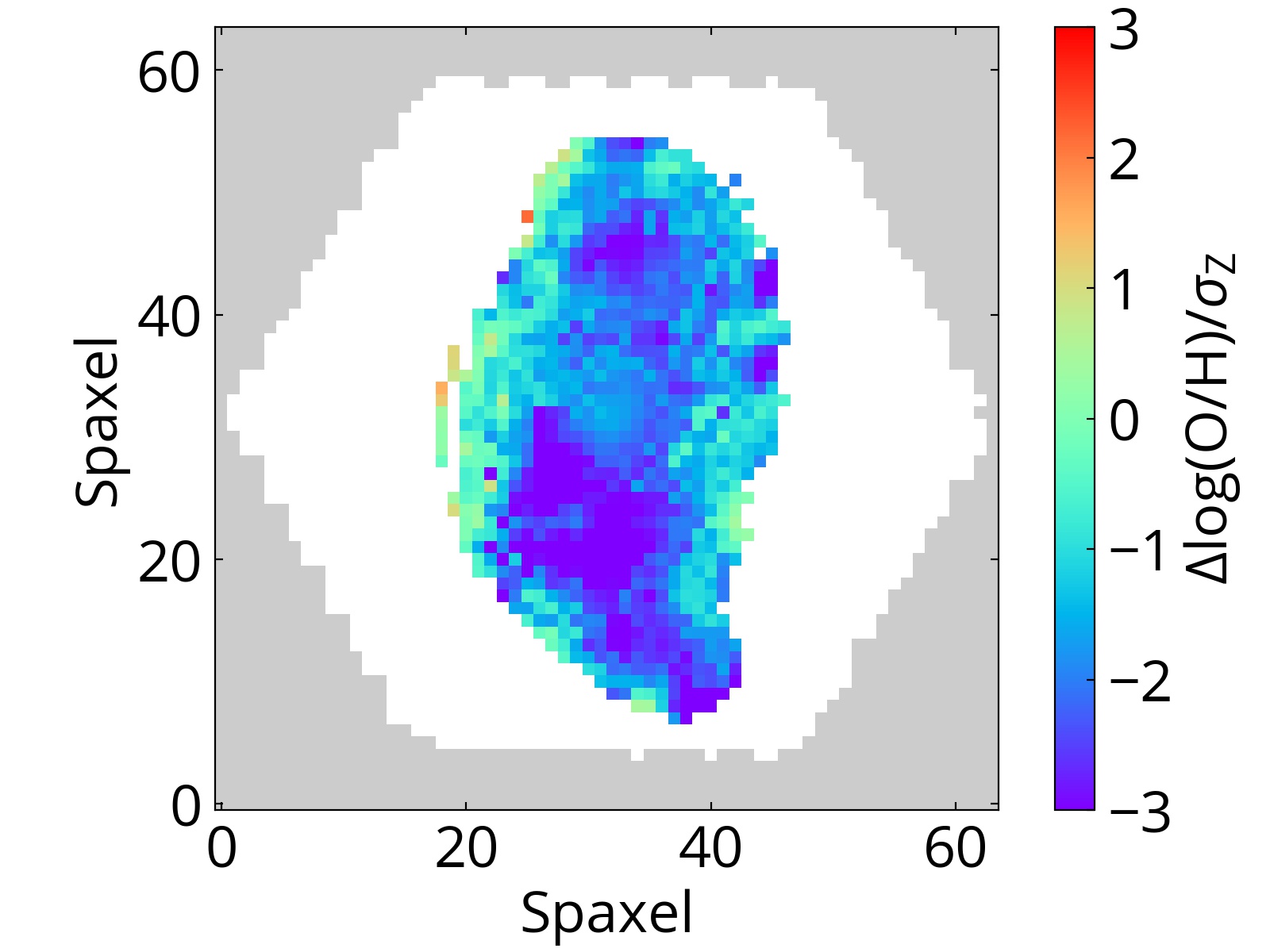}
   \includegraphics[height=1.8in]{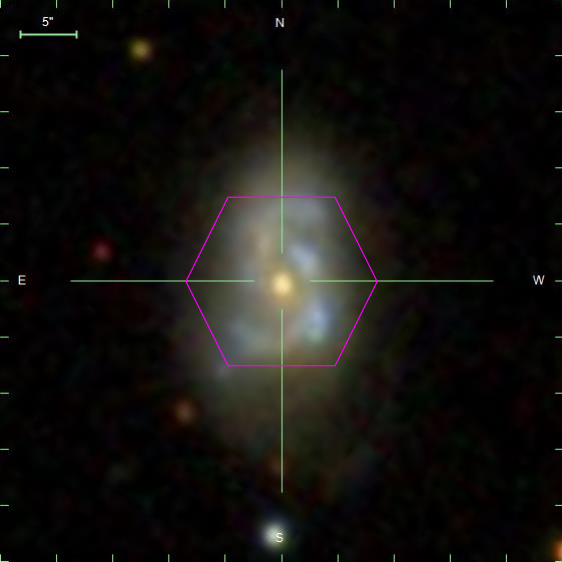}
   \includegraphics[height=1.8in]{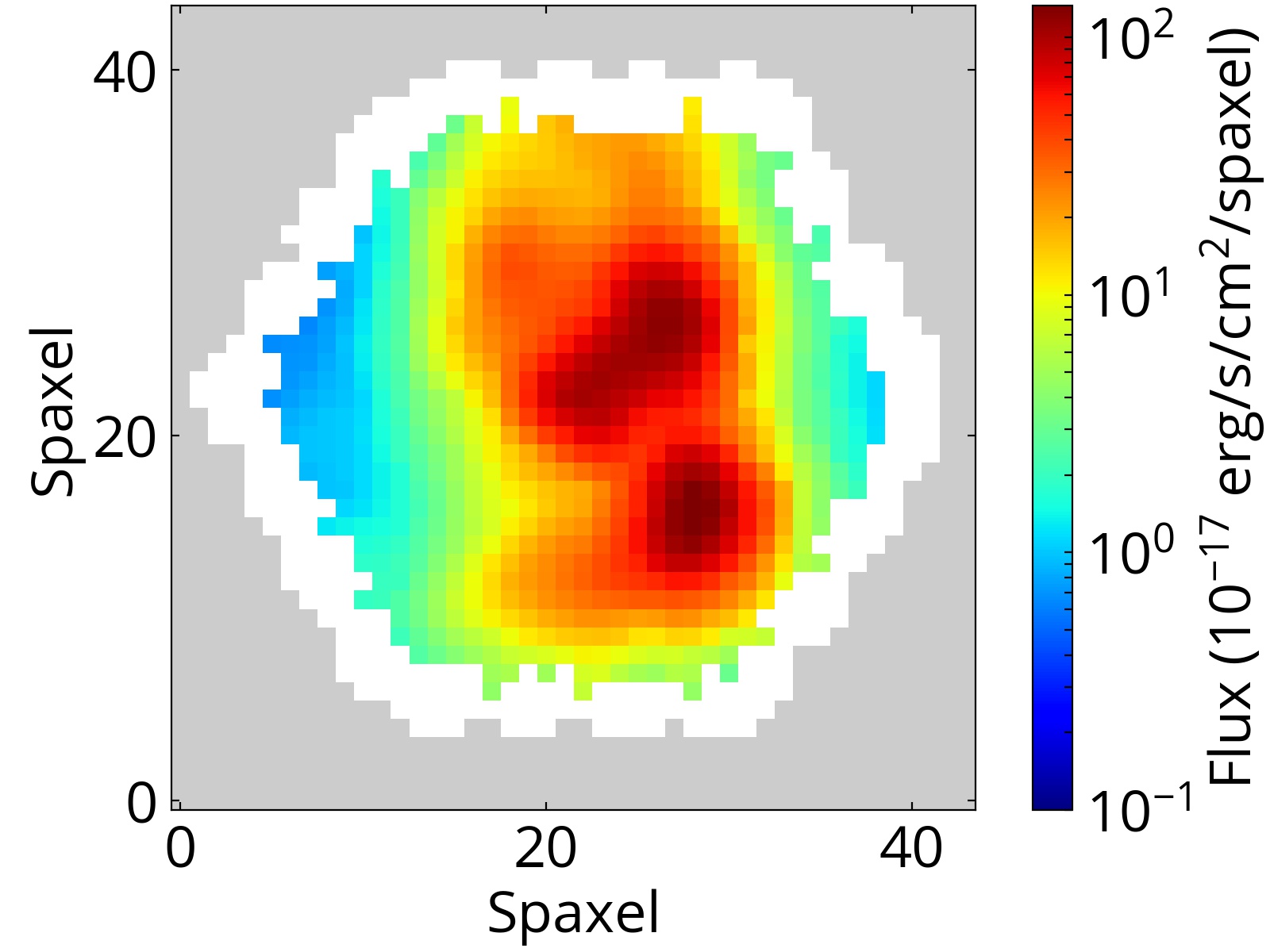}
   \includegraphics[height=1.8in]{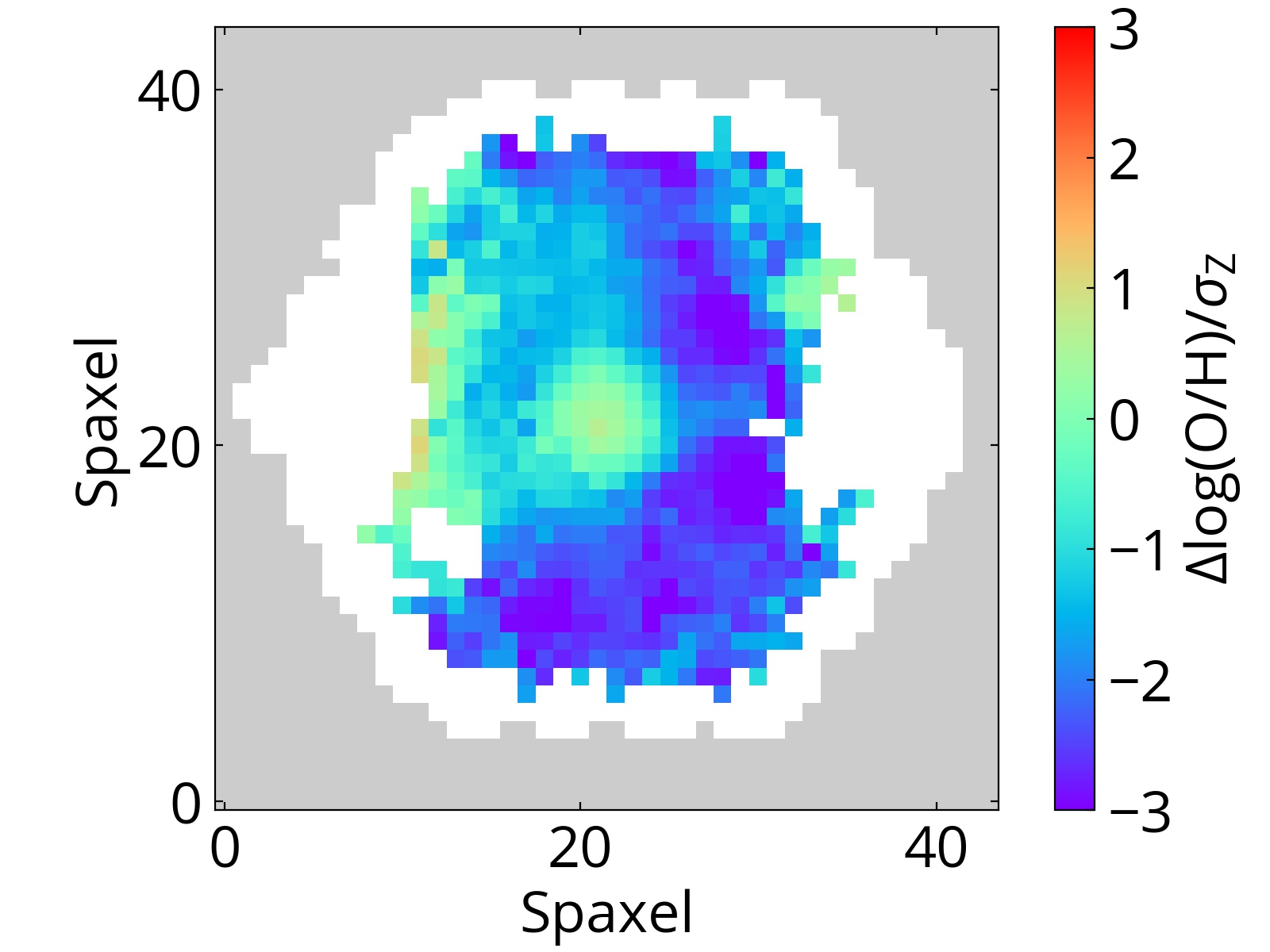}
   \includegraphics[height=1.8in]{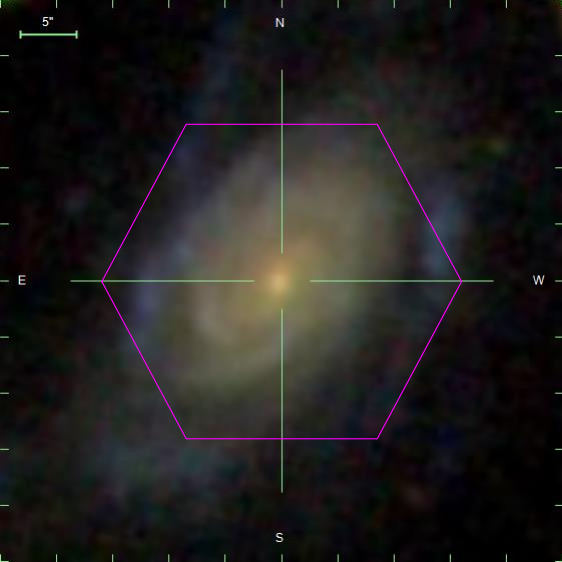}
   \includegraphics[height=1.8in]{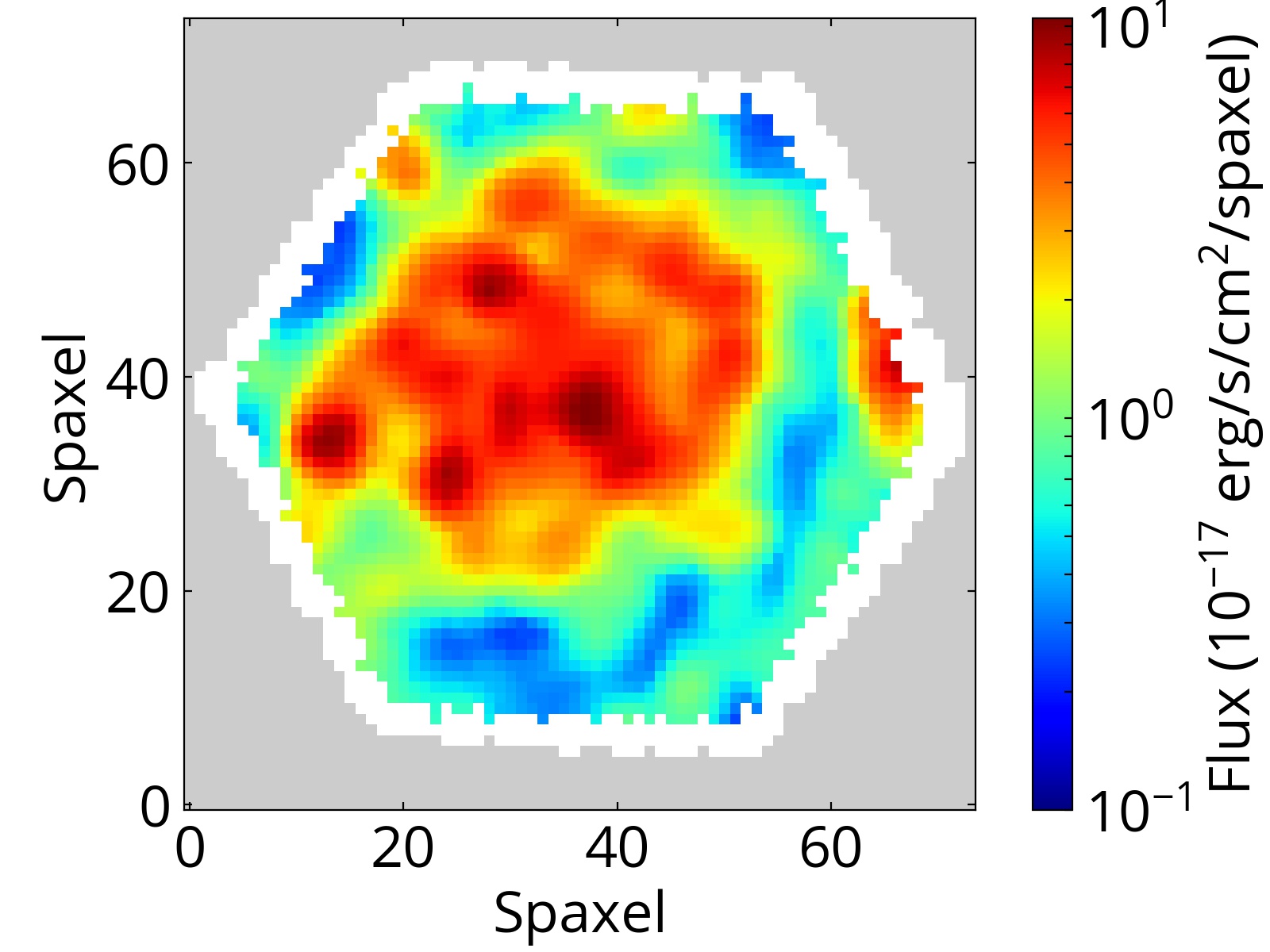}
   \includegraphics[height=1.8in]{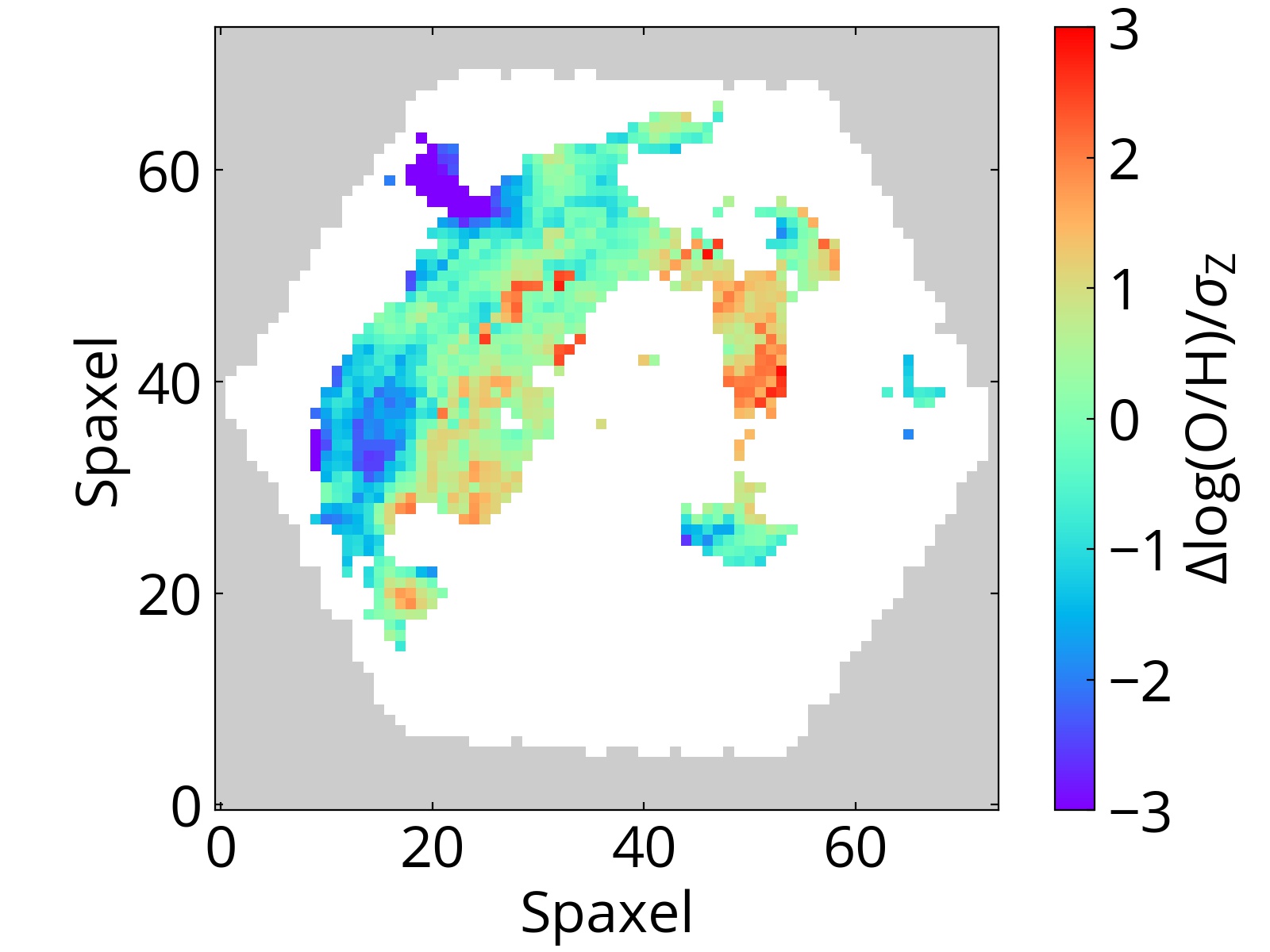}
      \includegraphics[height=1.8in]{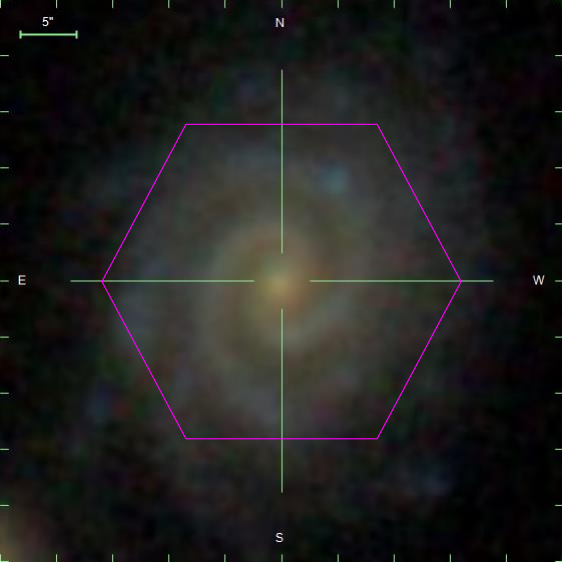}
   \includegraphics[height=1.8in]{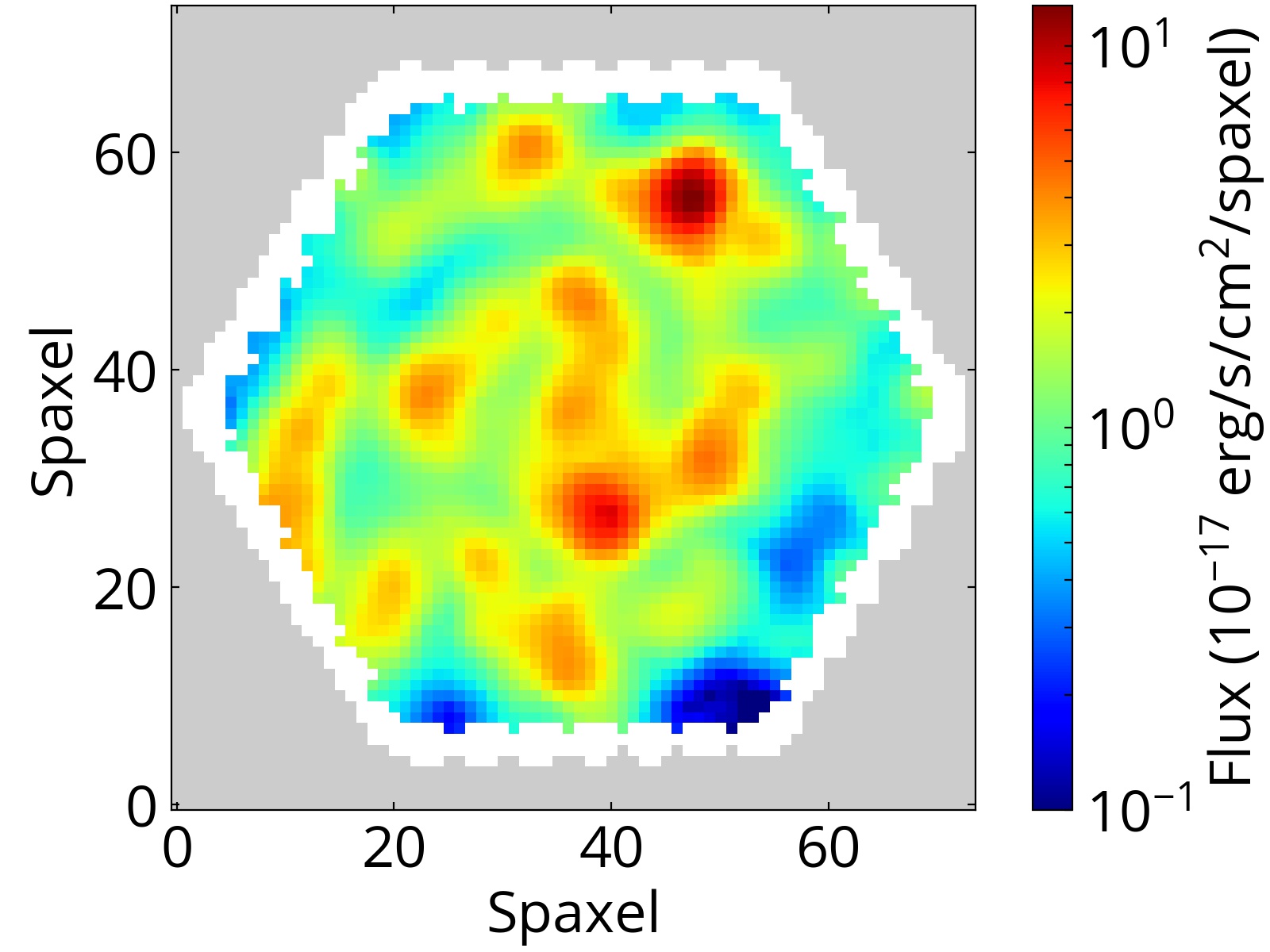}
   \includegraphics[height=1.8in]{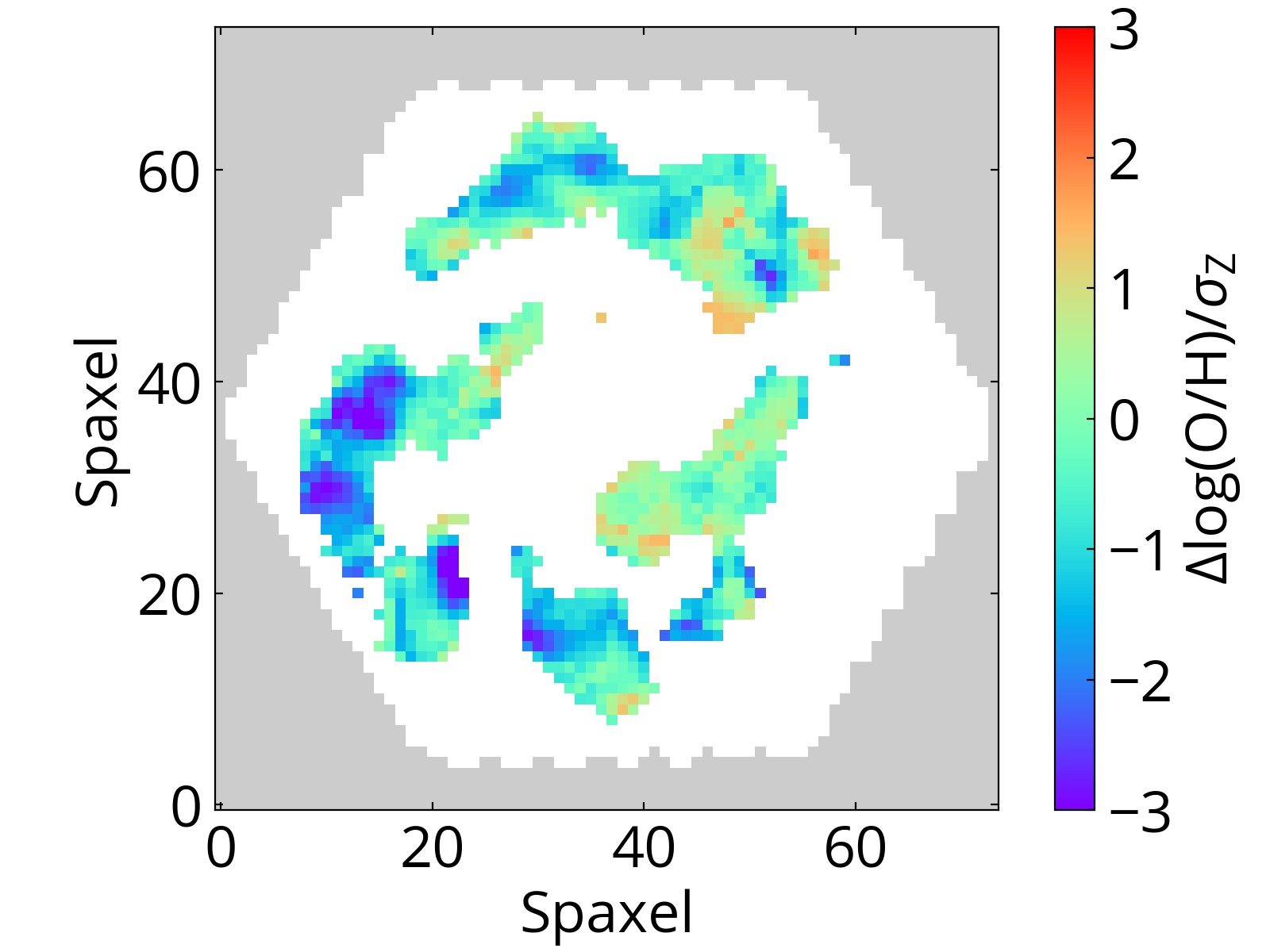}
   \includegraphics[height=1.8in]{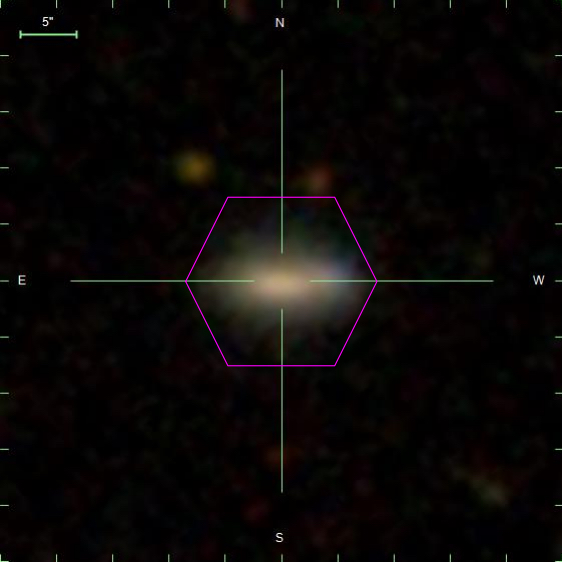}
   \includegraphics[height=1.8in]{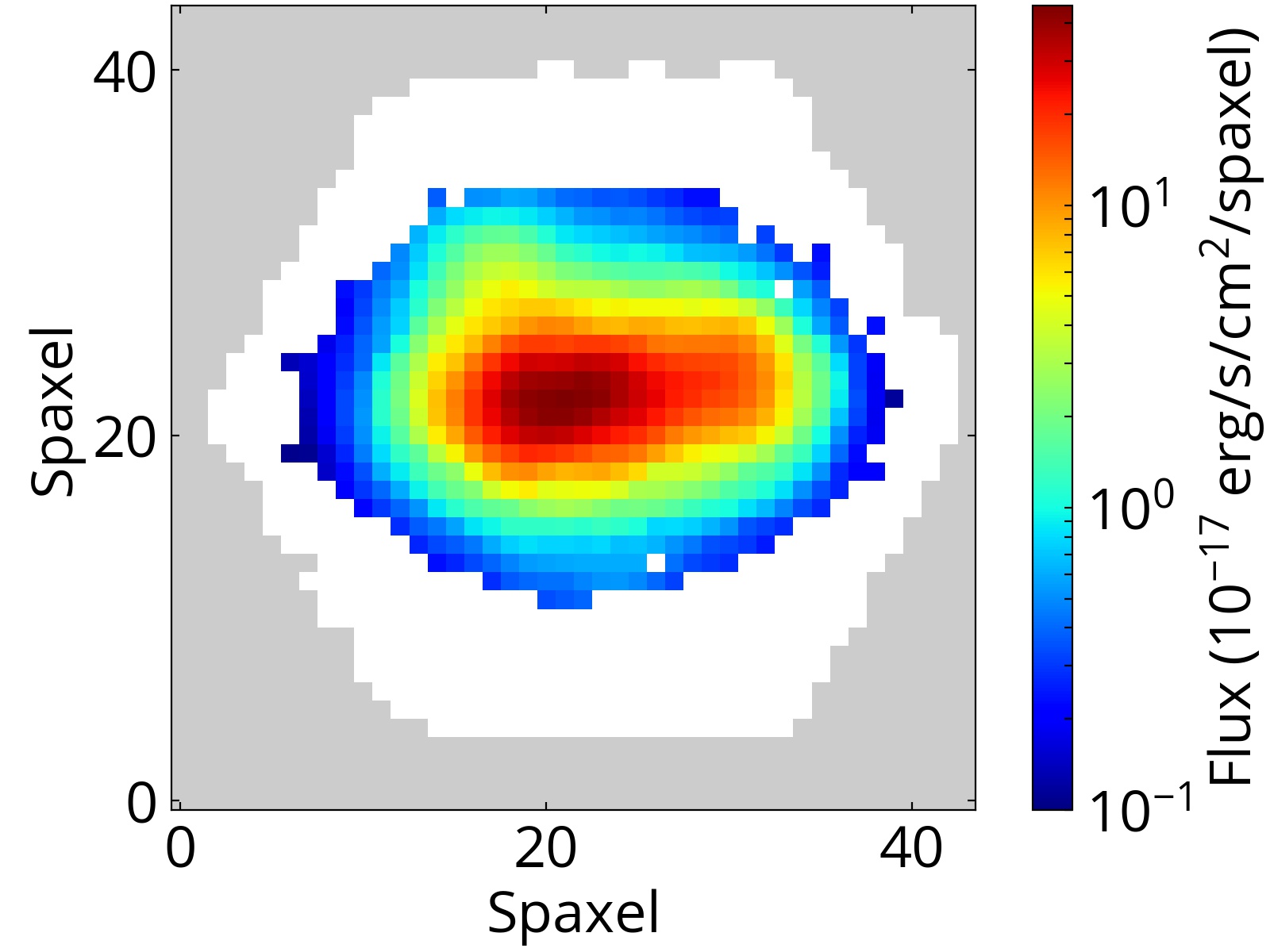}
   \includegraphics[height=1.8in]{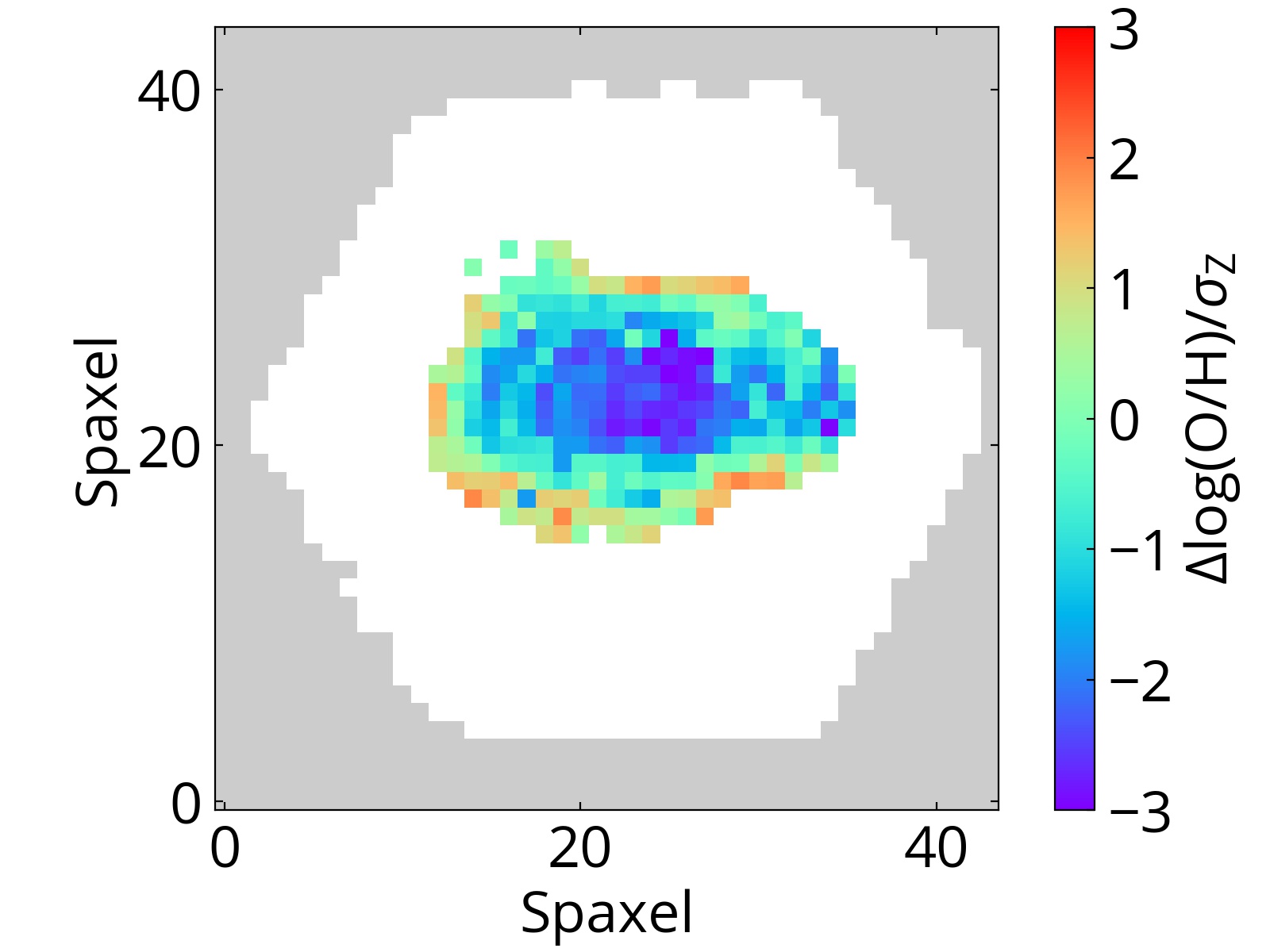}
   \caption{Examples of galaxies having ALM regions. MaNGA plate-IFU ID from top: 8252-9102, 9883-3701, 8466-12702, 8945-12705, 8551-3701. From left to right: SDSS optical images, \Ha\ maps, and metallicity deviation maps where $\sigma_{\rm Z}=0.07$\,dex. Some galaxies are associated with strong interaction, while some are isolated. The size of the spaxels is $0.5$\,arcsec on each side. Spaxels without metallicity measurement are either not purely star-forming regions, or have emission lines below the adopted S/N ratio cuts.} 
   \label{fig:new-example}
\end{figure*}

\begin{figure}[htbp] %  figure placement: here, top, bottom, or page
   \centering
   \includegraphics[width=3.2in]{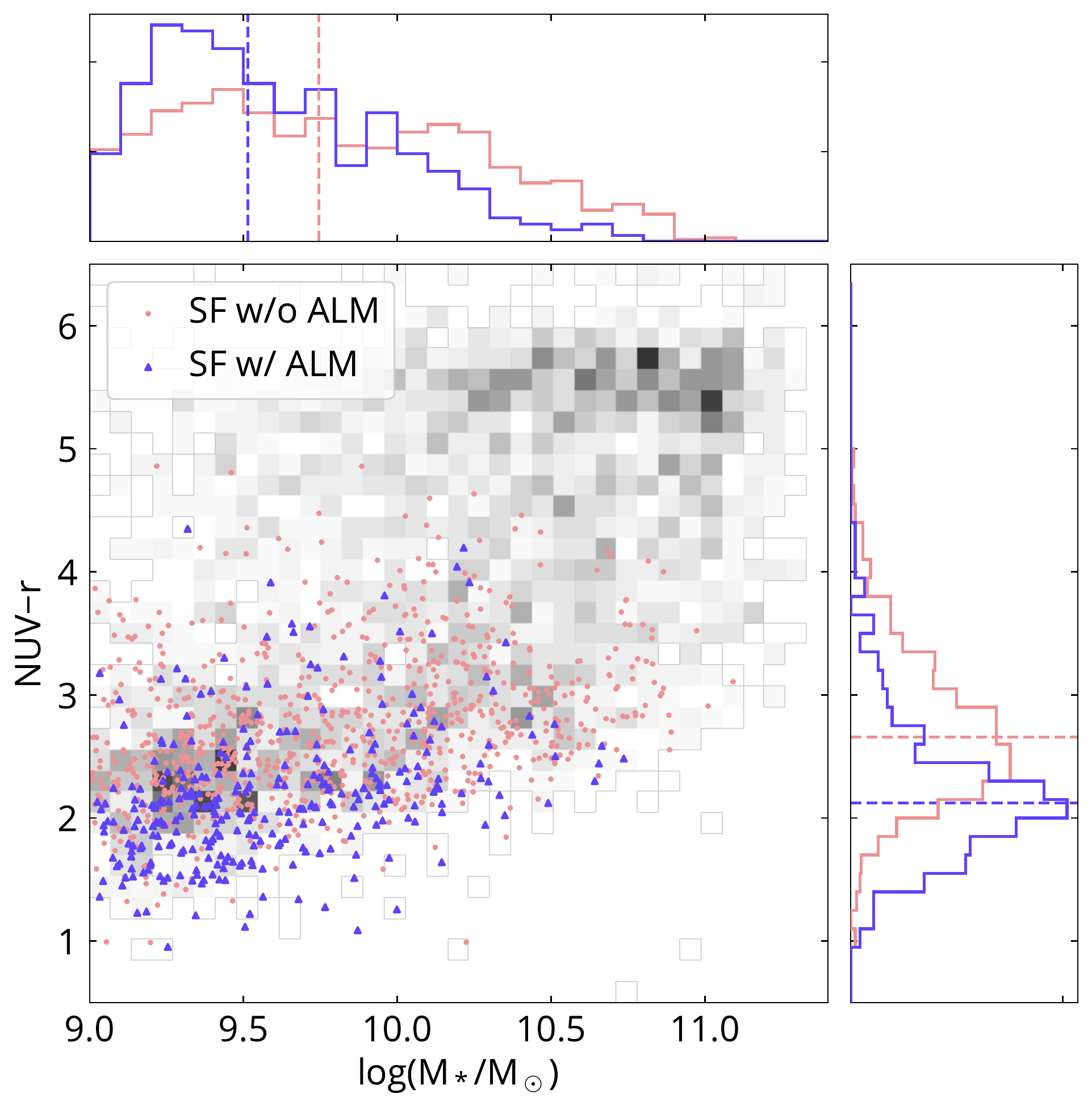} 
   \includegraphics[width=3.2in]{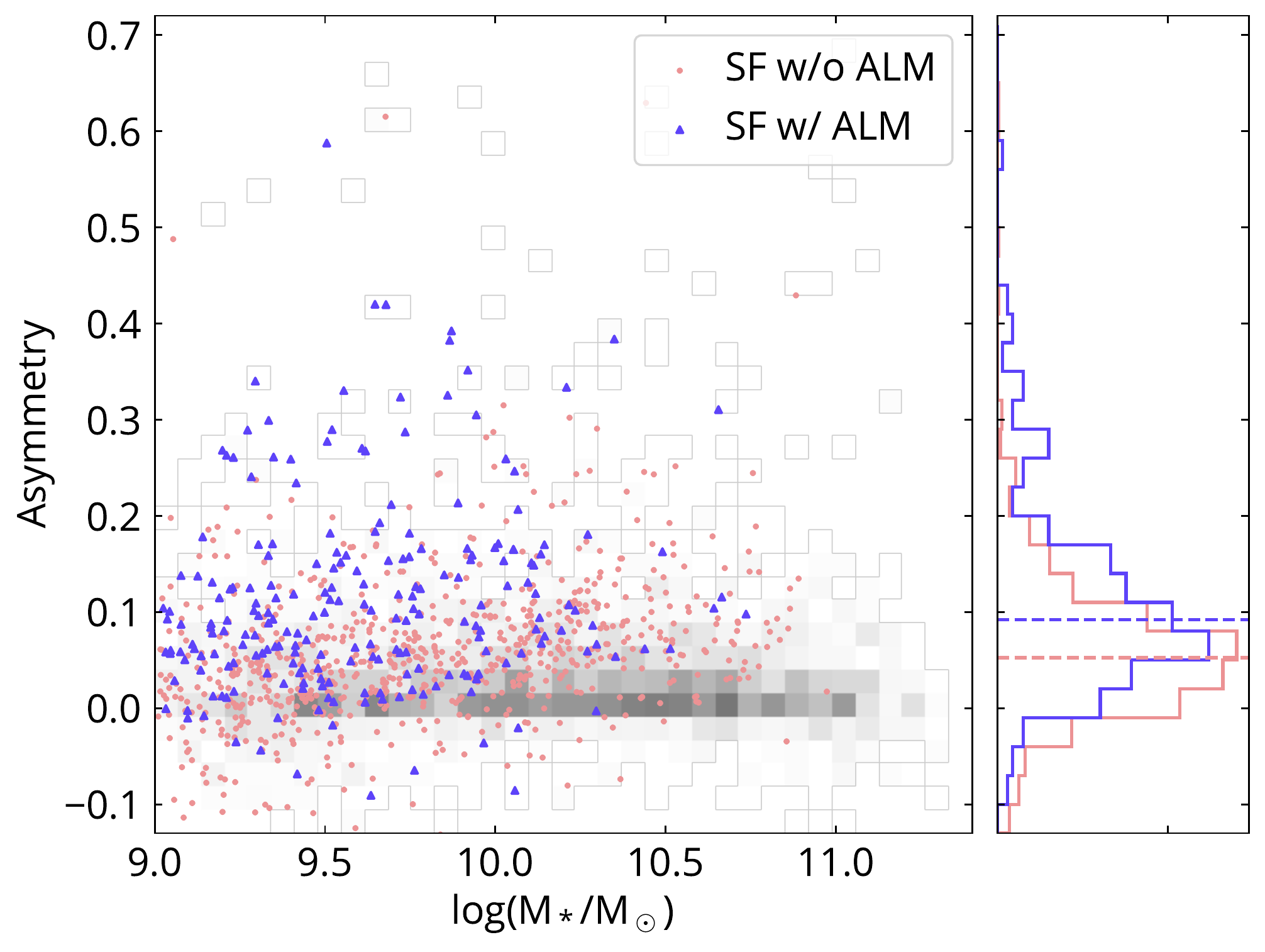} 
   \includegraphics[width=3.2in]{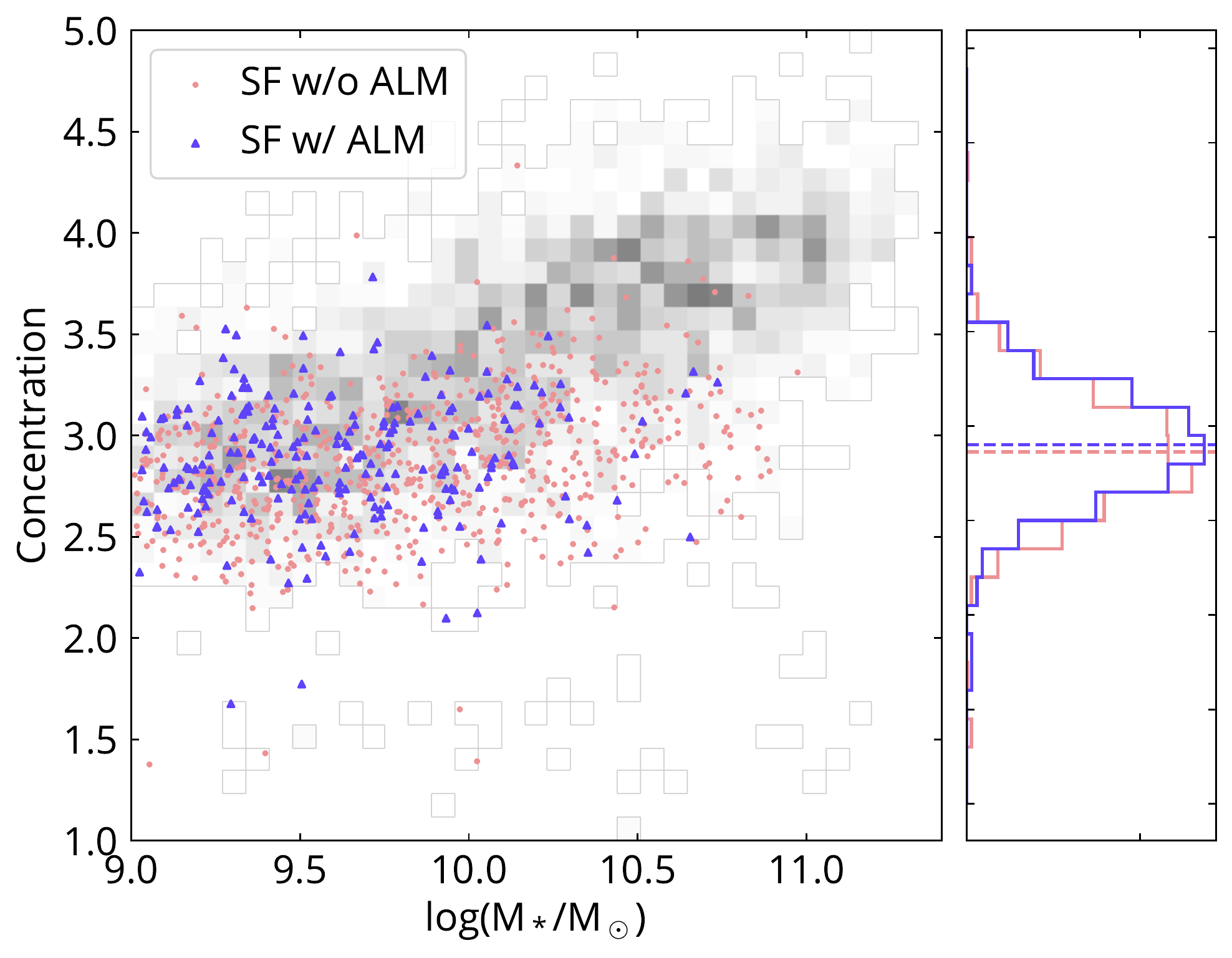} 
   \caption{Galaxy distributions of NUV$-r$ color, asymmetry index, and concentration index with respect to stellar mass. Blue triangles represent galaxies having ALM regions, and red dots are star-forming late-type galaxies without them. Background grey levels show the distribution of all MaNGA galaxies, including both late-type and early-type galaxies. For each parameter, we show the distributions of galaxies having ALM regions and the star-forming sample. The dashed lines represent the median. 
   } 
\label{fig:new-global-prop}
\end{figure}

\begin{figure*}[htbp]
\centering
   \includegraphics[width=3.2in]{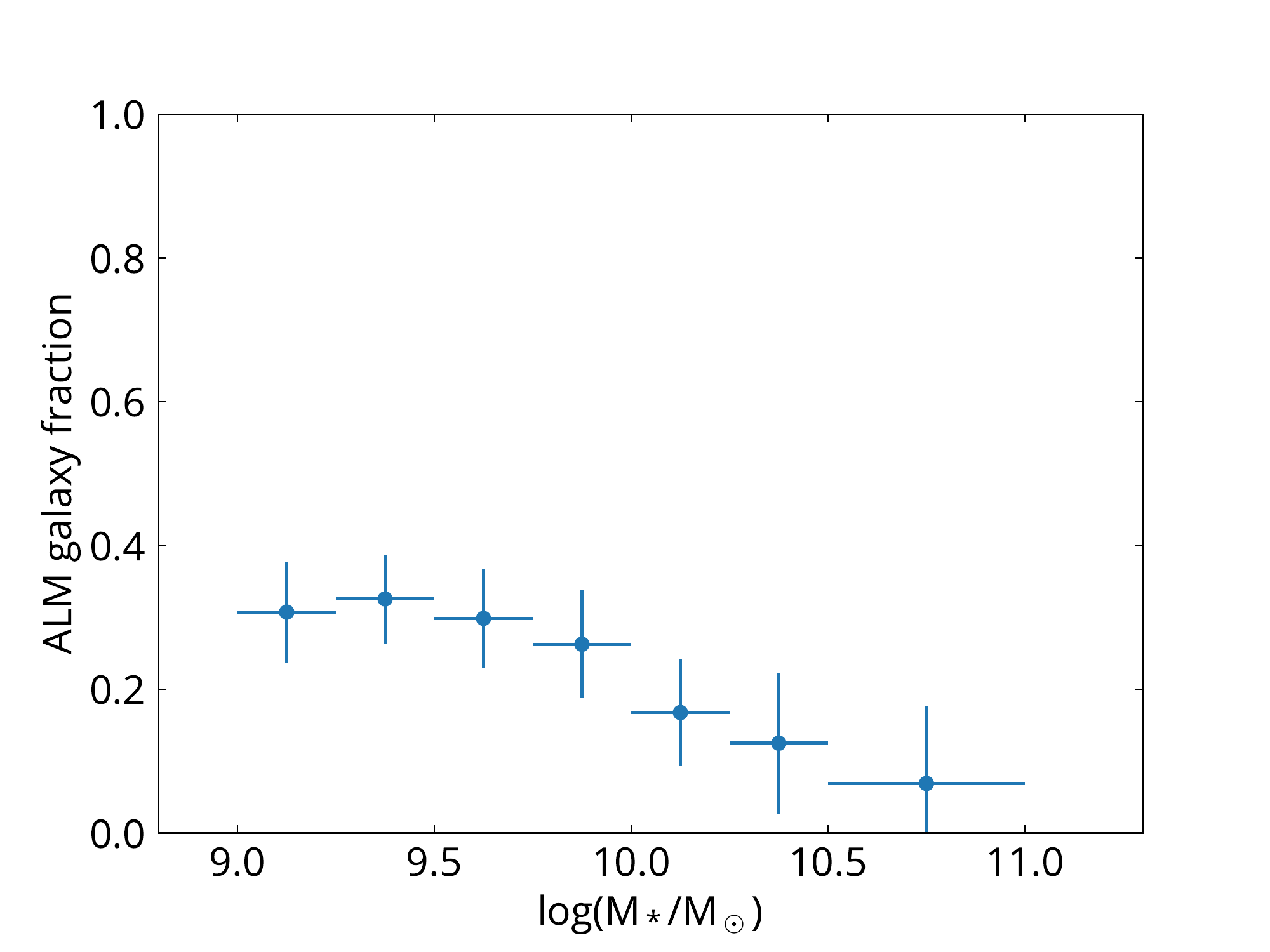} 
   \includegraphics[width=3.2in]{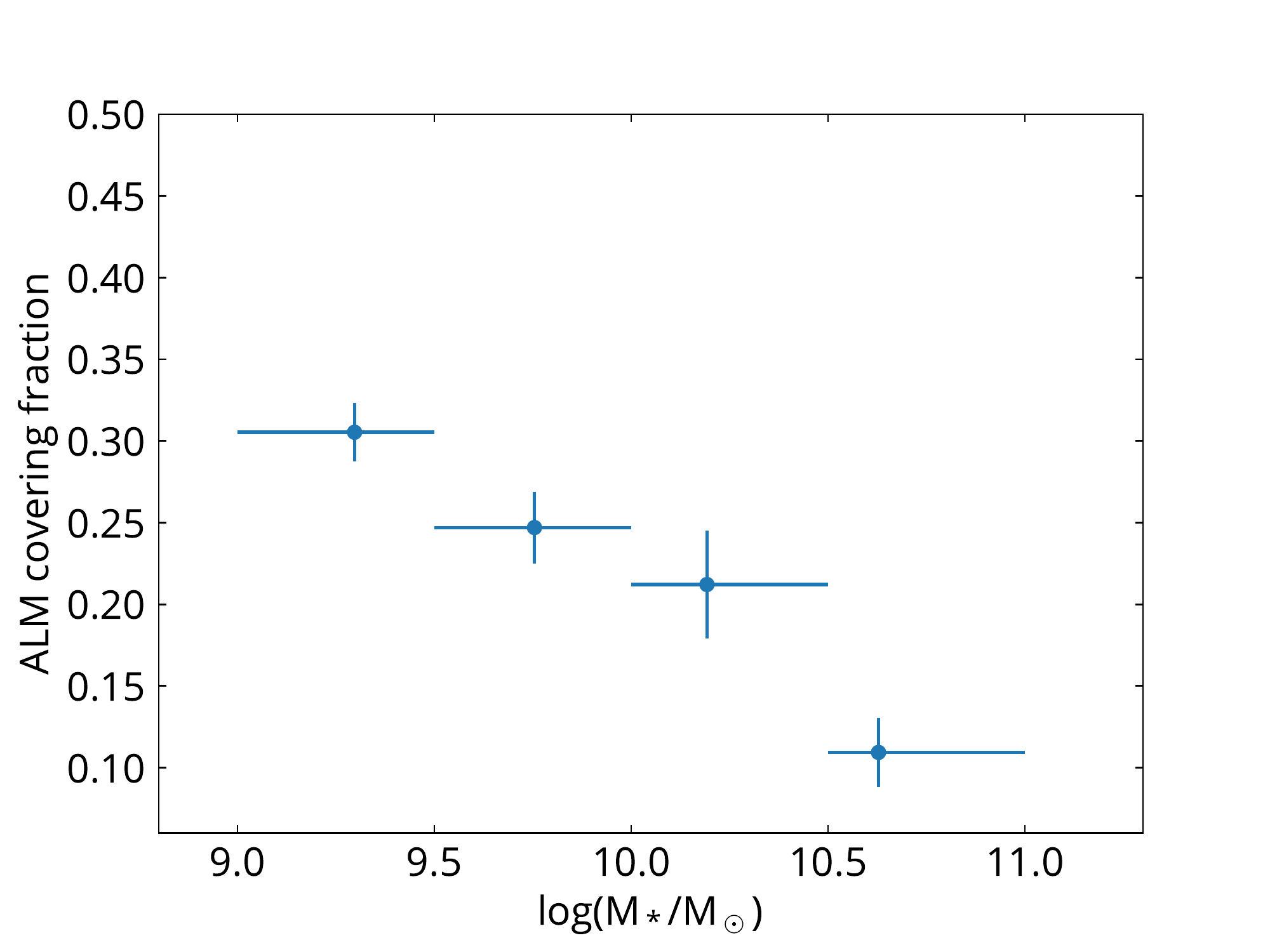} 
   \includegraphics[width=3.2in]{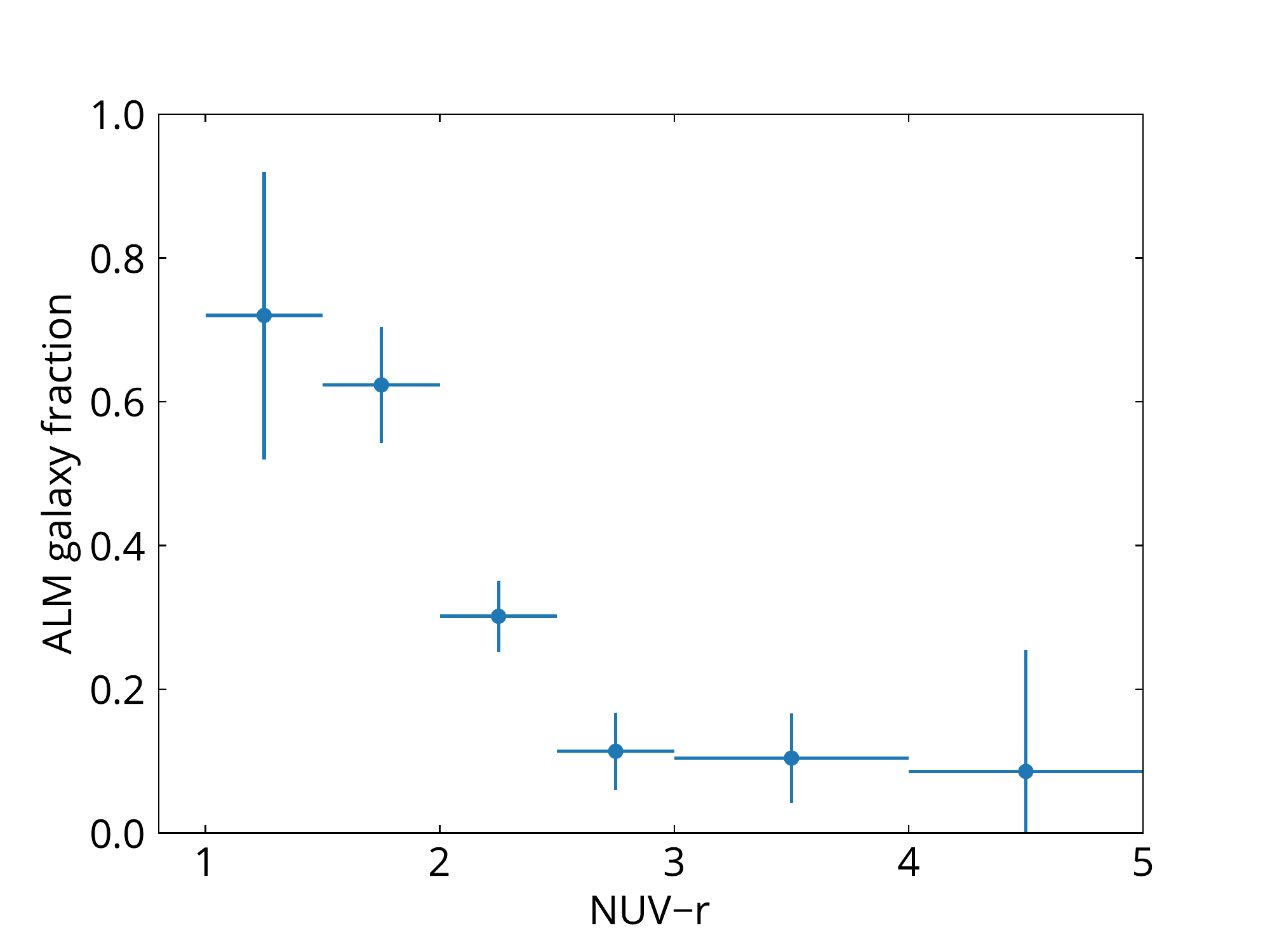} 
   \includegraphics[width=3.2in]{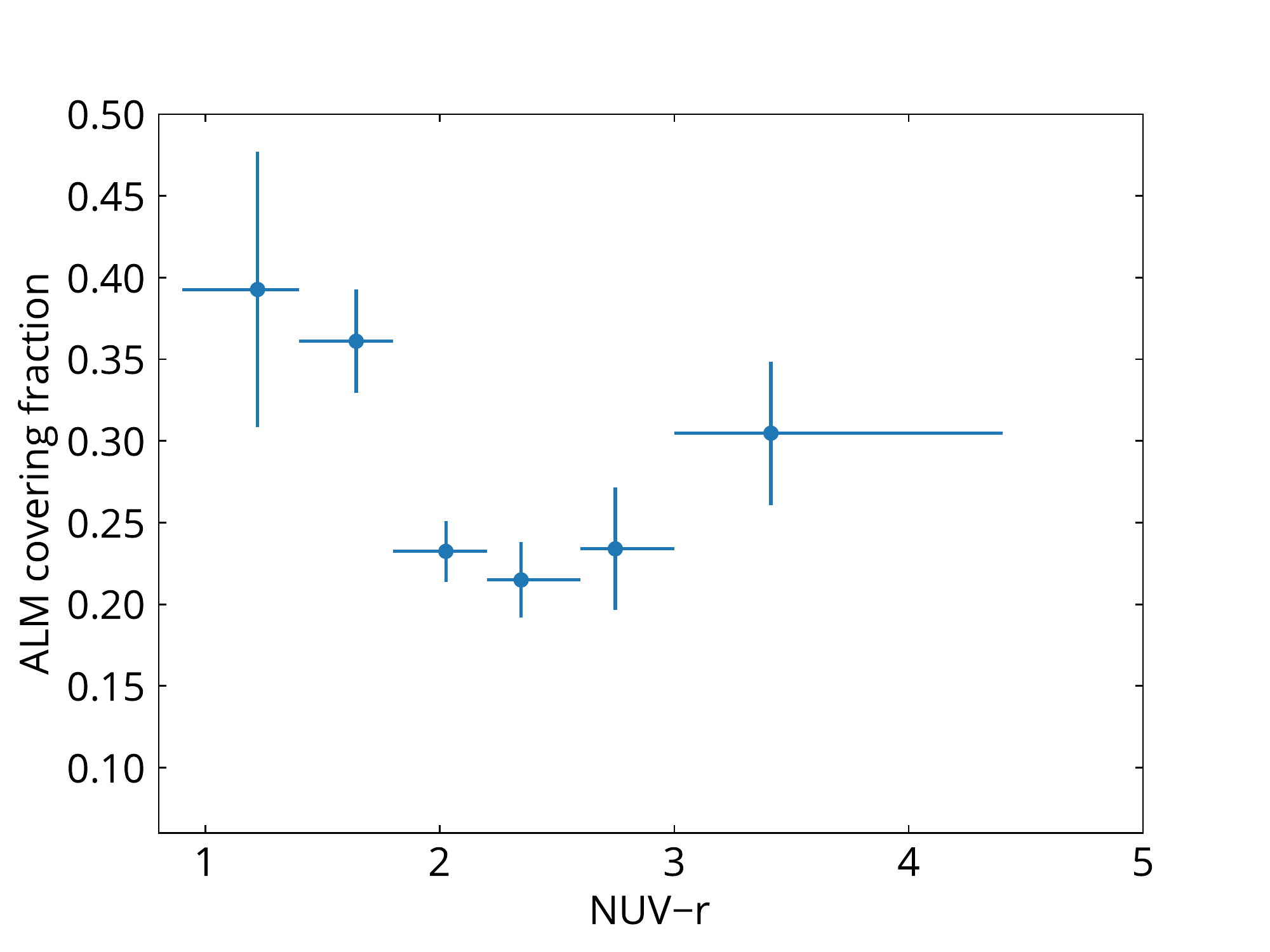} 
   \includegraphics[width=3.2in]{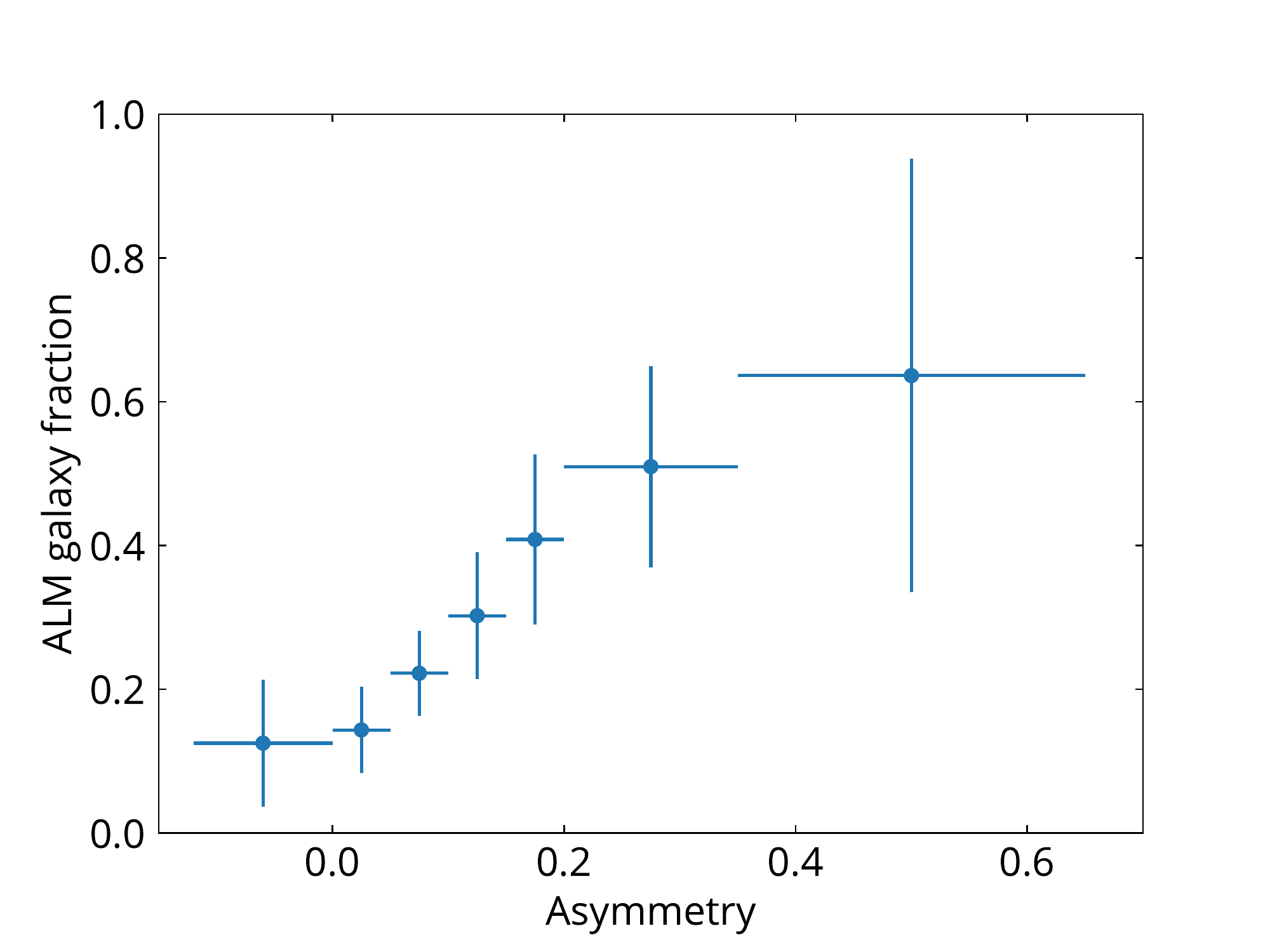} 
   \includegraphics[width=3.2in]{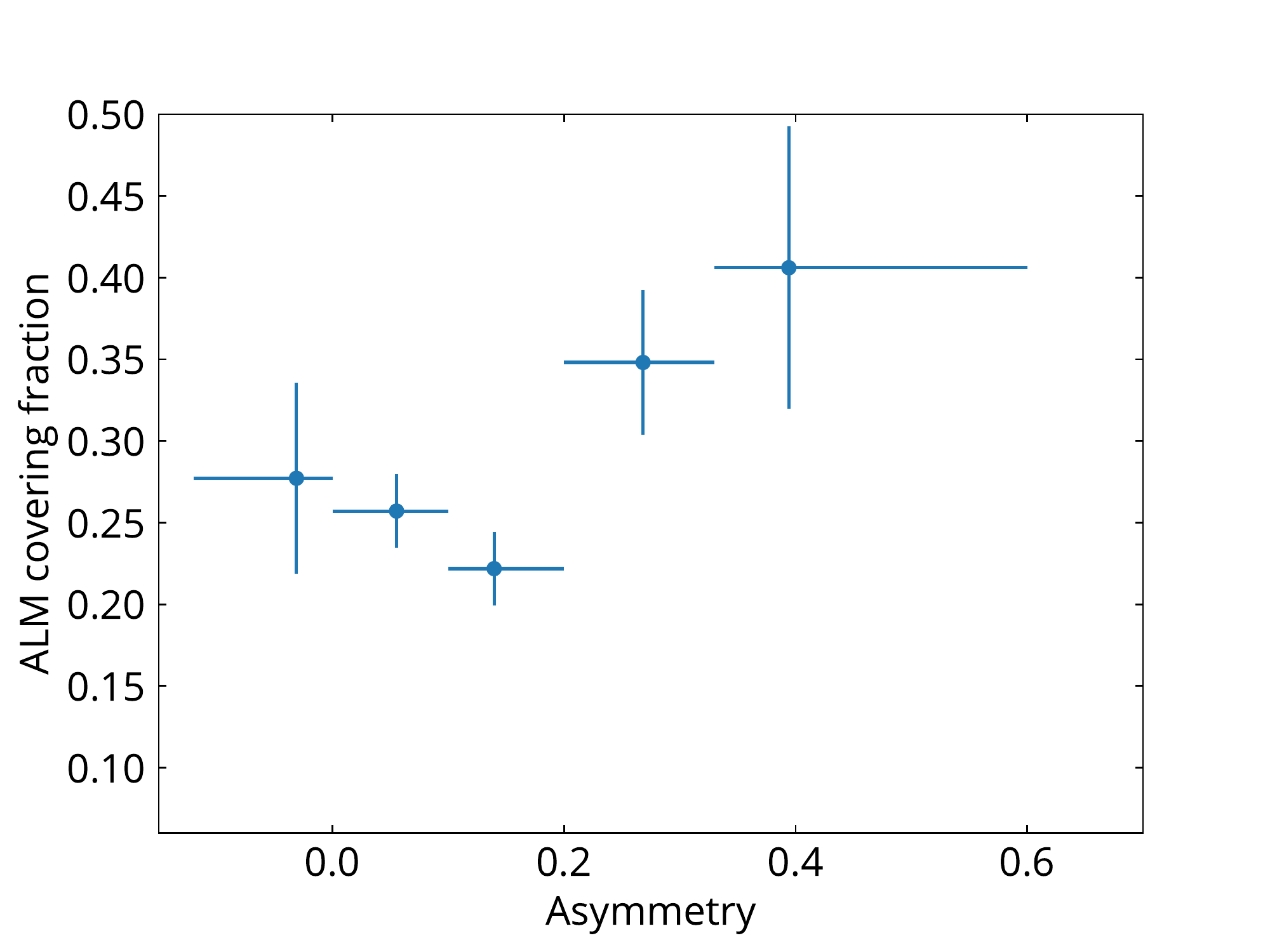} 
\caption{
Left: the fraction of star-forming galaxies having ALM regions (i.e. the occurrence rate) with respect to global properties. Right: the fraction of star-forming spaxels classified as ALM (the ALM covering fraction) in ALM galaxies. The x-axis error bars indicate the range of the bins. The y-axis error bars are Poisson errors (left) and standard deviation of the mean (right). The positions of the dots represent the median values in the bins.
} \label{fig:covering-frac}
\end{figure*}

Before any quantitative analysis, we begin this section by showing images that represent common examples of galaxies having ALM regions
(in Fig.~\ref{fig:new-example}). Such regions can be found in a variety of galaxies: interacting and non-interacting galaxies, both with and without bars. In the maps of metallicity deviation in Fig.~\ref{fig:new-example}, blue and purple spaxels represent ALM gas. Spaxels without a metallicity deviation measurement in the maps have either low S/N ratios in some emission lines, are not purely star-forming regions based on the BPT diagram, or have bad data quality flagged in the pipeline. A simple inspection of these examples shows that the ALM regions can cover relatively large parts of some interacting galaxies (8252-9102 and 9883-3701), while in other galaxies, the ALM regions are smaller and can be located in both the inner (8551-3701) and outer disk (8466-12702 and 8945-12705). Sometimes, especially in non-interacting galaxies, corresponding structures can be identified in the optical images, for example 8466-12702 and 8945-12705 in Fig.~\ref{fig:new-example}. These structures include: blue regions of ongoing star-formation and faint structures which might be accreting dwarf galaxies. In the following sections, we quantify these results and determine which global and local properties correlate most strongly with the presence of anomalously-low-metallicity regions.

\subsection{Correlations with global properties}
\label{sec:global-prop}
In Fig.~\ref{fig:new-global-prop}, we correlate the presence of ALM regions with the global properties of galaxies, including the total stellar mass, NUV$-r$ colors, asymmetry, and concentration. Compared to the star-forming sample, galaxies having ALM regions tend to be less massive, have bluer NUV$-r$ colors and higher asymmetry indices, but are not significantly different in terms of the concentration. We perform the Kolmogorov-Smirnov (K-S) test to quantify the significance of the difference between the star-forming sample and the galaxies having ALM regions. The p-values, the probability that the two distributions are sampled from the same parent distribution, are $1\times10^{-6}$, $2\times10^{-29}$, $2\times10^{-10}$, and 0.3, for total stellar mass, NUV$-r$, asymmetry index, and concentration, respectively. Therefore, the differences are significant in M$_*$, NUV$-r$, and asymmetry index, but not in concentration. 

In Fig.~\ref{fig:covering-frac}, we further investigate the occurrence rate of galaxies having ALM regions as a function of stellar mass, NUV$-r$ color, and the asymmetry index. The fraction of star-forming galaxies with ALM regions drops strongly with increasing stellar mass, from nearly 30\% at $10^9$ M$_{\odot}$ to nearly 0\% at $10^{11}$ M$_{\odot}$. A complementary view is provided by Fig.~\ref{fig:covering-frac} in the right panels. Here we select {\it only} the ALM galaxies, and then plot the fraction of star-forming spaxels that are classified as ALM in these galaxies as a function of the stellar mass. We define this as the `ALM covering fraction'. This shows that the ALM covering-fraction drops from about 30\% at $10^9$ M$_{\odot}$ to about 10\% above $10^{10.5}$ M$_{\odot}$. This means that ALM regions tend to cover larger fraction of SF regions in less massive galaxies. 

On average, galaxies with ALM regions are bluer in NUV$-r$ colors by $0.5$\,mag (Fig.~\ref{fig:new-global-prop}). Amongst the bluest galaxies (NUV$-r <$ 1.5), roughly 70\% have ALM regions (Fig.~\ref{fig:covering-frac}). This fraction drops to only about 10\% for galaxies redder than NUV$-r \sim$ 3.0. The results establish a connection between low-metallicity regions and higher global sSFRs. Also, the ALM covering fraction drops from about 35\% for galaxies bluer than NUV$-r =1.8$ to about 25\% for the rest. 

%The Kolmogorov-Smirnov (K-S) test shows that the p-value is $3\times10^{-36}$ (this is the probability that the two NUV$-r$ distributions are sampled from the same parent distribution). Fig.~\ref{fig:covering-frac} shows that the ALM covering fraction drops from about 35\% for galaxies bluer than NUV$-r =$ 1.8 to about 25\% for the rest. 
%{The ALM covering fraction increases for NUV$-r>3$. We visually inspected the galaxies in this bin, finding that the rise in the ALM covering fraction is due to the combination of strong extinction and highly inclined galaxies which have unreliable $b/a$ ratios caused by the foreground stars.}

Fig.~\ref{fig:covering-frac} shows that the fraction of ALM galaxies rises with increasing asymmetry index: from only about 10\% at Asymmetry $\sim$ 0, to $\sim 50$\% for Asymmetry $>$ 0.2 (strongly disturbed). Similarly, the ALM covering fraction in ALM galaxies is higher for strongly disturbed galaxies.  All these results are consistent with the conclusions in \citealt{Reichard2009} based on SDSS central fiber data that more metal-poor galaxies are more lopsided at fixed mass (see also \citealt{Michel-Dansac2008}). However, the majority of ALM galaxies actually do not have a strongly disturbed morphology. Specifically, 82\% of galaxies with ALM regions have an asymmetry index less than 0.2. In other words, while highly asymmetric galaxies are much more likely to have ALM regions, the majority of galaxies with ALM regions are not highly asymmetric. Some may be in interacting systems at early-stages without strong morphological disturbance, or be experiencing minor mergers with satellite galaxies. We will discuss this further in Sec.~\ref{sec:interaction}. 

Finally, Fig.~\ref{fig:new-global-prop} shows that there is no apparent difference in the concentration index between star-forming galaxies with and without ALM regions (with a p-value of $0.3$ from the K-S test).  The concentration index is a measure of bulge-to-disk ratios and the Hubble types of late-type galaxies \citep{Conselice2003}. Therefore, the presence of low-metallicity regions is nearly independent of the relative strength of the bulge for late type galaxies. However, we are pre-selecting star-forming galaxies, which will tend to have smaller bulges. Therefore, the lack of a trend could be due to the relatively small range of the concentration index.

\begin{figure*}[htbp] %  figure placement: here, top, bottom, or page
   \centering
   %\includegraphics[width=3.5in]{0407_dist_Dn4000} 
   %includegraphics[width=3.5in]{0407_dist_EWHa} 
   %\includegraphics[width=3.5in]{0407_dist_sigmaSFR} 
   %\includegraphics[width=3.5in]{0407_dist_Re} 
   \includegraphics[width=3.5in]{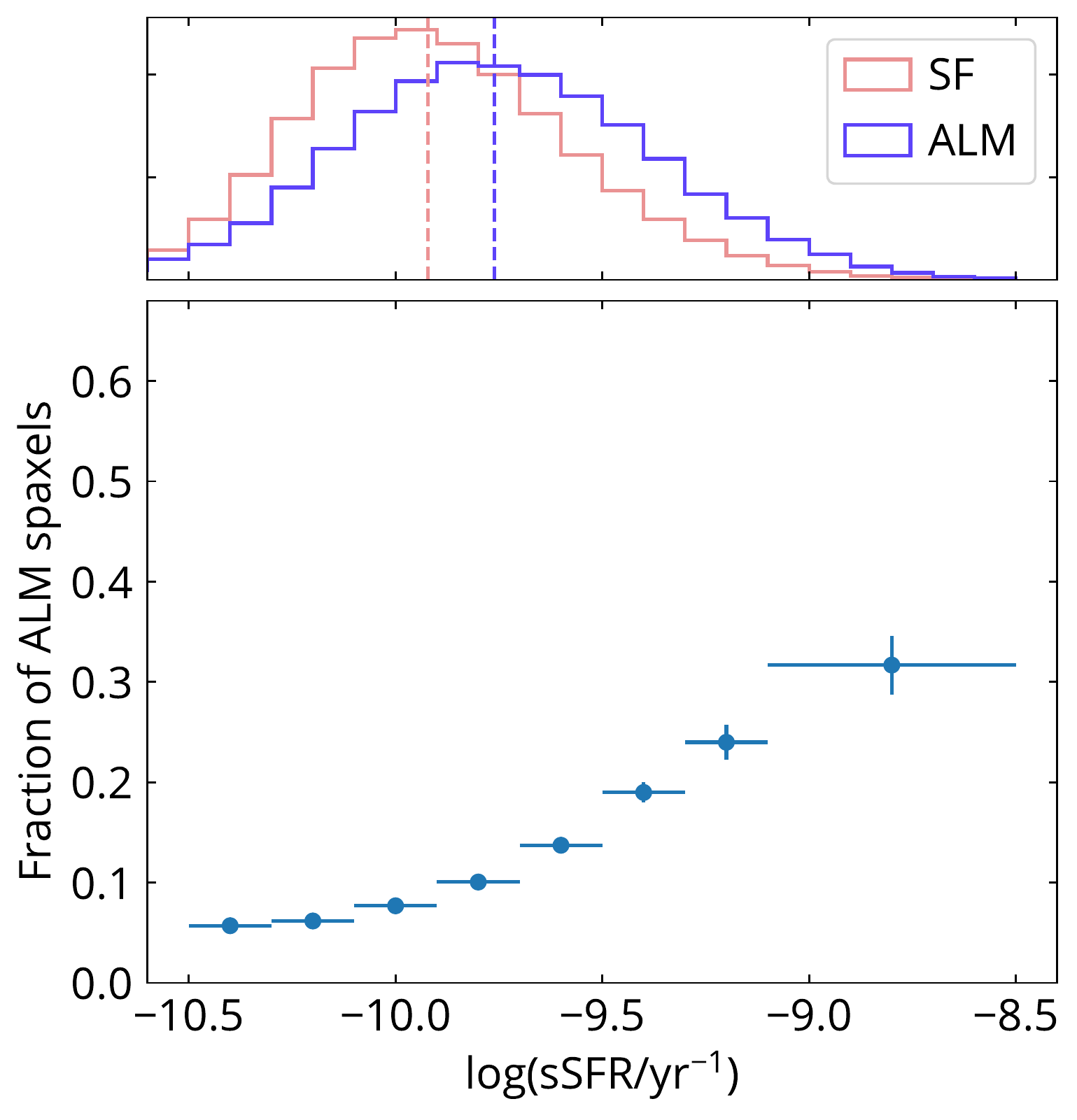} 
   \includegraphics[width=3.5in]{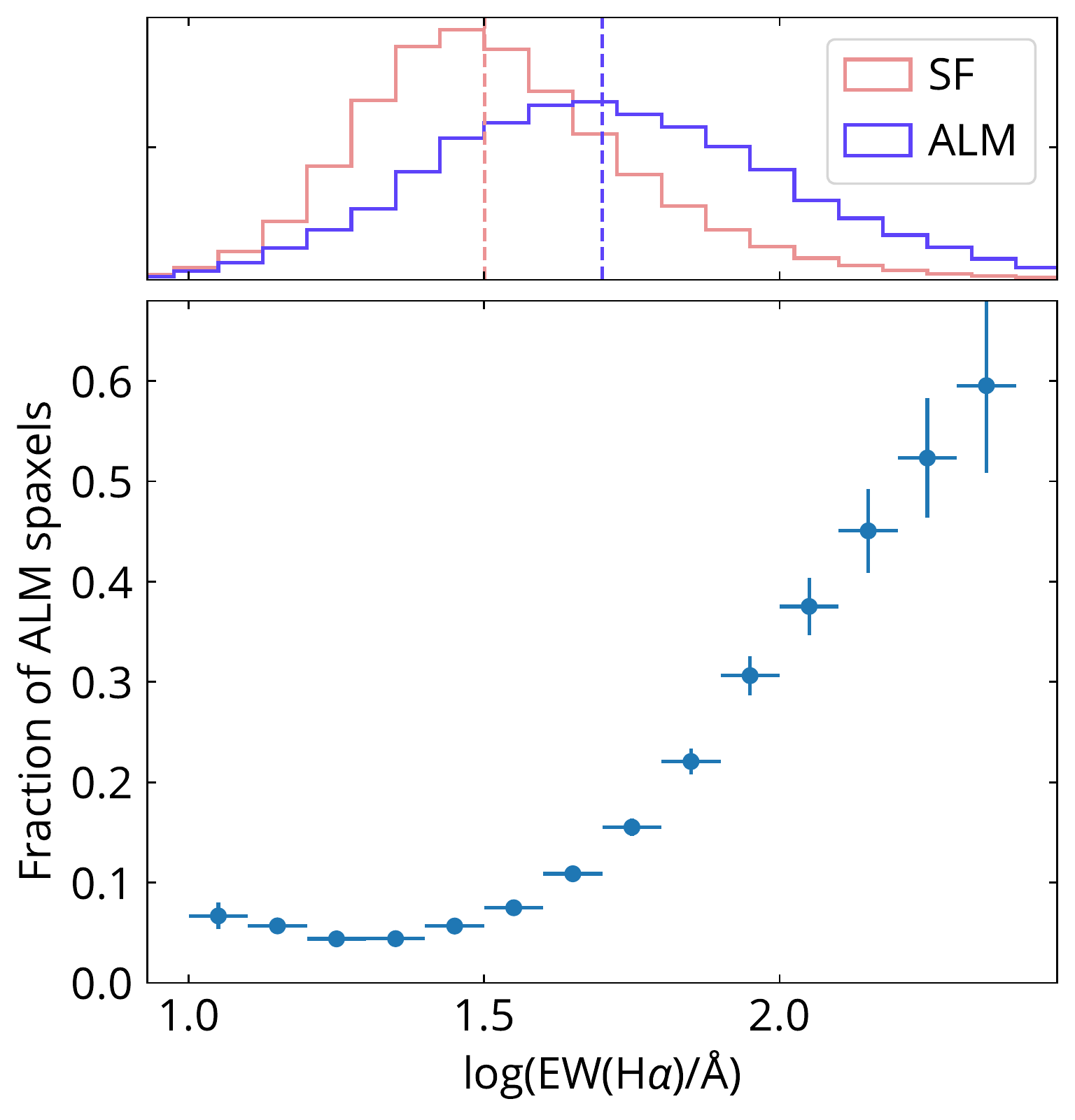} 
   \includegraphics[width=3.5in]{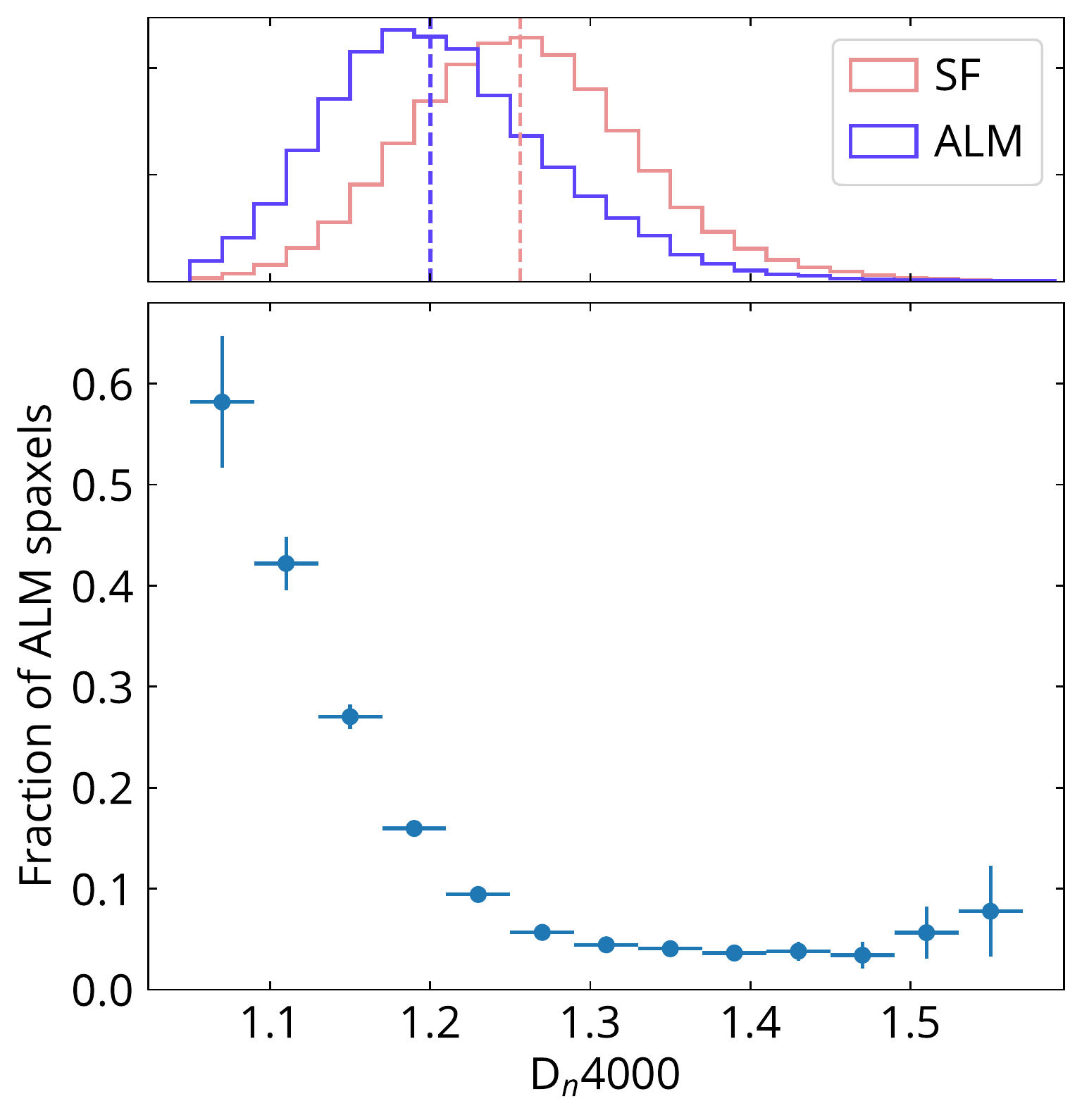} 
   \includegraphics[width=3.5in]{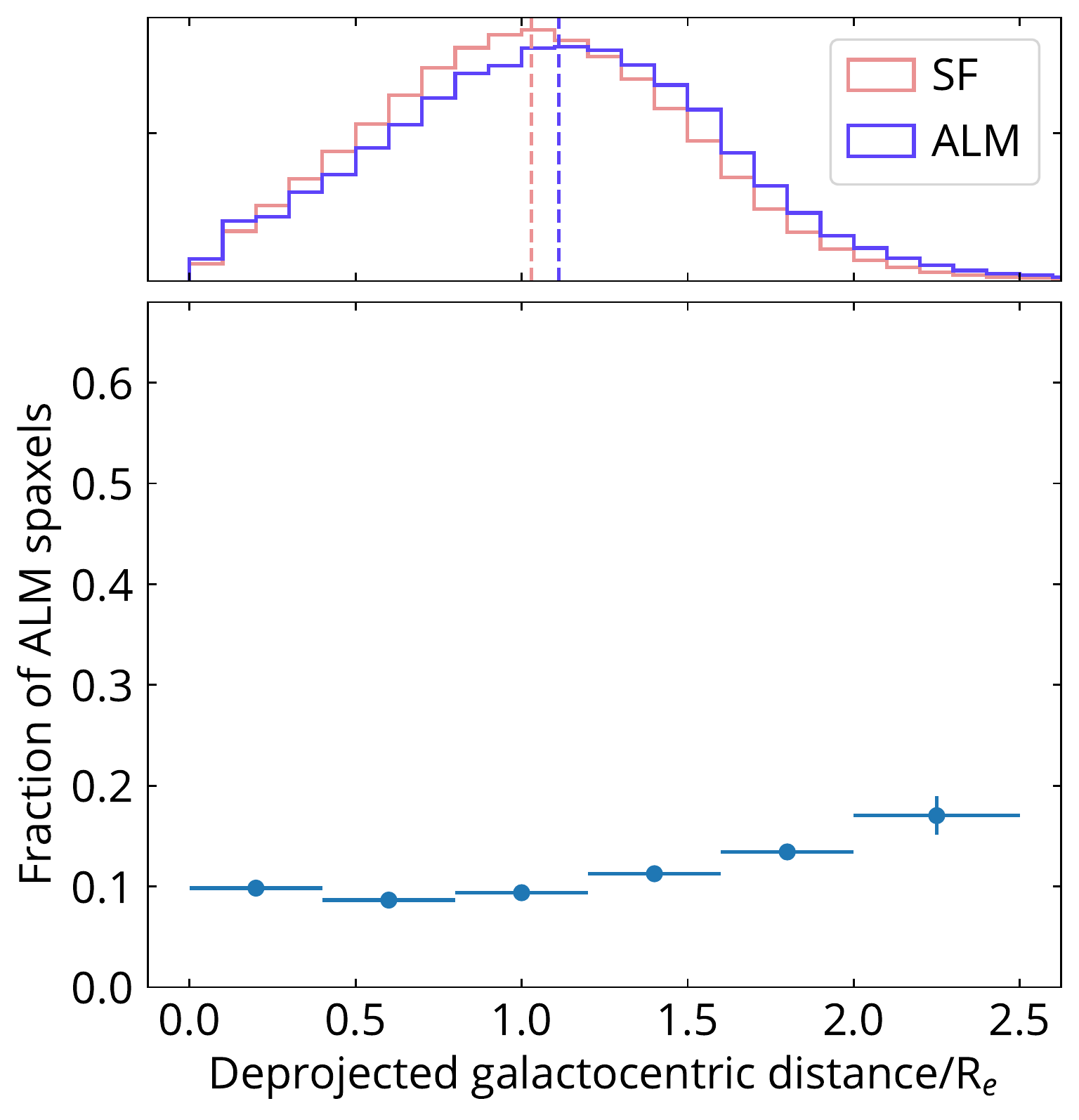} 
   \caption{The fraction of ALM spaxels amongst all star-forming spaxels, with the normalized distributions of star-forming and ALM spaxels at top. The fraction of ALM spaxels rises steeply with decreasing D$_n$4000, increasing sSFR and EW(\Ha), while slight increases at larger galactocentric radii.}
   \label{fig:new-local-prop}
\end{figure*}

%The differences of local properties between low-metallicity spaxels and normal star-forming spaxels. Normal spaxels are defined such that their metallicity deviations are within $2\sigma_{\rm Z}=0.14$\,dex. The dashed lines indicate the medians. The distributions are normalized so they do not represent the relative numbers between normal and low-metallicity spaxels.

\begin{figure}[htbp] %  figure placement: here, top, bottom, or page
   \centering
   \includegraphics[width=3.5in]{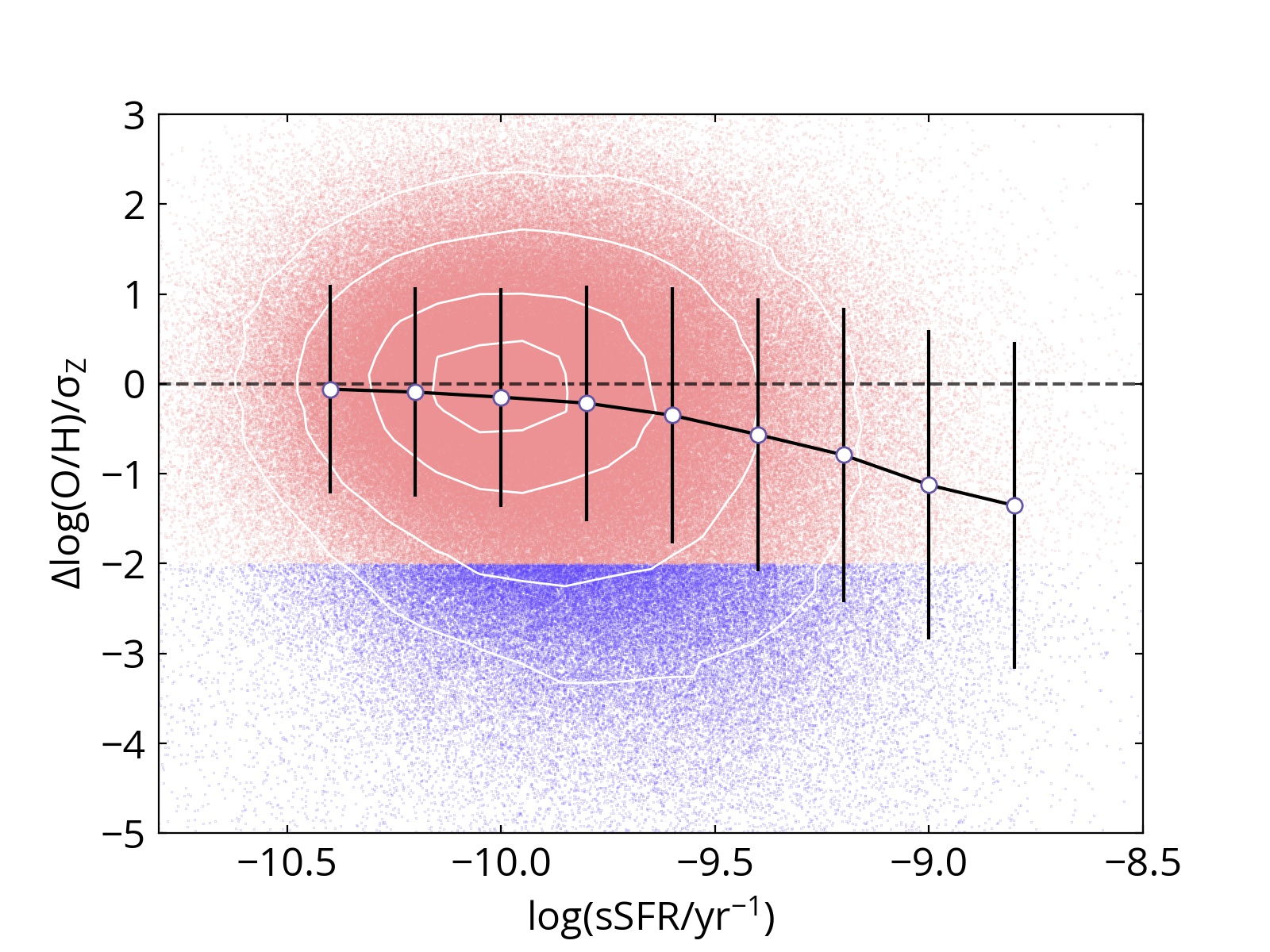} 
   \includegraphics[width=3.5in]{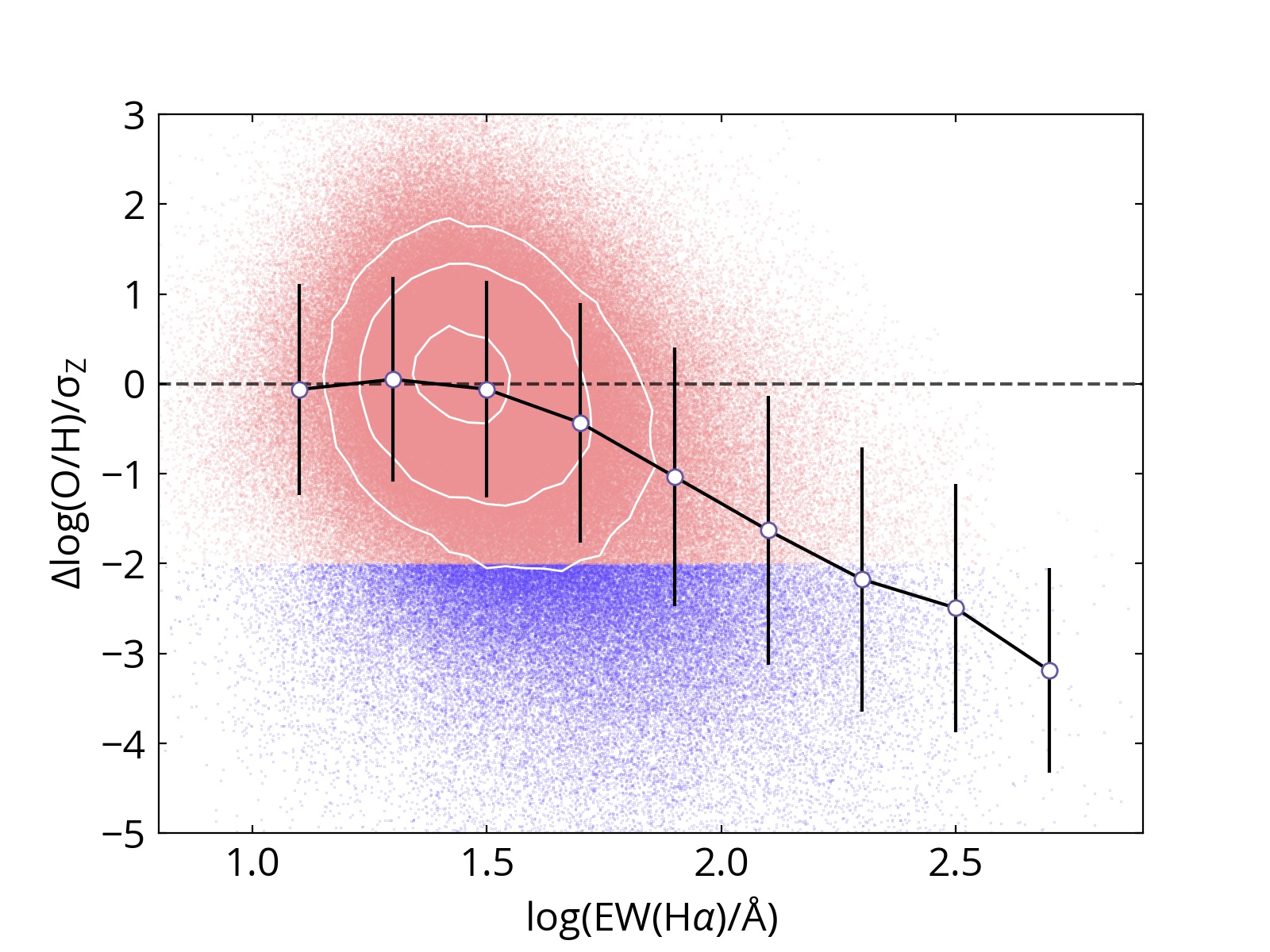} 
   \includegraphics[width=3.5in]{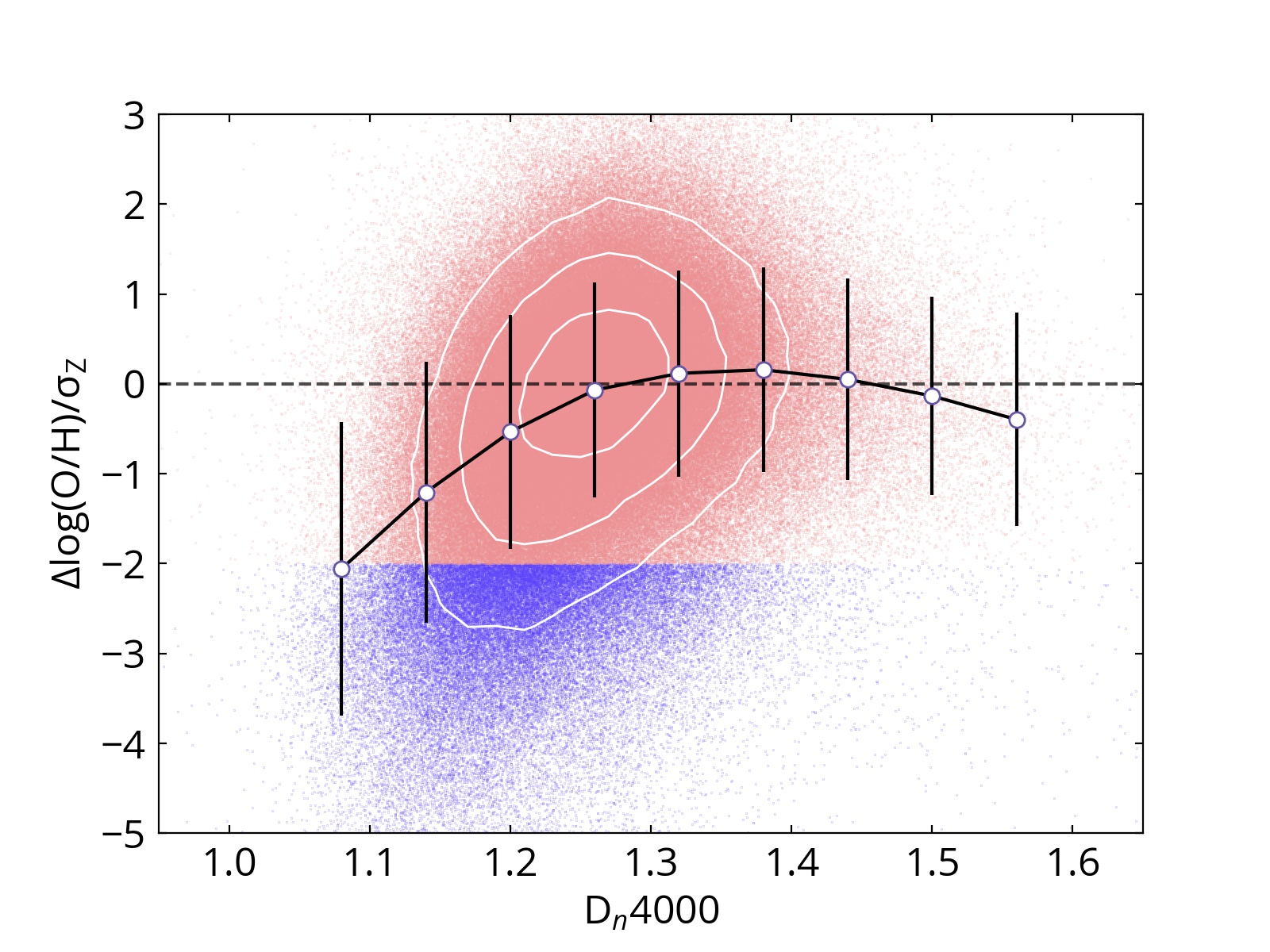} 
   
   \caption{Metallicity deviation versus specific SFR (top), EW(\Ha) (middle), and D$_n 4000$ (bottom) for star-forming spaxels, where $\sigma_{\rm Z}=0.07$\,dex. The circles indicate the median values and the error bars show the standard deviation. The blue dots are the ALM spaxels and the red dots are the rest of the spaxels. Spaxels having higher sSFR, higher EW(\Ha), and lower D$_n4000$ deviate more to the lower metallicity.  
   }
   \label{fig:new-EW-D4000}
\end{figure}

\subsection{Correlations with local properties}

In this section we will examine whether spaxels containing ALM gas differ systematically in terms of their other local properties. To probe a connection to the star formation we will use two local parameters. The first is the amplitude of the 4000\AA\ break or D$_n$4000 (the ratio of continuum fluxes on either side of the 4000\,\AA\ break caused by long-lived low-mass stars). This is an indicator of the specific SFR (sSFR) over timescales of $\sim10^8$ to $10^9$ years \citep{Kauffmann2003a}. We will also use the extinction-corrected H$\alpha$ emission-line luminosity to calculate the local star-formation rate per unit area ($\Sigma_{\rm SFR}$), and divide this by the local stellar surface mass-density ($\Sigma_*$) to determine the local specific star-formation rate (sSFR\,$= \Sigma_{\rm SFR}/\Sigma_*$). This measures the sSFR over much shorter timescales than D$_n$4000. We also use the equivalent width of \Ha\ (EW(\Ha)), which is a non-parametric indicator of sSFR. Although the relation between sSFR and EW(\Ha) is very tight \citep{Sanchez2013, Belfiore2018}, we decide to present the results of both of them because while sSFR has more direct physical meaning, EW(\Ha) is independent of stellar population models and free from the error propagation caused by the measurement of stellar surface mass density.

Fig.~\ref{fig:new-local-prop} shows the kpc-scale local properties of low-metallicity regions. The error bars are Poisson errors, and when we compute them, we divide the number of spaxels in each bin by 20 because the size of the PSF means that not all spaxels are independent. The fraction of ALM spaxels amongst all star-forming spaxels in all galaxies is a strong function of the local indicators of stellar age/star-formation. The fraction is only about 5\% for D$_n4000 > 1.3$, but rises steeply to almost 60\% for D$_n4000 <1.1$. It is consistent with \cite{Sanchez2015} where they show that lower metallicity is correlated with younger stellar ages, but here we further demonstrate that it still holds after controlling for the dependence of metallicity on the host galaxy mass and local stellar surface mass density. Similarly, the fraction is about 6\% for an sSFR\,$= 10^{-10.5}$\,yr$^{-1}$, but rises to about 30\% for sSFR\,$> 10^{-9}$\,yr$^{-1}$, and about 5\% for an EW(\Ha)$=20$\AA\ to 60\% for EW(\Ha)$\sim300$\AA.

These dependences imply that the ALM gas has a strong connection to the underlying stellar population and the recent star-formation history. This is consistent with other recent results. \cite{Rowlands2018} used the MaNGA data to classify the stellar population in  spaxels based on a principal component analysis of the continuum \citep{Wild2007}. They found that spaxels classified as starbursts have significantly lower gas-phase metallicity than other spaxels. Similarly, \cite{SanchezAlmeida2018} studied the local relation between SFR and metallicity using a sample of 14 dwarf galaxies with stellar masses ranging from M$_* = 10^{7.0-9.5}$\Msun, showing that spaxels with locally higher SFRs tend to have lower metallicity (see also \citealt{SanchezAlmeida2015}).

Fig.~\ref{fig:new-local-prop} also shows that the fraction of ALM star-forming spaxels rises slowly with increasing radius, from about 10\% inside 1 $R_e$, to 17\% outside 2 $R_e$. It means that ALM regions slightly prefer the outer regions. We will show in Sec.~\ref{sec:interaction} that the radial distribution of ALM regions is related to the interaction stage.

%\red{Fig.~\ref{fig:new-local-prop} also shows that the fraction of ALM star-forming spaxels rises slowly with increasing radius, from about 10\% inside 1 $R_e$, to 17\% outside 2 $R_e$. This is consistent with the increase in the fraction of ALM spaxels with decreasing local stellar surface mass density, from about 10\% at $\Sigma_* \sim 10^{9}$\,M$_{\odot}$\,kpc$^{-2}$ to 15\% at $\Sigma_* < 10^{7.5}$\,M$_{\odot}$\,kpc$^{-2}$. }

We note that while we have chosen a specific metric to define ALM regions (a metallicity deviation of more than 2 $\sigma_{\rm Z}$, equivalent to an offset of more than 0.14\,dex in O/H), the connection of metallicity deviation and star formation is independent of this particular choice. This is shown in Fig.~\ref{fig:new-EW-D4000}, where we plot the metallicity deviation of star-forming spaxels as a function of sSFR, EW(\Ha), and D$_n4000$. There are clear and systematic trends for larger metallicity deviations for young (more strongly star-forming) regions in both plots that are independent of any specific definition of ALM.

The strong connection between ISM metallicity and a young stellar population suggests that the star formation in the ALM regions may be triggered by the arrival of low-metallicity gas. The young stellar ages in low-metallicity regions also constrain the lifetime of these regions. We will return to this point in Sec.~\ref{sec:timescale} below.

\subsection{Interactions and Interaction Classes} 

\label{sec:interaction}

In Sec.~\ref{sec:global-prop}, we found that galaxies having higher asymmetry index tend to have ALM regions, suggesting that galaxy interactions play an important role in producing these ALM regions. Since the asymmetry index is mainly sensitive to the strongest morphological disturbances and later interaction stages, and does not reflect the entire interaction process, we further consider the visual classification of three interaction classes: mergers, close pairs, and isolated galaxies (see details in Sec.~\ref{sec:inter-class}). 

%\red{(Hsi-An: It would be better to give the numbers of  galaxies in each types so that readers can know the number of galaxies being averaged in each profiles in Fig 12.)}

In our star-forming galaxy sample, 52\% (29/55) of mergers and 30\% (49/161) of close pairs have ALM regions, while only 23\% (229/1006) isolated galaxies have them. The higher occurrence rate in stronger interacting systems suggests that interaction may be able to trigger ALM regions.

\begin{figure*}[htbp] %  figure placement: here, top, bottom, or page
   \centering
   \includegraphics[height=2.5in]{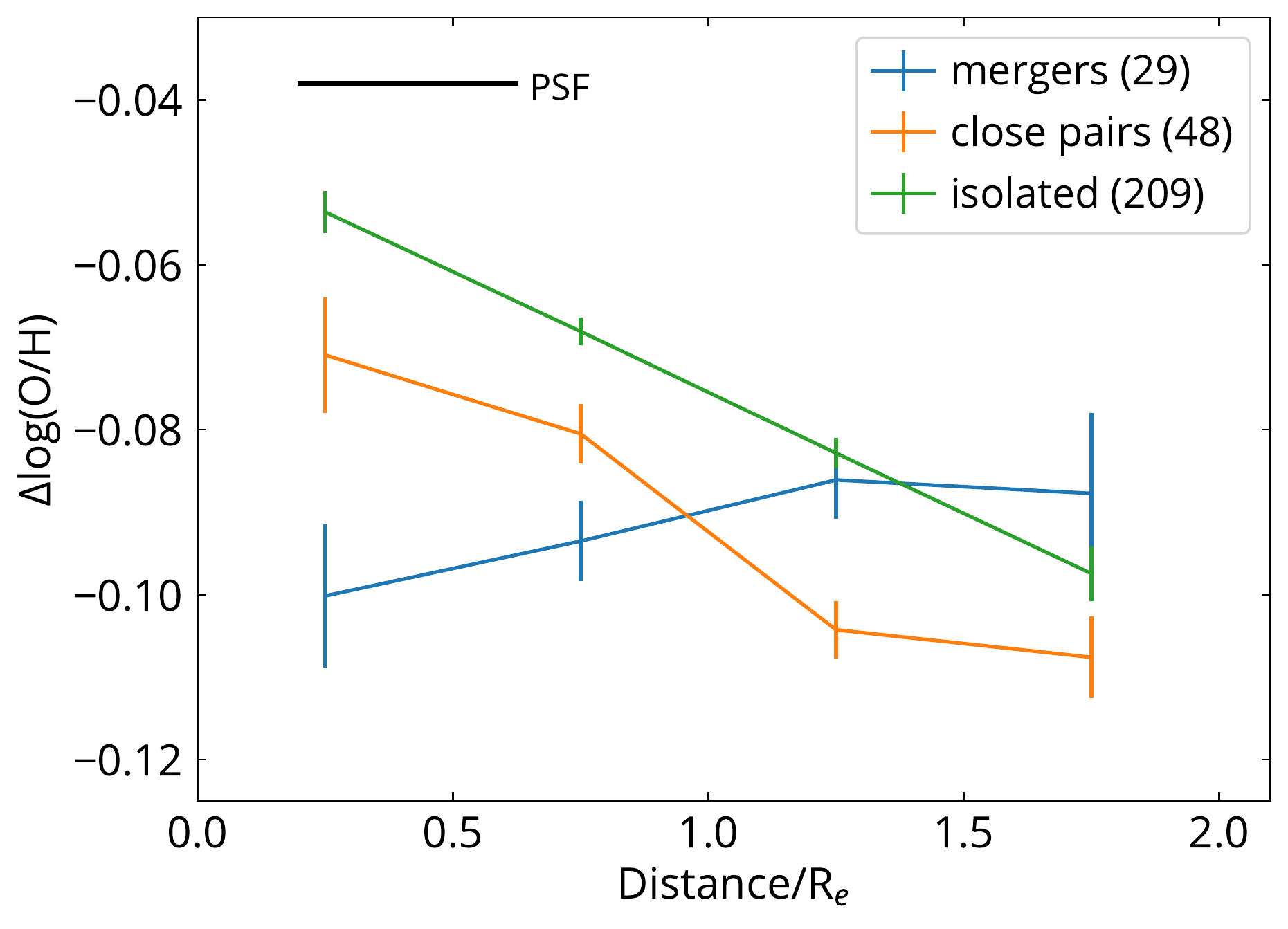} 
   \includegraphics[height=2.5in]{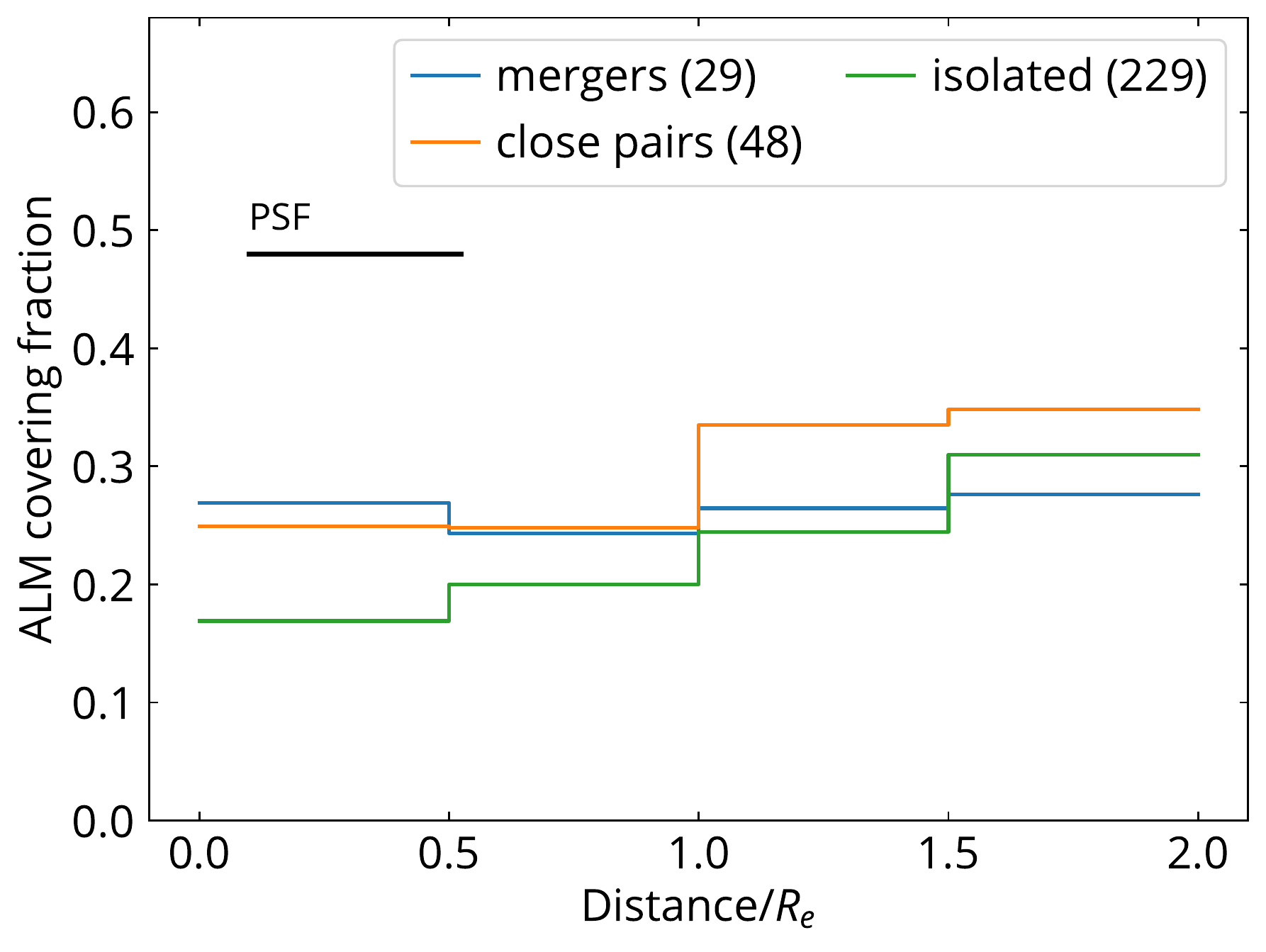} 
   \caption{Metallicity deviation (left) and covering fraction of ALM spaxels (right) versus galactocentric distances for galaxies having ALM regions. The metallicity deviations in the left panel show the median and the error bars show the standard deviation of mean. The horizontal segments show the median FWHM of PSF in units of $R_e$. Spaxels in mergers deviate to lower metallicity the most in the inner regions, and the covering fraction of the ALM gas shows little radial variation. In close pairs and isolated galaxies, both the amplitude of the metallicity deviation and the covering fraction of ALM spaxels increase with increasing radius. }
   \label{fig:new-dZ-dist}
\end{figure*}

Fig.~\ref{fig:new-dZ-dist} demonstrates how ALM regions distribute radially in the ALM galaxies. In each radial annulus, we collect all star-forming spaxels for a certain class of ALM galaxies, and compute the median metallicity deviation and the standard deviation of the mean (represented as vertical bars in the left panel of Fig.~\ref{fig:new-dZ-dist}). When computing the standard deviation of the mean, we take the PSF into account by dividing the total number of the spaxels by 20. Here we exclude a close pair galaxy 8312-12701 because it has an unreliable effective radius measurement in the NSA catalog. Fig.~\ref{fig:new-dZ-dist} shows that different interaction classes have distinct radial distributions of ALM gas, and different ALM covering fractions. Mergers have lower metallicity deviations over all radii, particularly in the inner regions ($<0.5 R_e$). In close pairs, the absolute metallicity deviations increase monotonically with radius, both in terms of the median deviation and the covering fraction of ALM spaxels. The radial distributions in mergers and close pairs suggest that, during the interaction, inflows lower the metallicity more significantly in the outer regions at the early stage, and then the low-metallicity material flows into the innermost region at the final coalescence phase.

One can argue that the ALM regions identified in interacting systems are due to the inward radial transport of outer disk gas (which is more metal-poor), instead of the accretion of gas from outside of the galaxy. While this scenario is possible, it is at odds with our finding that ALM regions are more likely to be found in the outskirts in close pairs and isolated galaxies.

%Because the $\Sigma_*$-Z relation in Fig.~\ref{fig:new-MZrelation} is mainly determined by non-interacting systems, the ALM regions seen in interacting galaxies might be that the gas motion is decoupled from the stars, instead of the metal-poor inflow. However, it is the case, we should have roughly the equal chance to see both anomalously high metallicity regions and ALM regions in interaction galaxies. What about radial transport?

%Mergers have the strongest median metallicity deviations and the largest covering fractions of ALM spaxels. This is true at all radii, but particularly so in the inner regions ($<1.5 R_e$). In close pairs, the metallicity deviations increase monotonically with radius, both in terms of the mean deviation and the covering fraction of ALM spaxels. The radial distributions in mergers and close pairs suggest that, during the interaction, inflows lower the metallicity more significantly in the outer regions at an early stage, but the low-metallicity material is mixed throughout during the final coalescence phase.

In isolated galaxies, the metallicity deviation as a function of galactocentric radii is similar to that in close pairs, with larger deviation in the outer regions $\gtrsim 1R_e$, while the amplitude is weaker than close pairs on average. This similarity hints that the ALM regions in isolated galaxies might be related to weaker interaction, like accreting satellite galaxies or CGM/IGM gas.

%depends on the galaxy stellar mass. Fig.~\ref{fig:new-dZ-dist} shows that isolated galaxies more massive than $\sim10^{9.6}$\Msun\ tend to preferentially have low-metallicity regions in the outer regions $\gtrsim 1R_e$, with the overall radial trend  is similar to close pairs (but with lower amplitude). On the other hand, there are no strong monotonic radial trends for low-metallicity regions in isolated galaxies with stellar masses $10^{9-9.6}$\Msun. 

%\red{In isolated galaxies, the metallicity deviation as a function of galactocentric radii depends on the galaxy stellar mass. Fig.~\ref{fig:new-dZ-dist} shows that isolated galaxies more massive than $\sim10^{9.6}$\Msun\ tend to preferentially have low-metallicity regions in the outer regions $\gtrsim 1R_e$, with the overall radial trend  is similar to close pairs (but with lower amplitude). On the other hand, there are no strong monotonic radial trends for low-metallicity regions in isolated galaxies with stellar masses $10^{9-9.6}$\Msun. }

\begin{figure*}[htbp] %  figure placement: here, top, bottom, or page
   \centering
   \includegraphics[width=2.1in]{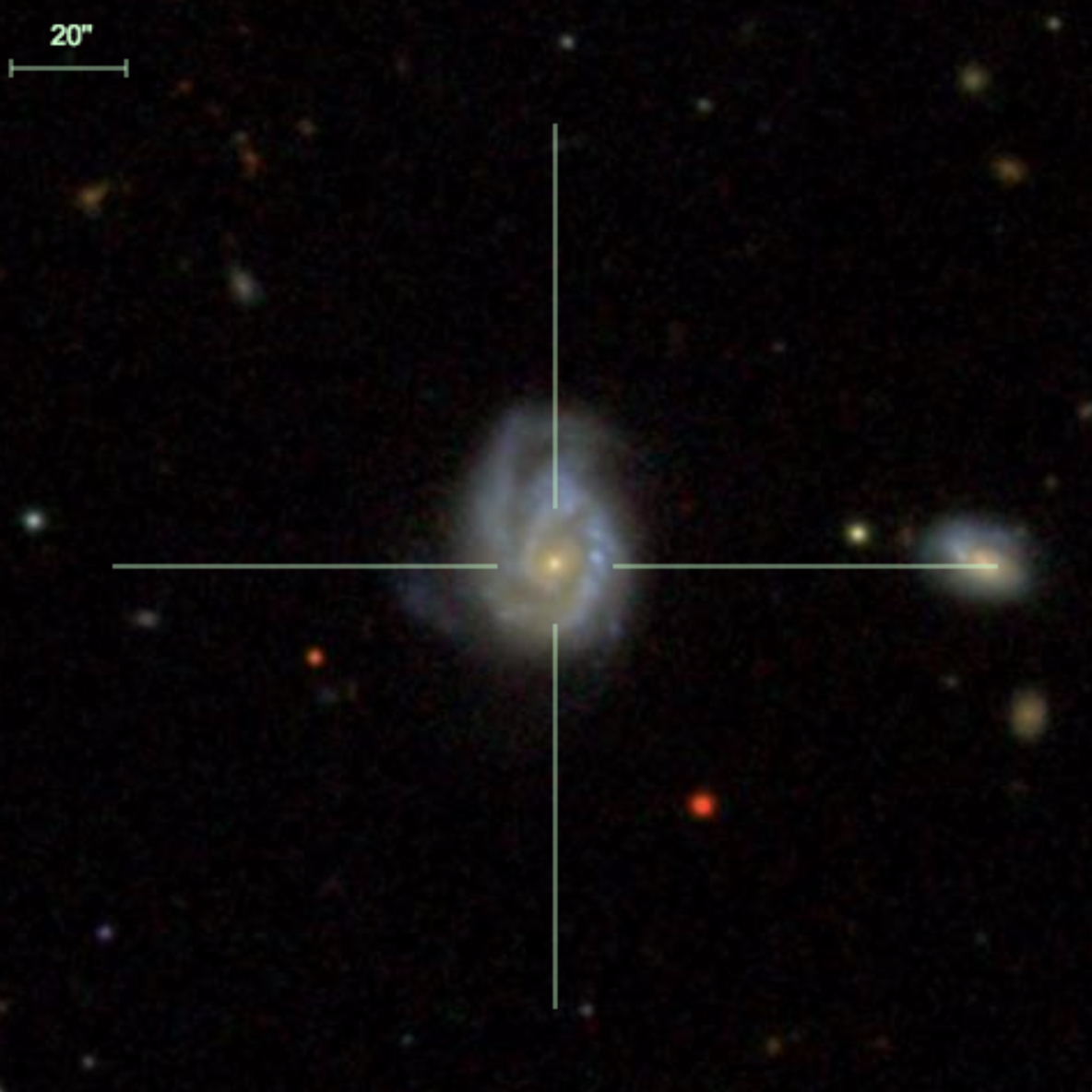}
   \hspace{0.8in}
   \includegraphics[width=2.1in]{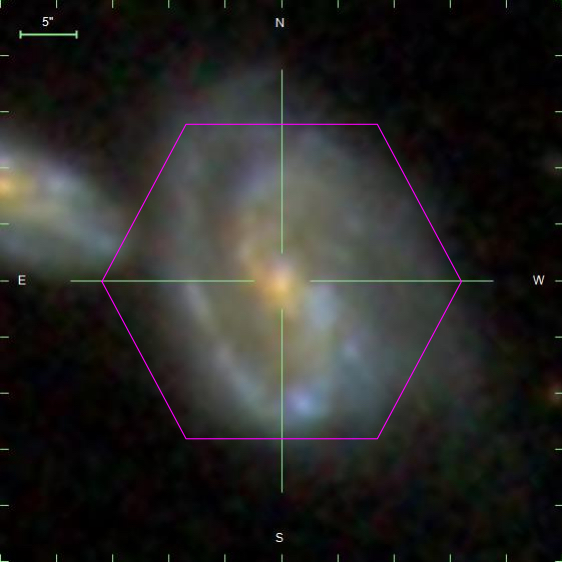}
   \\
   \vspace{0.15in}
   \includegraphics[width=2.7in]{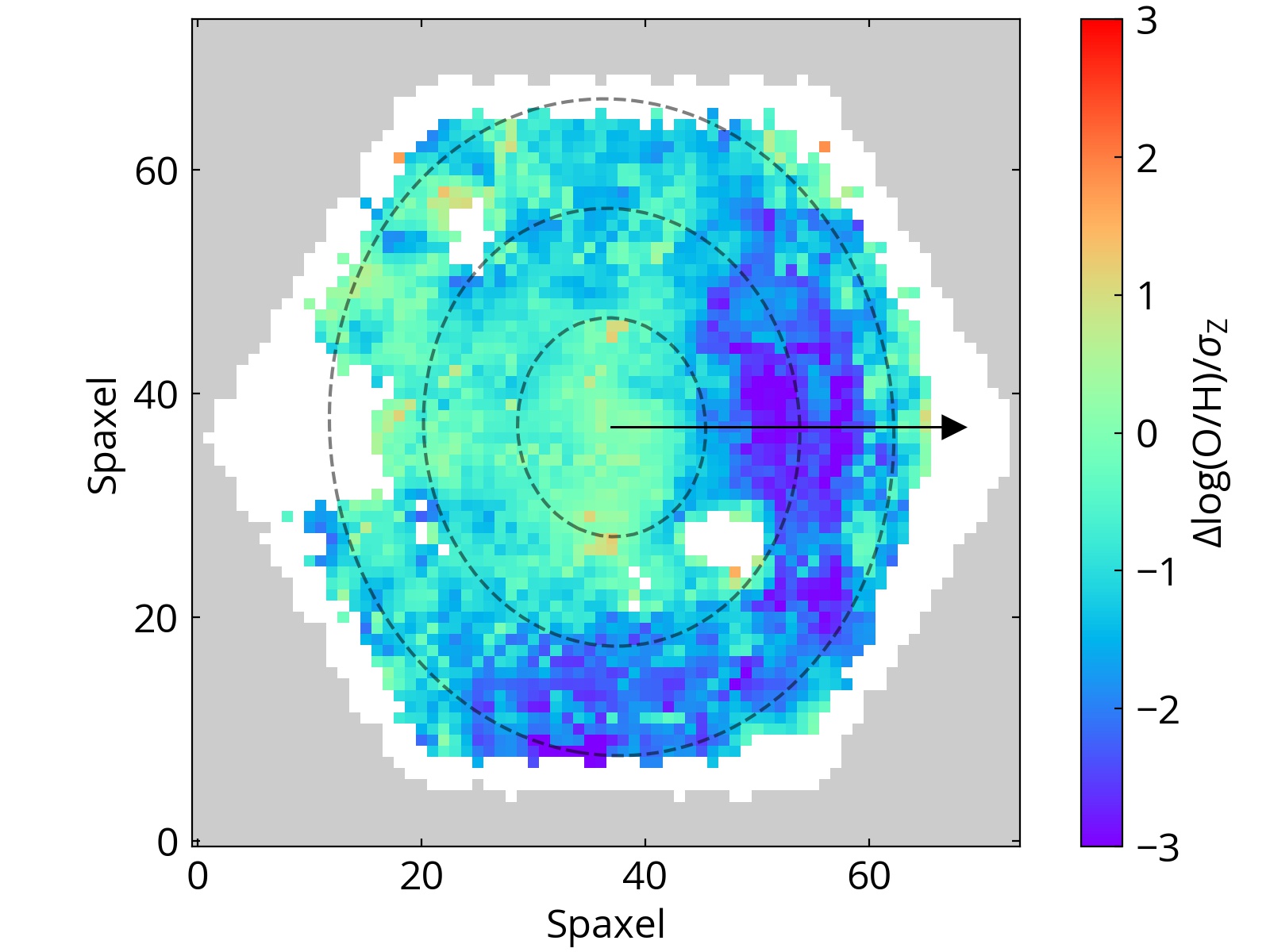} 
   \hspace{0.2in}
   \includegraphics[width=2.7in]{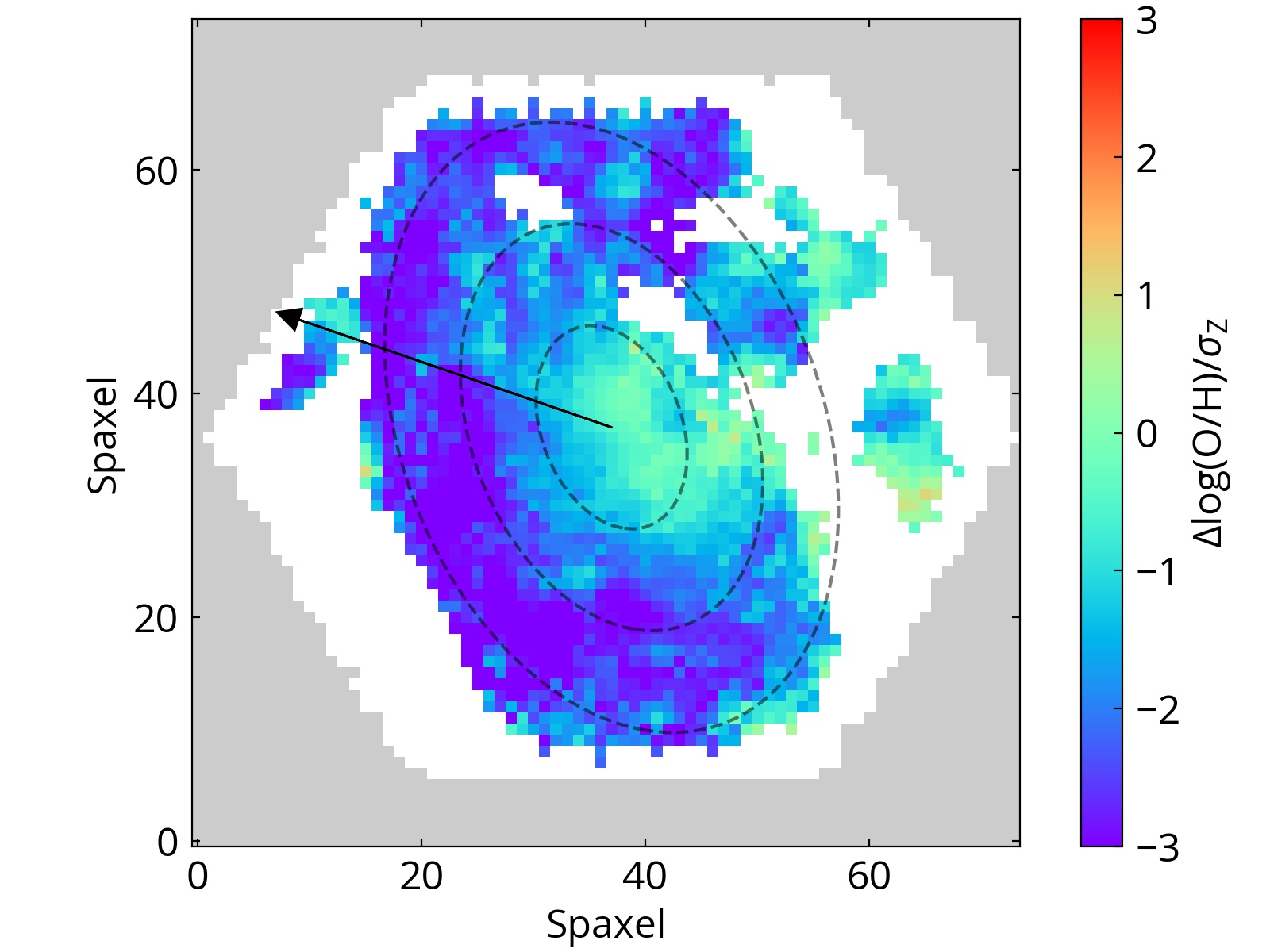} 
   \\
   \includegraphics[width=2.9in]{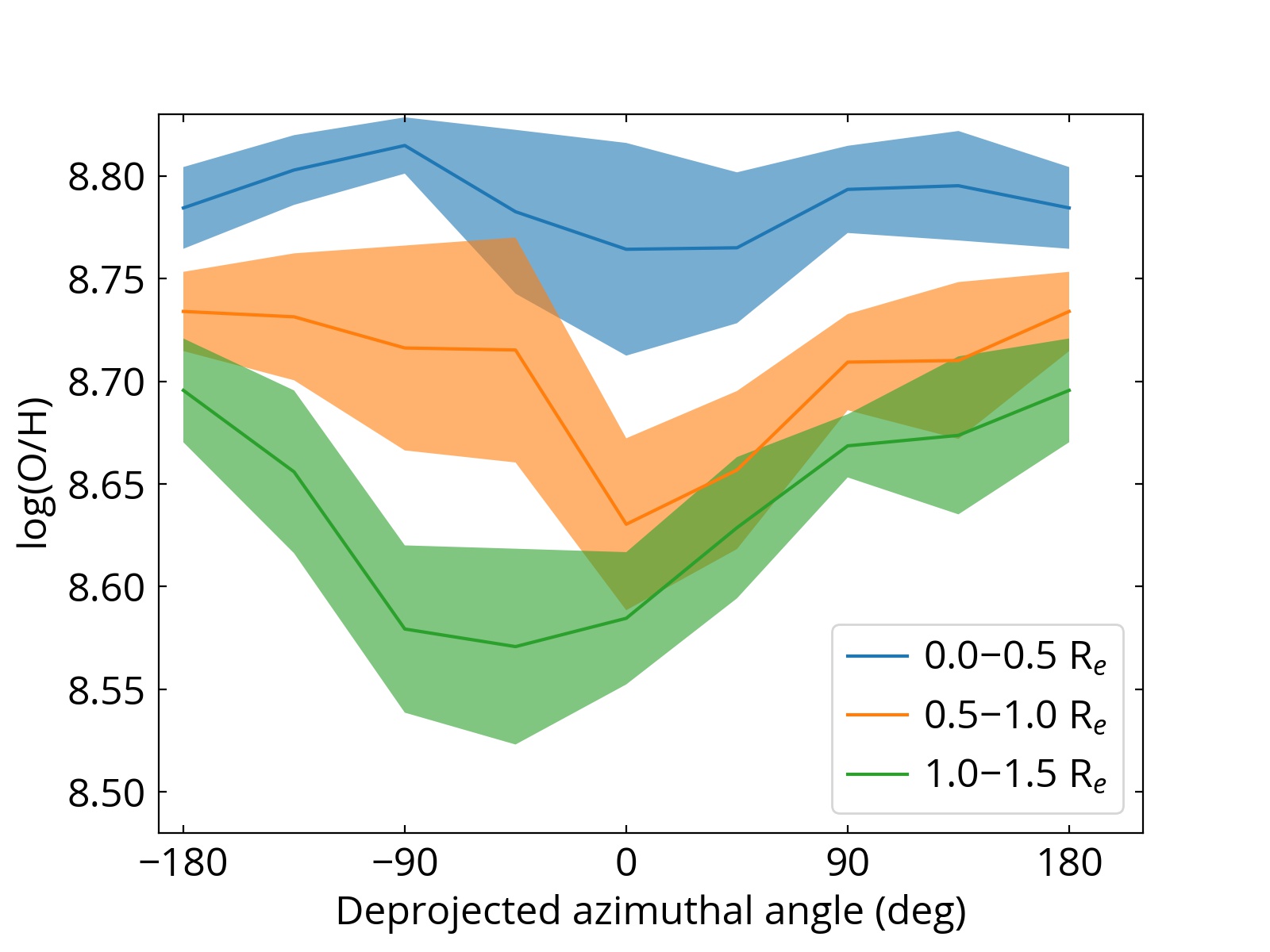} 
   \includegraphics[width=2.9in]{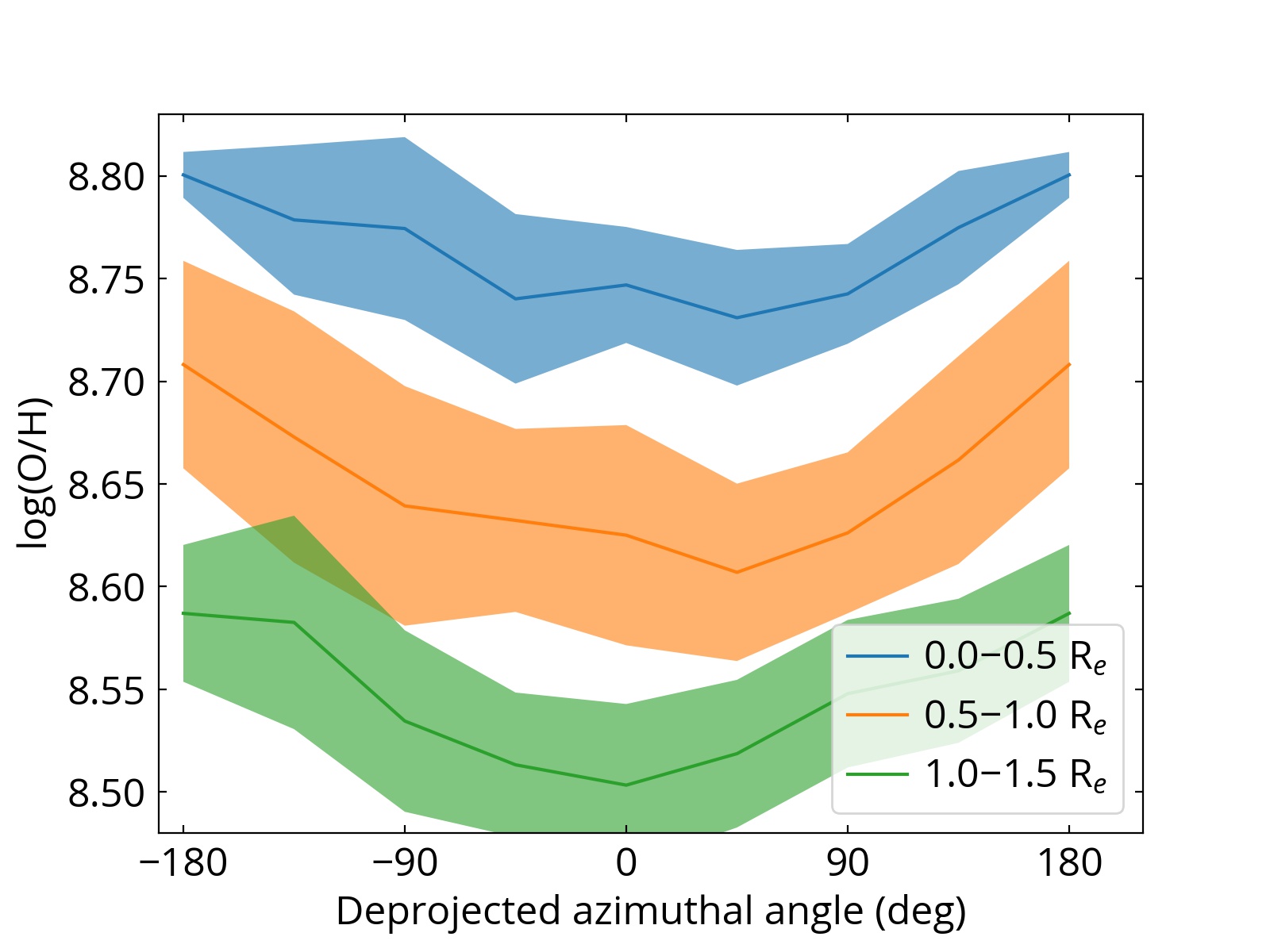} 
   
   \caption{Examples of azimuthally asymmetric metallicity distributions (left: 8454-12703; right: 8313-12702). From top to bottom: SDSS optical images, maps of metallicity deviation, and the metallicity distributions with respect to de-projected azimuthal angle. The dashed ellipses indicate the elliptical effective radii of 0.5$R_e$, $1R_e$, and $1.5R_e$. The arrows show the direction of the companions. In the plots of metallicity distribution versus azimuthal angle (bottom), the lines represent the mean while the shaded regions represent the standard deviation. The de-projected azimuthal angles are corrected for inclination, with 0 degree set to the direction of companions. The metallicities on the sides closer to the companions are lowered by $0.1-0.15$\,dex at distances $>0.5R_e$.}
   \label{fig:new-close-pair}
\end{figure*}

In addition to radial variations, some close pairs have intriguing azimuthally asymmetric metallicity distributions. Fig.~\ref{fig:new-close-pair} presents two of the clearest examples, 8454-12703 (left) and 8313-12702 (right). They have projected separations of 45 and 17\,kpc ($h=0.7$) respectively from their companions. The regions of ALM are found mainly in the outer regions ($>0.5 R_e$) in both cases. Most interestingly, the metallicity deviations are much stronger on the sides closer to their companions. This azimuthal dependence can also be seen in the measured metallicity, as shown in the bottom row of Fig.~\ref{fig:new-close-pair}, where the $x$-axes are the azimuthal angles corrected for inclination, with 0 degrees pointing toward the companions. 

The fact that the side closer to the companion has lower metallicity suggests that these azimuthally asymmetric metallicity distributions may originate from the interaction with the companion. In 8454-12703 and 8313-12702, the low-metallicity regions have sizes on galactic scales. If they were from the infalling satellite galaxies, the galactic-scale infalling structures would have been seen in the optical images, and the morphologies of 8454-12703 and 8313-12702 would be strongly disturbed. However, 8454-12703 and 8313-12702 do not show any such discrete structures and their disks are not strongly disturbed. Therefore, they may be the most convincing cases that their ALM regions are due to the metal-poor gas inflow, instead of accreting satellite galaxies or minor mergers. Out of 48 close pairs with ALM regions, we find that 6 of them, including the two in Fig.~\ref{fig:new-close-pair}, show such azimuthal metallicity dependences. The occurrence of such metallicity distributions may depend on the interaction configurations, mass ratios, primary or secondary galaxy in close pairs, and/or gas content.

%In our sample, not all close pairs have such asymmetric metallicity distribution \blue{QUANTIFY??}. 

%The dashed grey lines represent all late-type galaxies, including those with and without ALM regions.

\begin{figure*}[htbp] %  figure placement: here, top, bottom, or page
   \centering
   \includegraphics[height=1.5in]{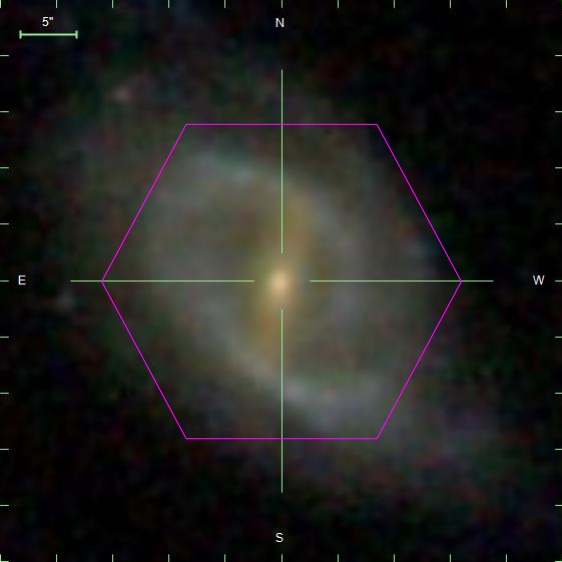}
   \includegraphics[height=1.5in]{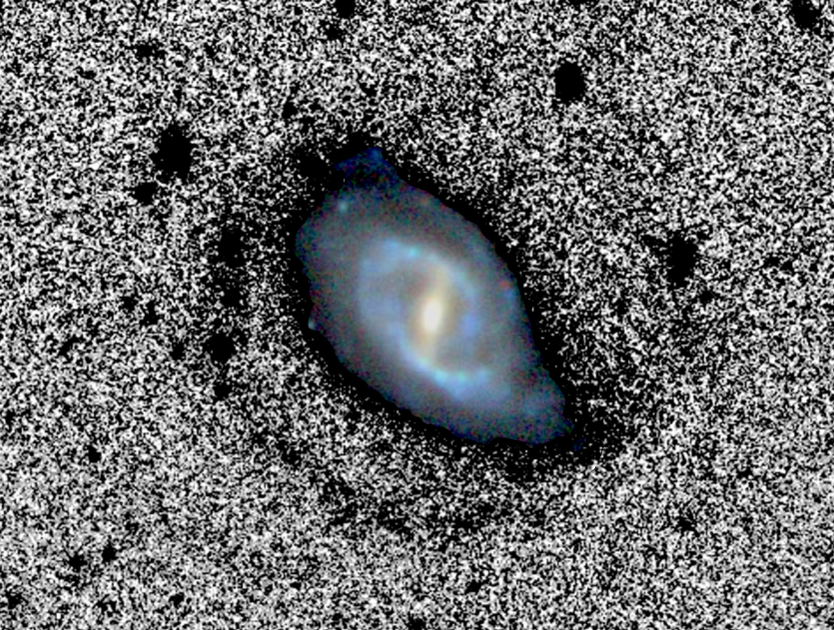}   
   \\
\includegraphics[height=1.5in]{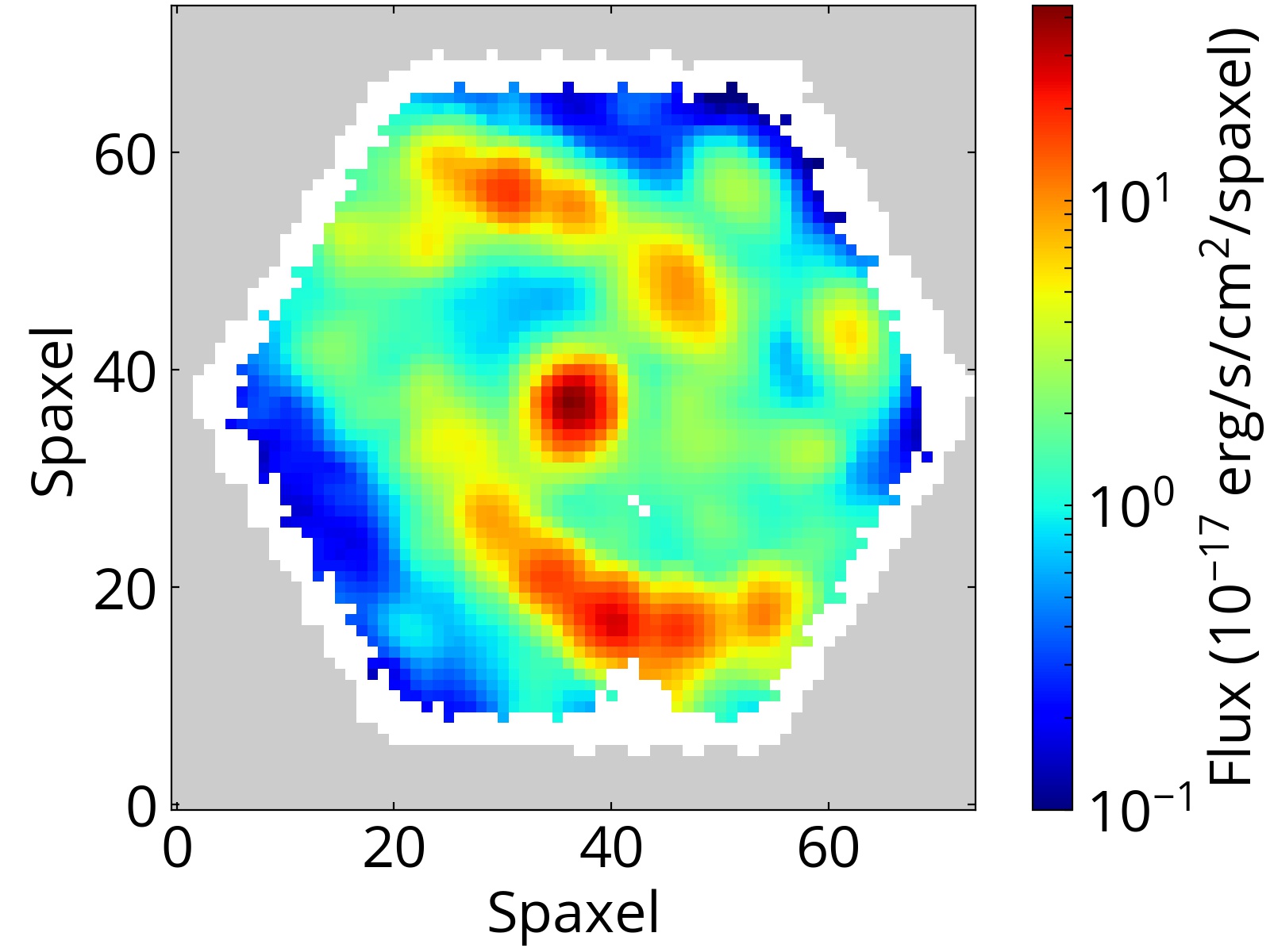}
\includegraphics[height=1.5in]{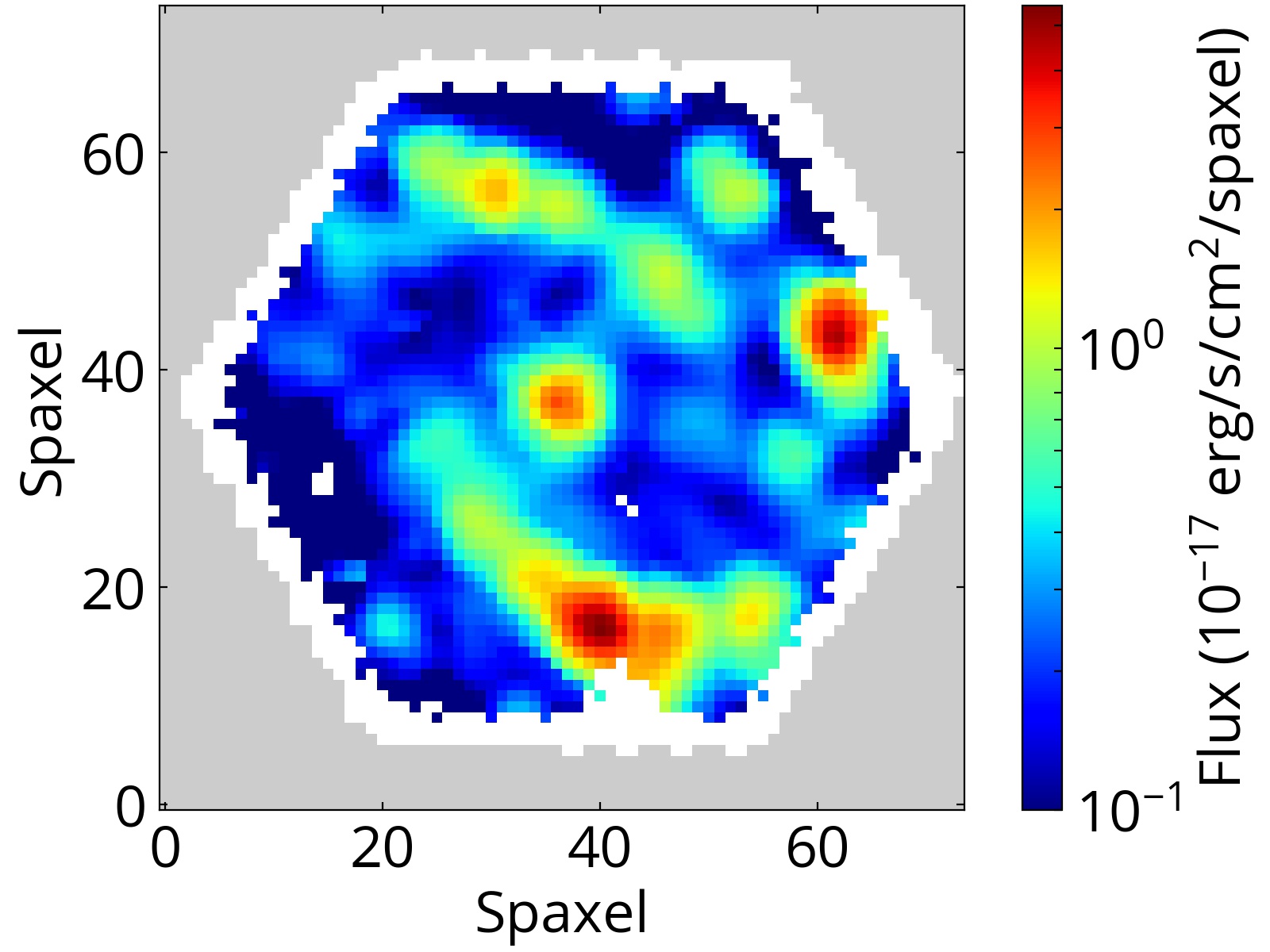}
\includegraphics[height=1.5in]{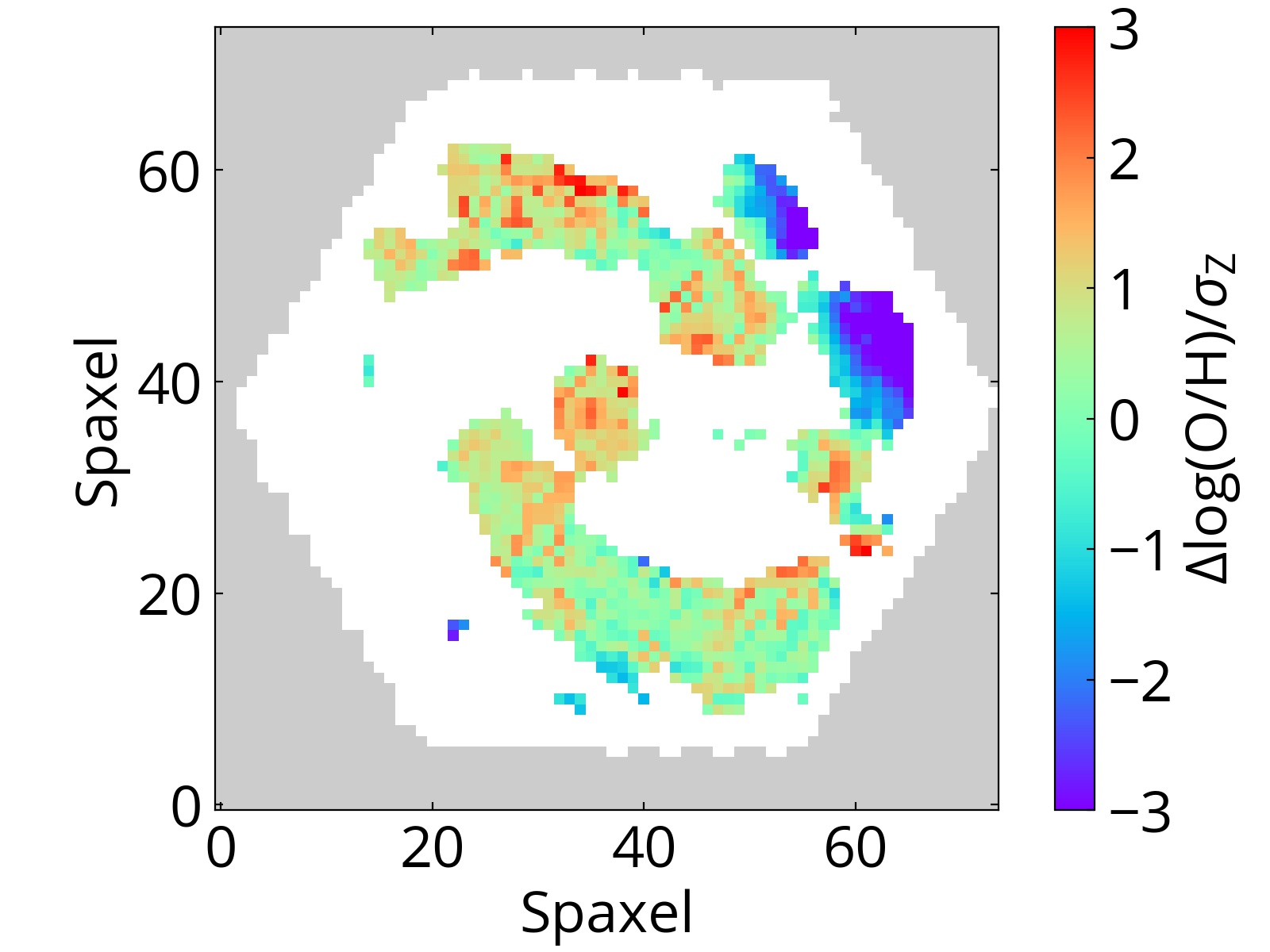}  

\caption{A galaxy (8931-12702) with a `great circle' tidal stream. The top row shows the optical images from SDSS (left) and from a deeper optical survey DECalLS DR7 (right) where the tidal stream can be easily seen. The second row, from left to right, shows its \Ha, \oiii, and metallicity deviation maps. With the distorted spiral arms, bar probably induced by interaction, and the tidal stream, the ALM region in this galaxy is due to a infalling dwarf galaxy. }
\label{fig:new-remnant}
\end{figure*}

While the majority of galaxies with ALMs are not strongly interacting or merging systems, they may be interacting with small satellite galaxies that do not significantly perturb their morphology. As an example, we consider here the isolated ALM galaxy 8931-12702 (Fig.~\ref{fig:new-remnant}). The galaxy has a strong bar and two distorted spiral arms. In addition, a faint third arm appears north-west to the galaxy without any symmetric counterpart on the other side. Furthermore, the third arm has much lower metallicity than the rest of the galaxy. A much deeper optical image from DECalLS DR7 shows that there is a faint `great circle' structure around the galaxy. This structure is similar to the tidal streams like the Sagittarius stream around the Milky Way \citep{Belokurov2006} and NGC\,5907 \citep{Martinez-Delgado2008}. $N$-body simulations show that such tidal streams are due to the disruption of a satellite galaxy in the past $6-10$\, Gyr \citep{Law2005, Johnston2008}. The ALM region is connected to the great circle structure, suggesting they share the same origin. 

\subsection{Metal-poor Outer Rings}
\label{sec:ring}
\begin{figure*}[htbp] %  figure placement: here, top, bottom, or page
   \centering
\includegraphics[height=1.3in]{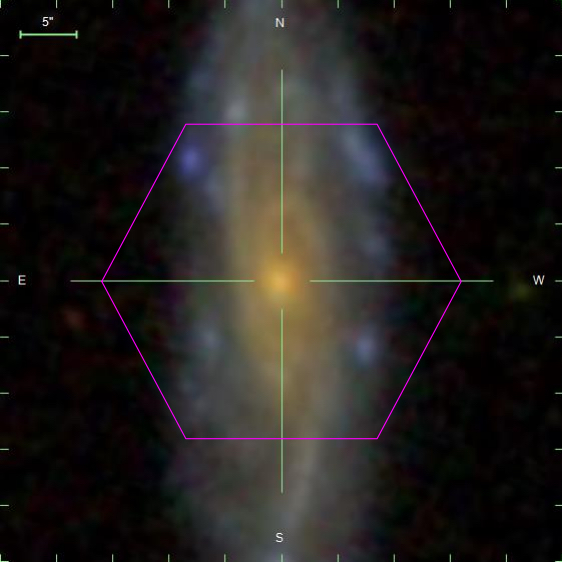}
\includegraphics[height=1.3in]{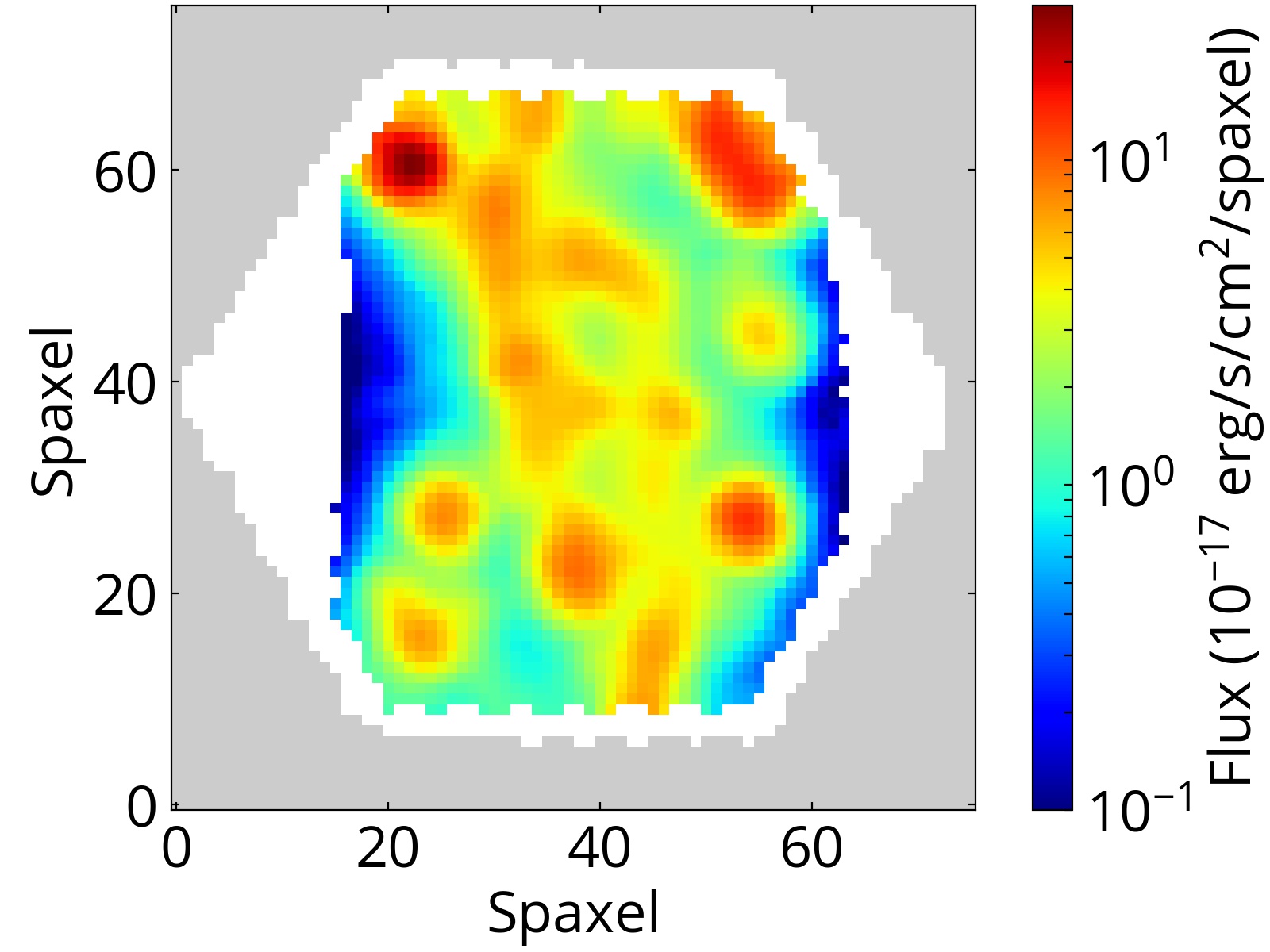}
\includegraphics[height=1.3in]{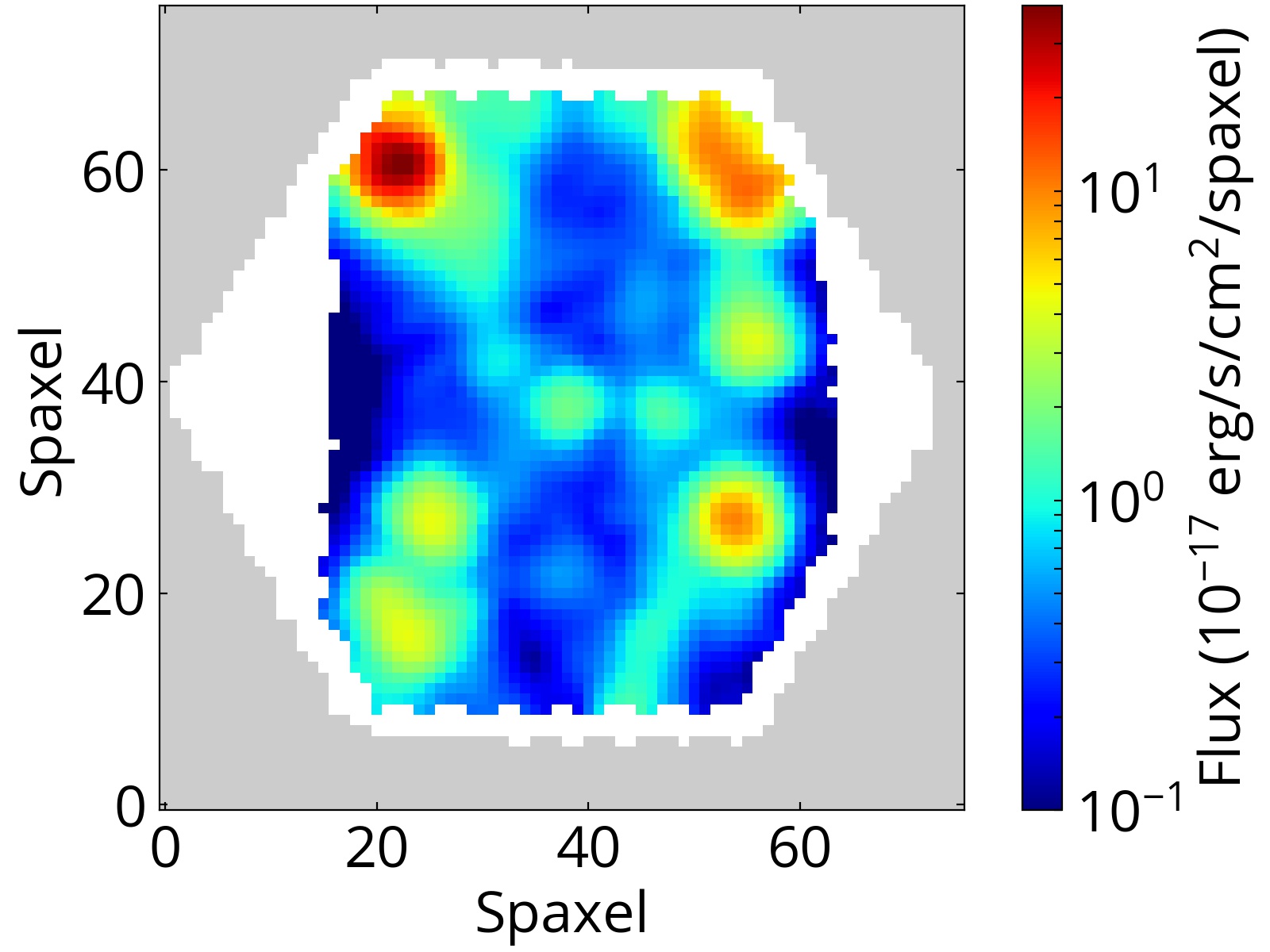}
\includegraphics[height=1.3in]{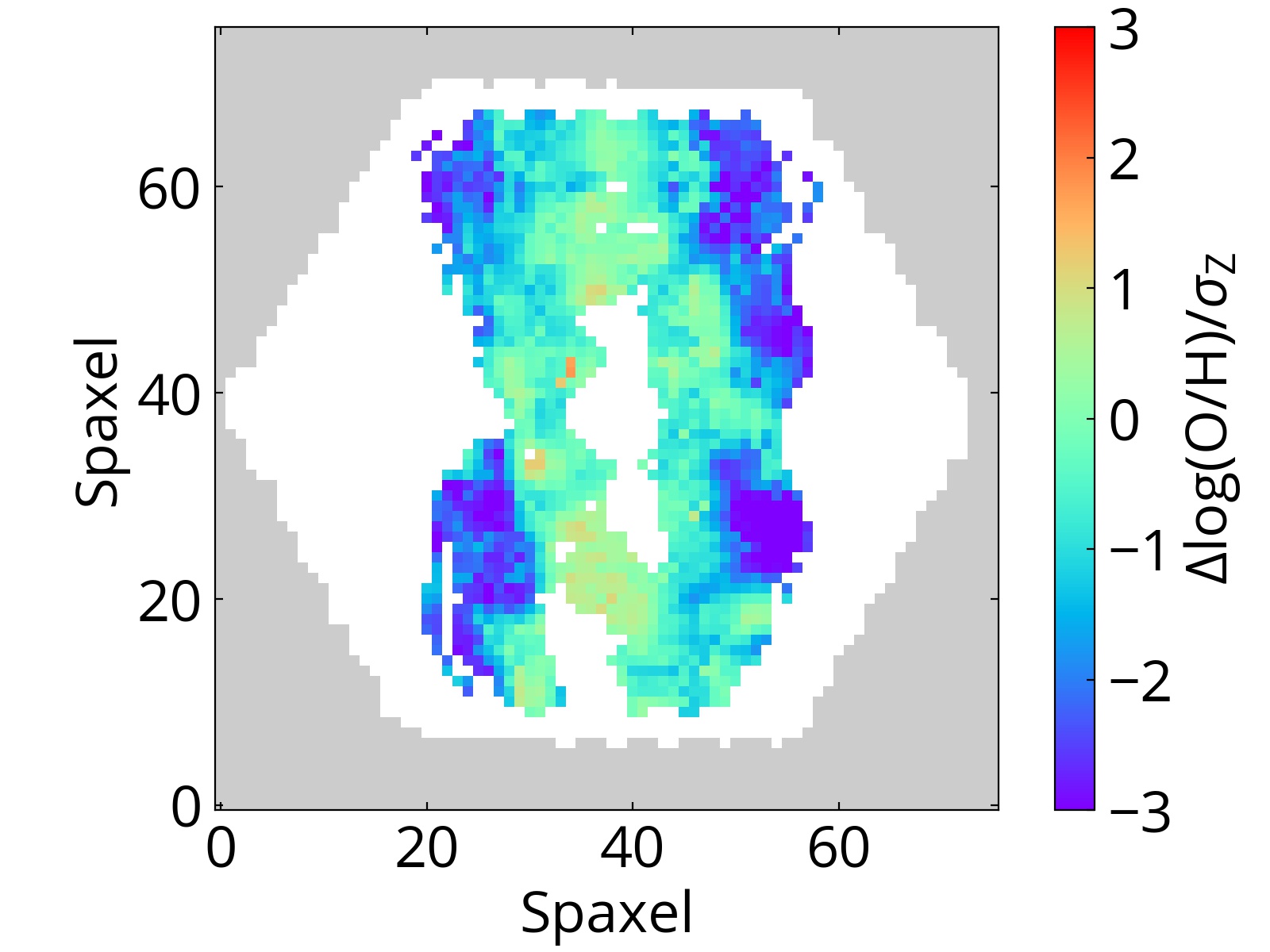}\\
\includegraphics[height=1.3in]{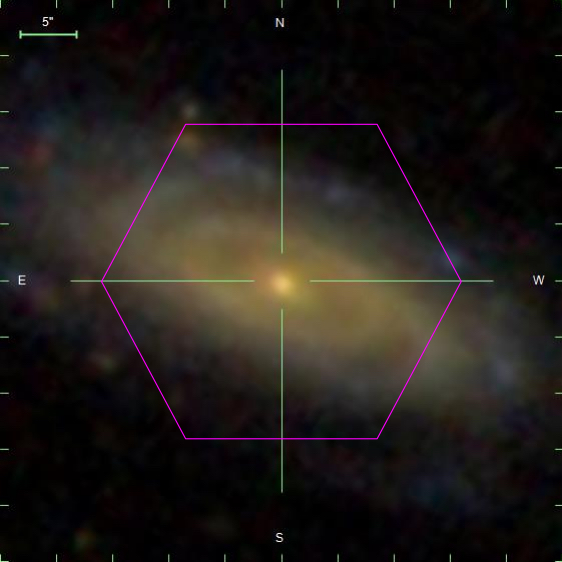}
\includegraphics[height=1.3in]{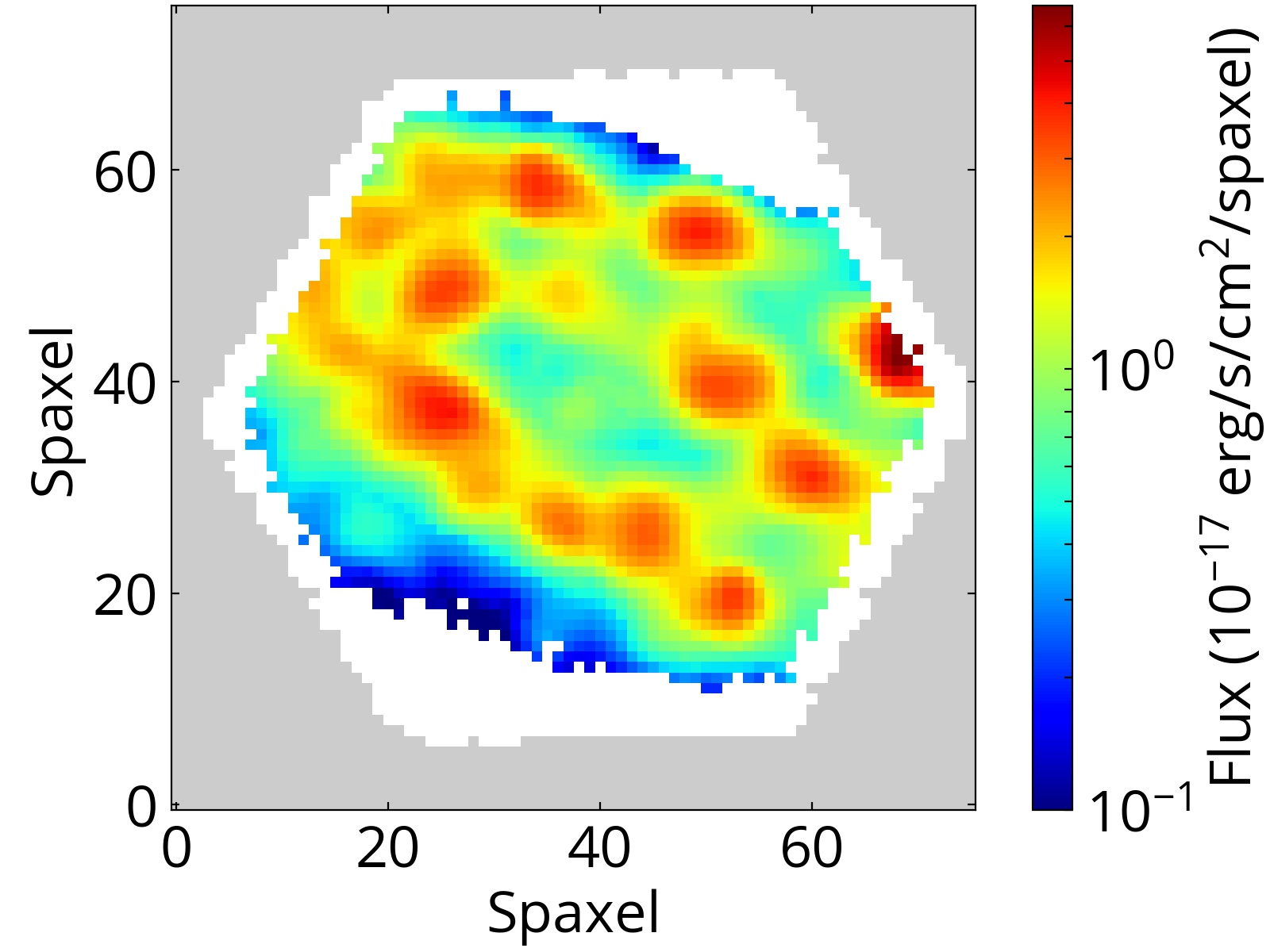}
\includegraphics[height=1.3in]{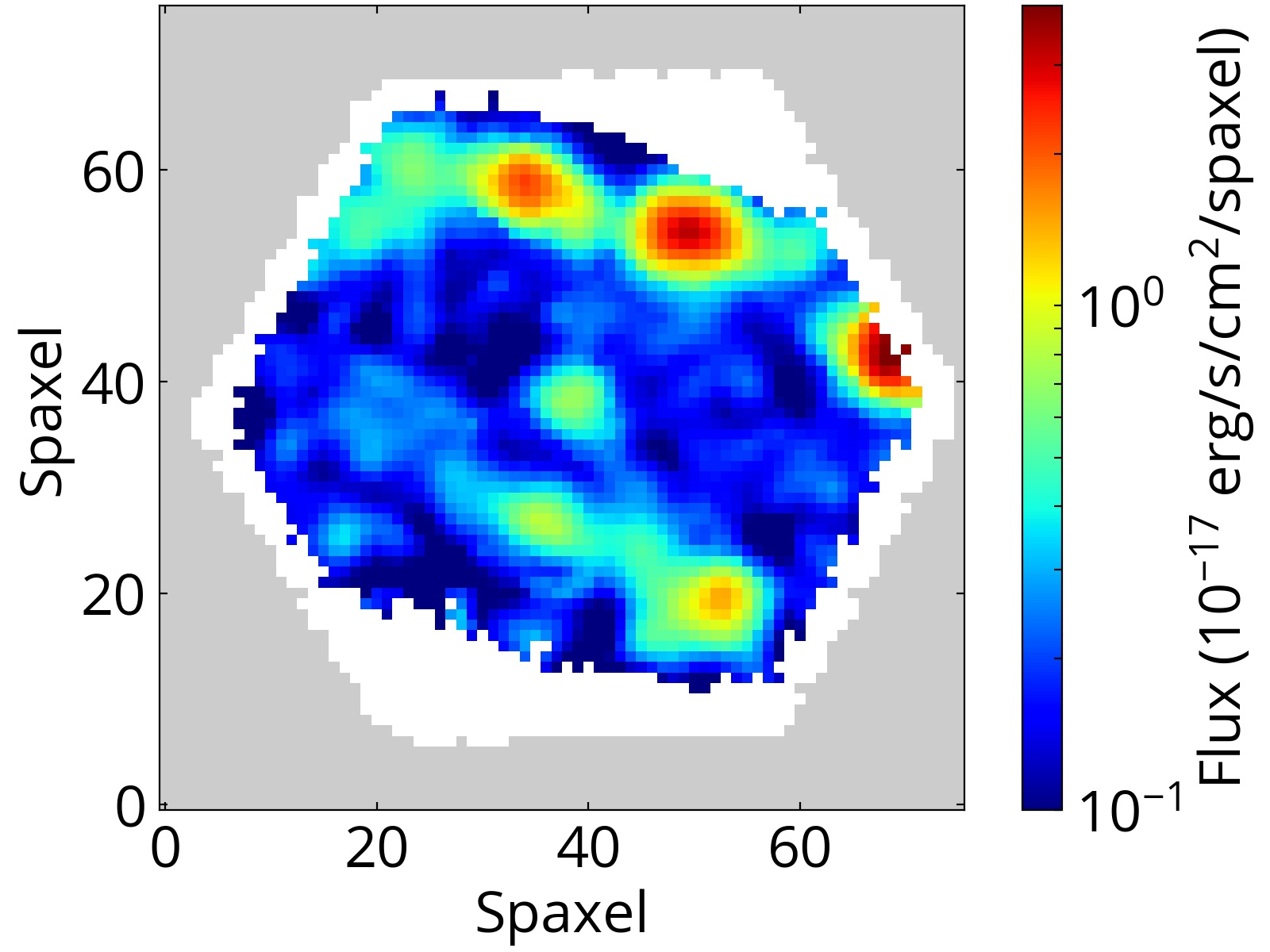}
\includegraphics[height=1.3in]{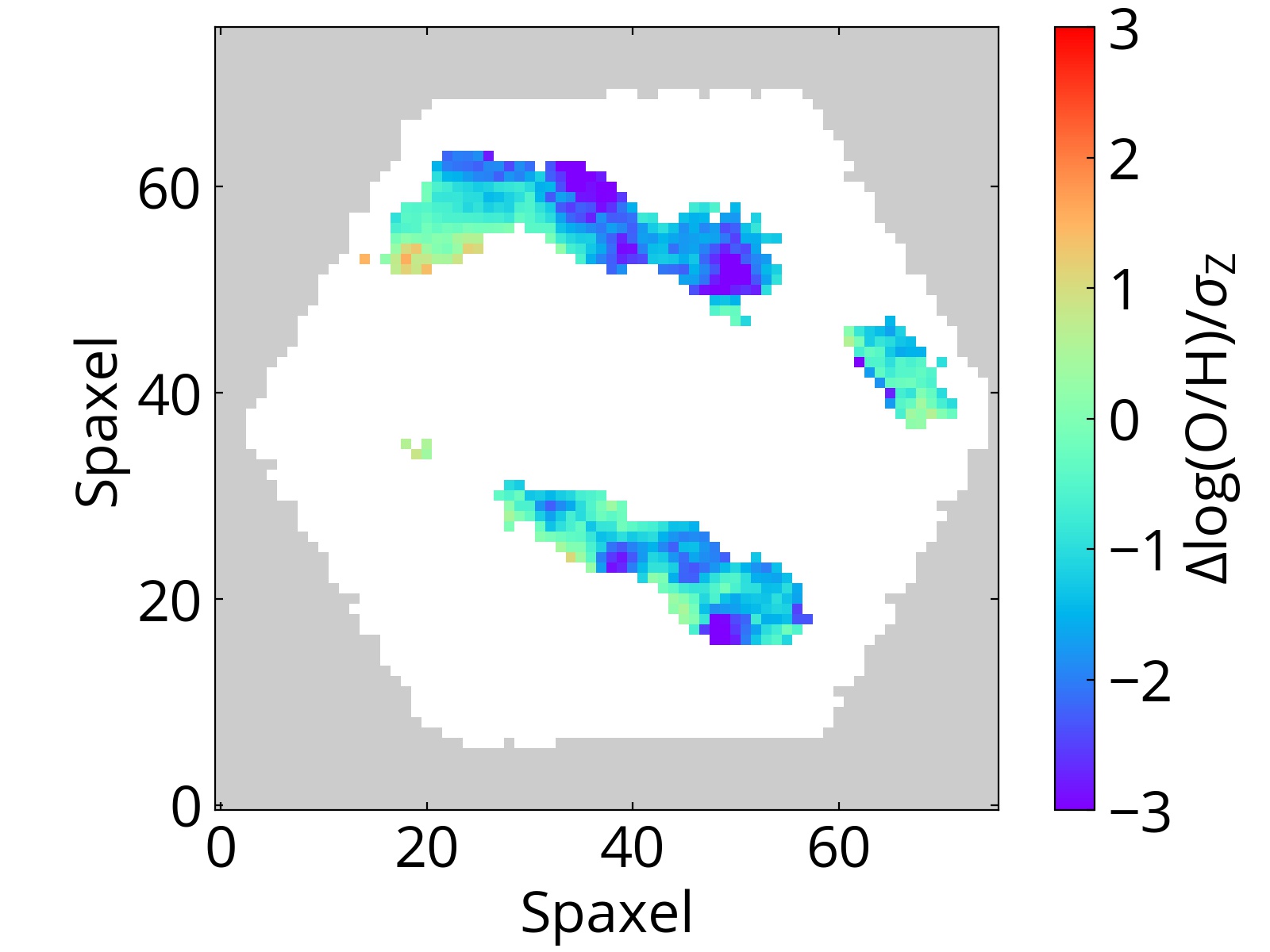}\\\includegraphics[height=1.3in]{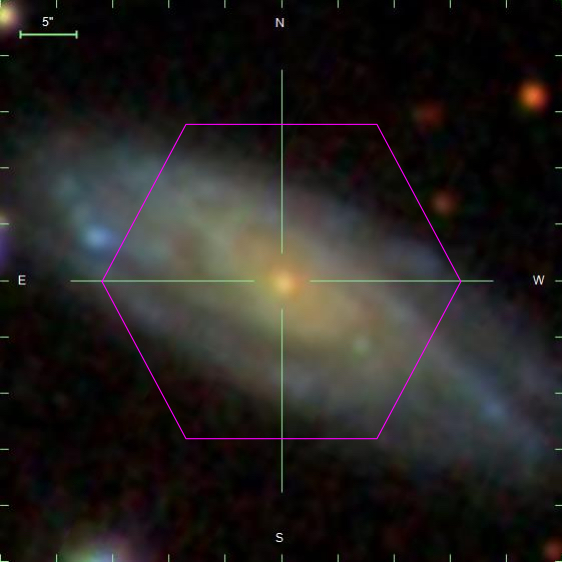}
\includegraphics[height=1.3in]{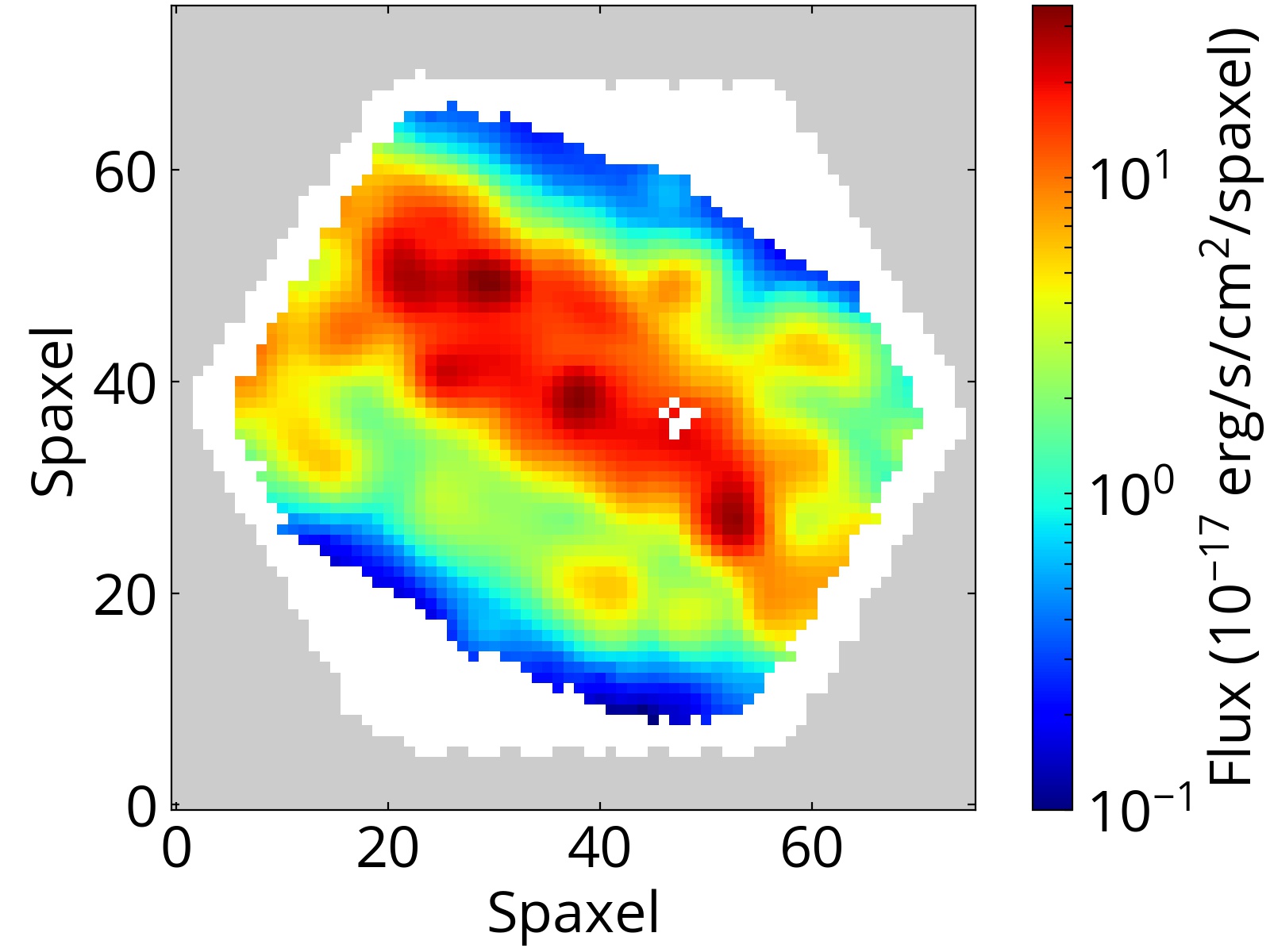}
\includegraphics[height=1.3in]{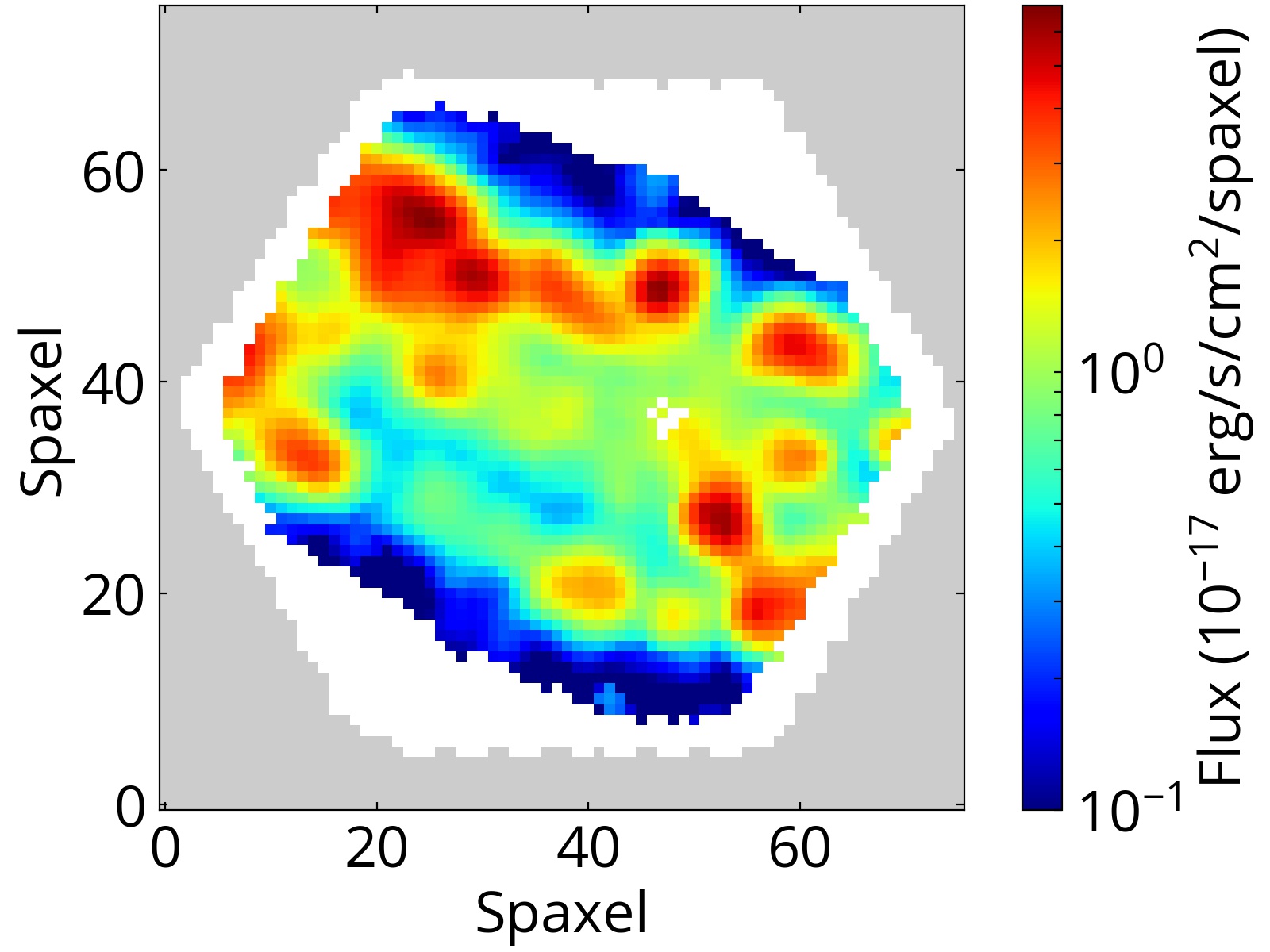}
\includegraphics[height=1.3in]{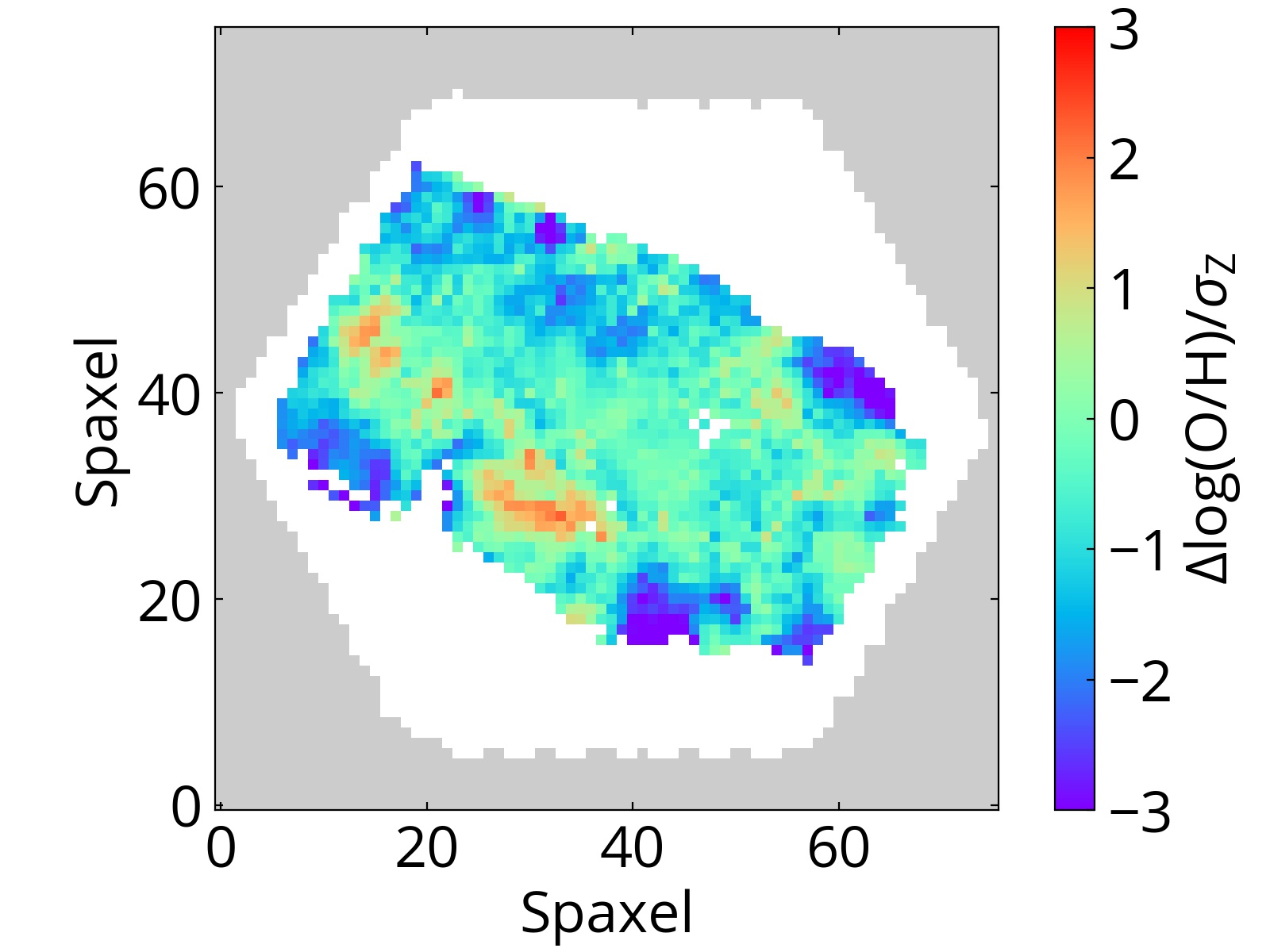}\\
\includegraphics[height=1.3in]{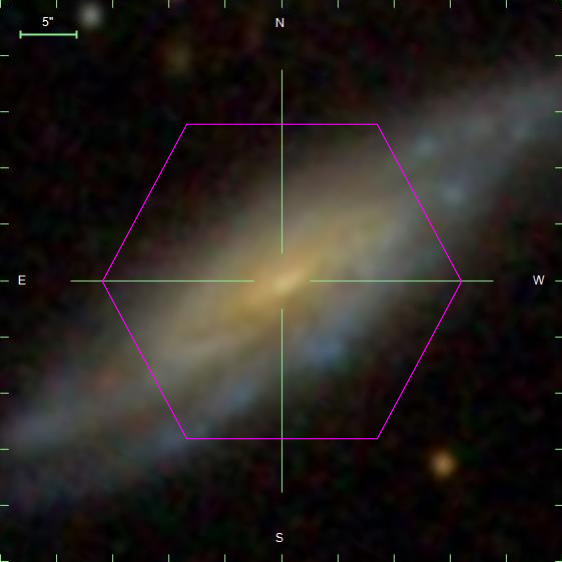}
\includegraphics[height=1.3in]{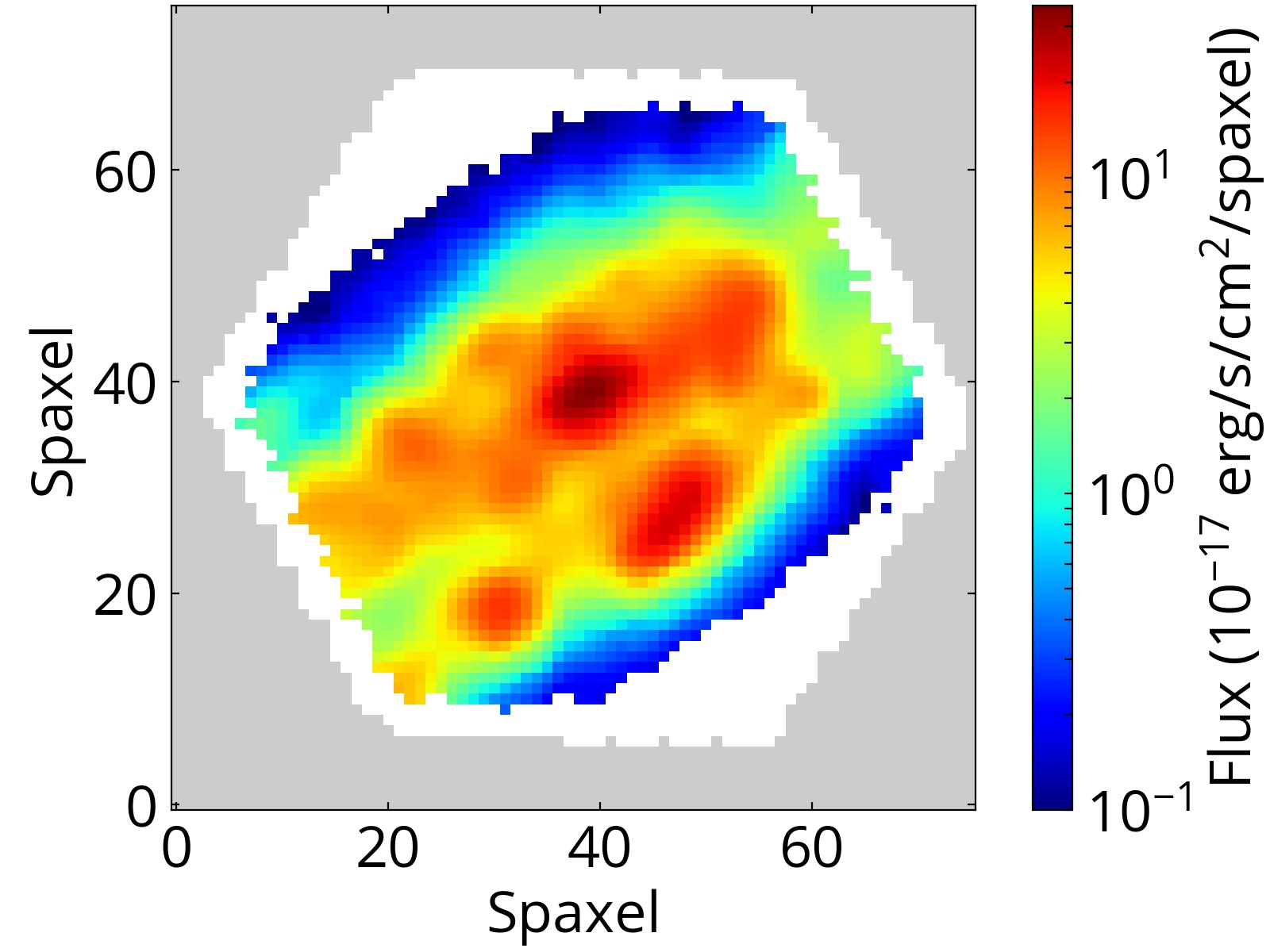}
\includegraphics[height=1.3in]{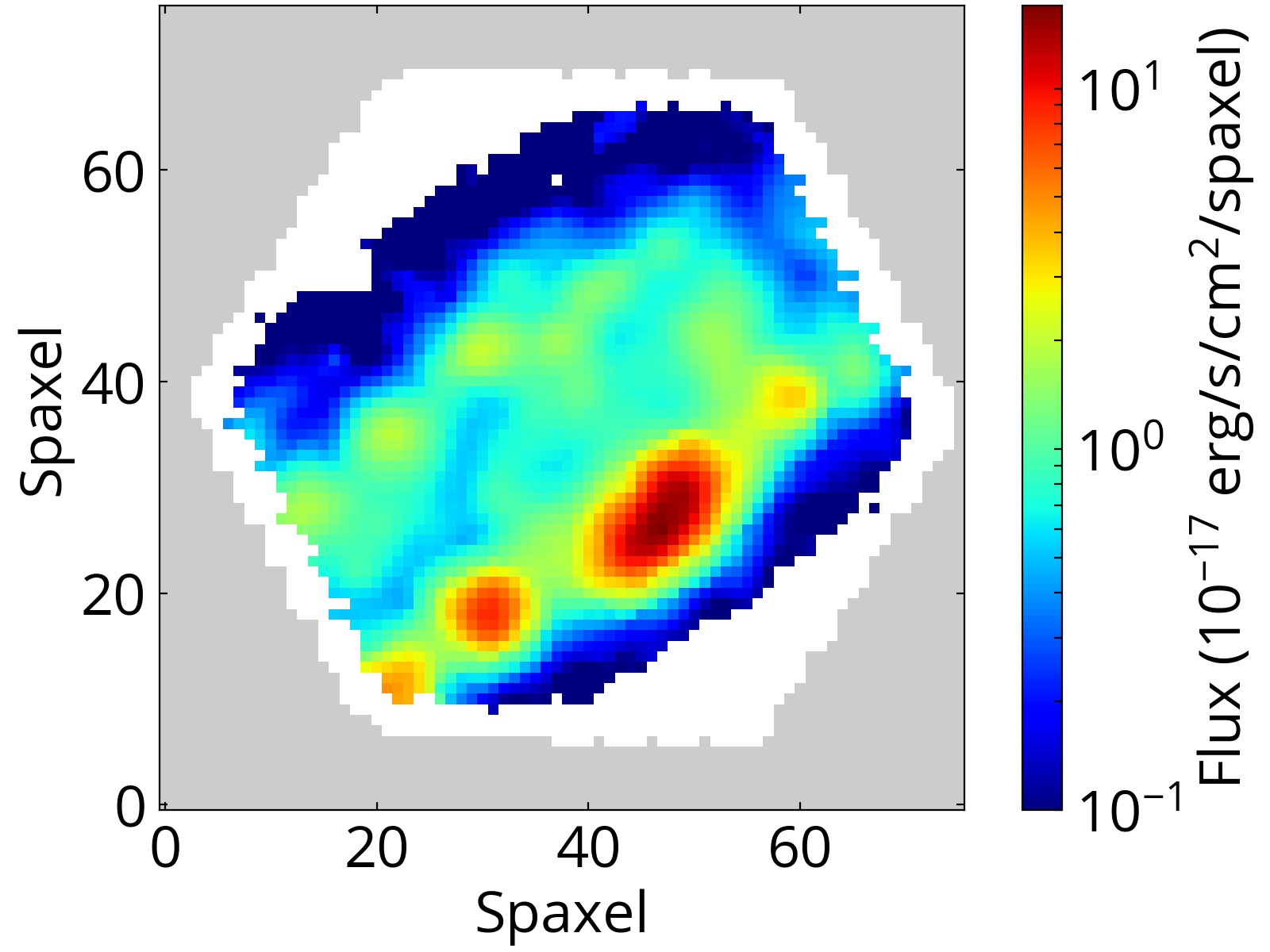}
\includegraphics[height=1.3in]{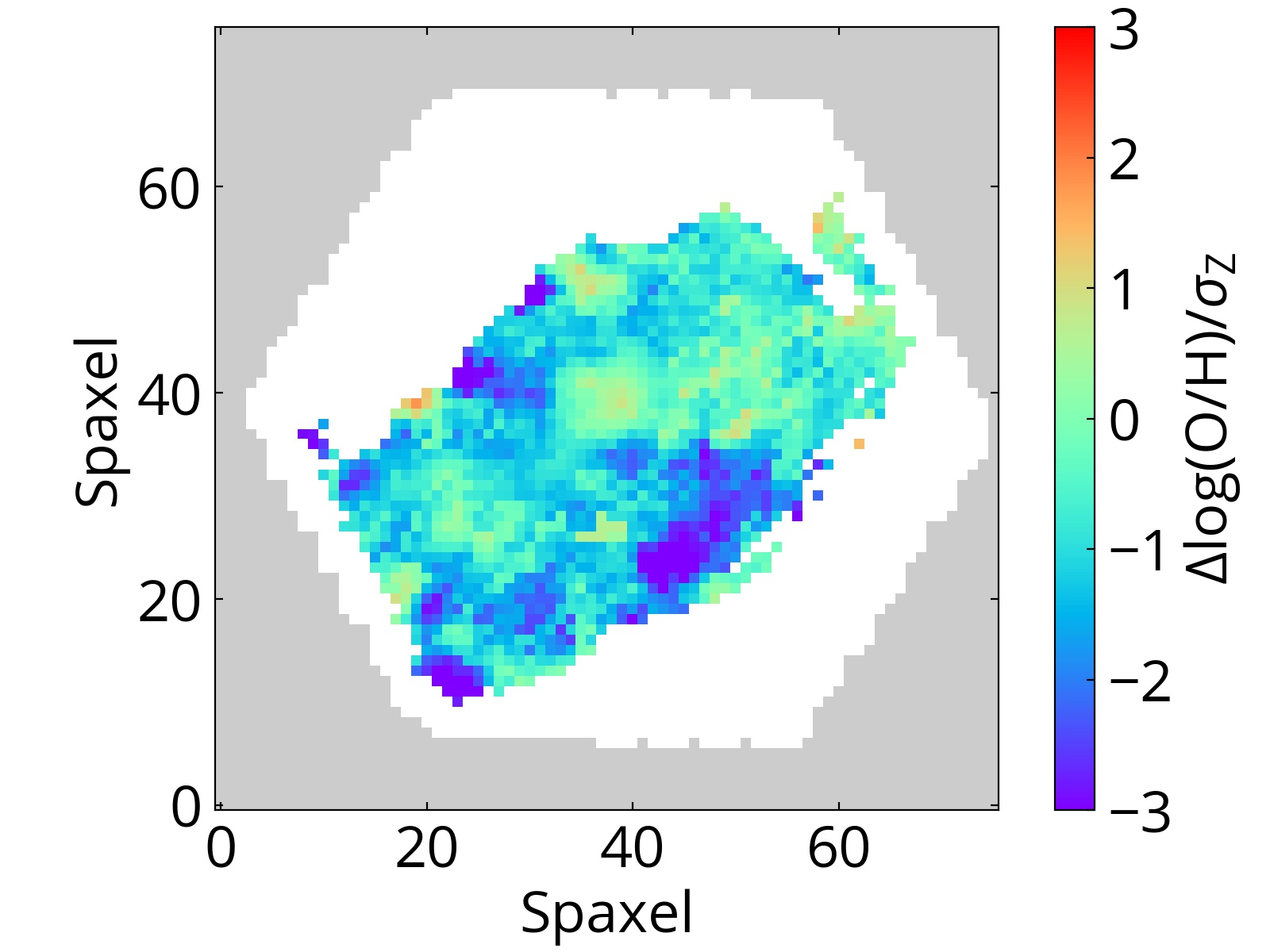}
   \caption{Examples of galaxies with ALM rings in the outskirts. The MaNGA ID from top: 8257-12705, 8613-12702, 7960-12704, 7957-12702. From left to right: optical SDSS images, \Ha\ maps, \oiii\ maps, and metallicity deviation maps. The ALM rings can be easily seen in the \oiii\ maps. All these galaxies are isolated and non-interacting, and they have higher stellar mass ($>10^{9.7}$\Msun), older stellar population at the centers, with ongoing star-forming regions on the edges of the stellar disks. }
   \label{fig:new-ring}
\end{figure*}

We found above that the strongest radial trends for ALM regions are that they are more prevalent in the outer disks of isolated massive star-forming galaxies. Here we show some examples of the morphology of these outer ALM regions. Fig.~\ref{fig:new-ring} shows objects that have ALM regions in their outskirts, forming an outer `ring' (see their \oiii\ maps). All these galaxies are more massive than $\sim10^{9.6}$\Msun, with quite old stellar populations in their centers. 

If these ring-like ALM regions were due to the disruption of infalling satellites, based on the sizes of the regions, the satellites should have been able to disturb the galaxy structure. However, in these galaxies, we do not see tidal streams, interaction-induced bars, distorted spiral arms, nor any extra structure. Thus, it is still not clear how to form such large-scale ring-like structures, and why ring-like low-metallicity regions are only seen in more massive galaxies. It can be due the inside-out disk formation when the outer gaseous disk satisfies the Toomre instability to form stars and the resulting stars have not yet replenished the ISM with metals. The stellar mass dependence might be related to different dominant modes of accretion in different mass regimes \citep{Rees1977}, or due to the recycled wind mode \citep{Oppenheimer2010}.

\section{Discussion}
\label{sec:discussion}

\subsection{The sources of low-metallicity gas}
\label{sec:gas-source}

First, we note that our methodology searches for low-metallicity outliers, so it is mainly sensitive to individual events that have significantly lowered the ISM metallicity relative to expectations. Examples of such events would include enhanced radial inflow induced by interactions, accretion of dwarf galaxies, and infalling gas clouds from CGM and IGM. On the other hand, if there is gas inflow feeding galaxies constantly and spatially uniformly, its effect on metallicity would be incorporated into the local $\Sigma_*$-Z relation and these regions would not be identified as ALM regions with our method. 

Previous work on gas-phase metallicity mainly focuses on the properties of the general population of star-forming spaxels. Some studies show that there is no strong correlation between the local $\Delta \log({\rm O/H})$ with local SFRs \citep{Sanchez2013, Barrera-Ballesteros2018}, while some report such a relation \citep{SanchezAlmeida2015, SanchezAlmeida2018}. In terms of radial distribution, the metallicity gradients of MaNGA star-forming galaxies are reported in \cite{Belfiore2017}. Similarly, the correlation between metallicity and local stellar surface mass density has been presented in several studies \citep{Moran2012, Rosales-Ortega2012, Barrera-Ballesteros2016}. While it is still under debate whether galactocentric distance or surface mass density is more fundamental to the metallicity distribution (e.g. \citealt{Pilyugin2017}), empirically, the local stellar surface mass density is strongly related to the distance (for a given stellar mass). Thus, when we select ALM spaxels by using the local surface mass density, we are effectively incorporating the radial dependence of metallicity. Our approach is different from these aforementioned studies, as we focus on the anomalously-low-metallicity outliers rather than the mean.  Because ALM spaxels only constitute $\sim10$\% of the star-forming spaxels, they make very little contribution to the correlations seen for the entire star-forming population and may be easily missed. Here we are specifically targeting them to study their properties and origin. 

% do not find a strong correlation of the $\Delta \log({\rm O/H})$ with local SFRs. Also, considering the oxygen abundance at $1R_e$, studies based on IFU surveys do not see a significant relation between metallicity and SFRs on the global scales \citep{Sanchez2013, Sanchez2017, Barrera-Ballesteros2017}. On local kpc scales, local metallicity only shows very weak dependence on local SFR and sSFR \citep{Barrera-Ballesteros2018}. \red{(Sanchez Almeida does not like this.)}

What is the origin of this ALM gas? One important clue is that the incidence rate and spatial distribution of this material is significantly larger in strongly asymmetric galaxies. This can be seen in Fig.~\ref{fig:new-global-prop} and Fig.~\ref{fig:covering-frac}, as well as in Fig.~\ref{fig:new-dZ-dist}. This implies that in these cases, the low-metallicity inflow occurs as a result of strong interactions. This is further supported by the difference in the radial distributions of the ALM material in close pairs vs. mergers (Fig.~\ref{fig:new-dZ-dist}). This implies that during the early stages of the interaction, the ALM material has been delivered to the outer part of the galaxy. During the subsequent merger the ALM material is redistributed throughout the galaxy. The delivery of this gas leads to enhanced star-formation on local and global scales in these systems.

However, it is important to emphasize that most of the galaxies with ALM regions are not in strongly interacting systems. While in principle, this gas could have an entirely different origin, the most straightforward interpretation is that it is delivered by qualitatively similar (but smaller-scale) mechanisms, such as the capture of small satellite galaxies (see Fig.~\ref{fig:new-remnant}) or accretion of gas clouds from the CGM or IGM. This is consistent with the association between ALM gas and strong star-formation that we see on local and global scales in the whole sample (not just mergers and close pairs). The CGM is a plausible source for such material: CGM clouds have a mean metallicity about three times lower than the ISM, and comprise at least as much gas mass as the ISM (\citealt{Tumlinson2017} and references therein). 

%\red{The asymmetric metallicity distribution found in close pairs (Fig.~\ref{fig:new-close-pair}) is caused by inflow gas. If it were a satellite causing the large area of lower metallicity, it would strongly disturb the galaxy. }

%Theoretical simulations show that inflow deposits gas more uniformly over a range of radii \citep{Brook2012, Christensen2016}. But here we show that the accretion is more prevalent in the outskirts?

\subsection{Timescales}
\label{sec:timescale}

In order to understand how frequently gas is accreted by galaxies, we begin by attempting to constrain the lifetime of the ALM regions we have identified. Below, we will consider several constraints: the lifetime of massive stars responsible for most of oxygen in the ISM, the mixing timescales of low-metallicity gas, and the self-enrichment of the gas by metal-rich stellar ejecta.

The strong local connection between low metallicity regions and lower D$_n4000$ in Fig.~\ref{fig:new-EW-D4000} suggests that the young stellar population is recently born from the low-metallicity gas. Therefore, the stellar age serves as a clock in the low-metallicity regions. With a median D$_n4000$ of 1.2 for low-metallicity spaxels, the corresponding stellar age is $\sim250$\,Myr \citep{Kauffmann2003a}, providing a timescale for the lifetime of low-metallicity regions. The presence of low-metallicity gas is also correlated with the global specific SFR (based NUV$-r$). On global scales, a simple causality argument implies that the SFR cannot vary significantly on time scales much shorter than a galaxy crossing time (of-order a hundred Myr). This is consistent with the timing constraints based on the local stellar population.

We can also consider how long it would take low-metallicity material to physically mix with the surrounding (higher-metallicity) ISM. Locally, typical non-circular velocities in the ISM are of-order 10 km s$^{-1}$, so gas would mix on spatial scales of a few kpc on time scales of a few hundred Myr. Differential rotation will lead to shear, and radial mixing on timescales of a galaxy rotation period (few hundred Myr). The azimuthally asymmetric metallicity distribution seen in close pairs (Figure~\ref{fig:new-close-pair}) also suggests that the mixing timescale is at least half of the rotation periods. Based on their stellar velocities, 8454$-$12703 and 8313$-$12702 have rotation periods of $\sim$200\,Myr, implying a mixing timescale of at least $\sim100$\,Myr. This is in agreement with our previous estimates. Numerical simulations based on the mixing driven by supernova \citep{DeAvillez2002} and thermal instabilities \citep{Yang2012} are also consistent with these timescales. 

Finally, we can consider how long it will take, once star formation begins, until the resulting metal-rich stellar ejecta self-enrich the ISM and erase the signature of the low-metallicity gas. Almost all oxygen in the ISM is produced by massive stars ($\gtrsim8$\Msun). Their lifetimes are short ($\lesssim30$\,Myr), so oxygen from the nucleosynthesis in massive stars can enrich the ISM via supernovae very quickly. If we consider a simple closed-box model given a gas depletion time $\tau_{\rm SF}$ and a constant SFR (so $\Sigma_{*} = \Sigma_{\rm SFR} \times t$), the mass fraction of oxygen in ISM $X_O (t)$ is:
\begin{equation}
X_O (t) =  \frac{\Sigma_{oxygen}}{\Sigma_{gas}}  = \frac{P \times \Sigma_{*}}{\Sigma_{tot} - \Sigma_{*}} = \frac{P \times t}{\tau_{\rm SF} - t},
\end{equation}
where $P$ is the dimensionless oxygen yield integrating over IMF, and $t$ is the time since the onset of star formation. Assuming $P=0.007$\,\citep{Kobayashi2006, Zahid2012} and the Salpeter IMF, the time it takes for ISM to reach solar from an initial half solar metallicity is $0.17\tau_{\rm SF}$. Based on this simple closed-box model, we have an order-of-magnitude timescale for massive stars to enrich ISM with oxygen. With the Kennicutt-Schmidt relation \citep{Schmidt1959, Kennicutt1998, Bigiel2008}, we estimate $\tau_{\rm SF} \sim 1$\,Gyr in the ALM regions, so this gas will be self-enriched on a timescale of a few hundred Myr.

We conclude that all these arguments yield consistent lifetimes for the ALM regions of a few hundred Myr. 

%\subsection{Hint for the local fundamental metallicity relation}

%(maybe talk about some relevant chemical models???)

\subsection{Masses and accretion rate}

ALM regions are commonly found across the population of late-type galaxies. While they are more common in the lower-mass galaxies, they can be found across the whole mass range. Moreover, they do not show any apparent dependence on concentration, a proxy for bulge-to-disk ratios and hence the Hubble types of late-type galaxies \citep{Conselice2003}. The ubiquity of these ALM regions suggests that they are part of the normal life-cycle of typical late-type galaxies. If indeed accretion events causing the low-metallicity regions happen occasionally for all late-type galaxies, then the fraction of late-type galaxies having low-metallicity regions indicates the duty cycle for such accretion events. Based on the occurrence rate of having ALM regions in the star-forming galaxies, the overall duty cycle is $\sim25$\% (307/1222). With the lifetime of ALM regions to be a few hundred Myr, this implies there is an accretion event once every $\sim$ Gyr.
  
We now make an order-of-magnitude estimate of the mass involved and the implied accretion rate. Let us consider a typical MaNGA galaxy with ALM regions. For the typical stellar mass of 10$^9$ to 10$^{10}$ M$_{\odot}$, the typical ISM (atomic and molecular) gas mass will be a few$\times10^{9}$\Msun\  \citep{Catinella2018}. Based on the covering fraction of ALM spaxels in ALM galaxies over this mass range, the implied mass of metal-poor gas in such a galaxy would be about $10^9$\Msun. With one event per \,Gyr, the accretion rate per late-type galaxy would be $\sim 1$\Msunyr.

We can now compare this to the rate at which gas is converted into stars. In the local Universe, the star-forming sequence has a mean SFR ranging from $\sim0.3$\Msunyr\ for $10^9$\Msun\ to $\sim1$\Msunyr\ for $10^{10}$\Msun\ galaxies \citep{Schiminovich2007}. Therefore, an accretion rate of $1$\Msunyr\ is able to supply gas for these low-mass late-type galaxies at a rate sufficient to offset their star formation. Of course, using ALM regions as sign-posts may miss some of the accretion, especially if it happens more smoothly (rather than in events that are discrete in time, space, and chemical composition).

%self-regulate model \citep{Tinsley1978, Koppen1999}

%\blue{ \cite{Chiappini2001, Chiappini2009} say that accretion is mainly low-metallicity gas. }

\section{Conclusions}
\label{sec:conclusion}

In this paper, we have used MaNGA data to identify regions of anomalously-low-metallicity (ALM) in the ISM in 1222 local star-forming late-type galaxies. We have chosen to use the \oii/\nii-based oxygen metallicity calibrator in order to minimize the effect of ionization parameter on the estimates. Only pure star-forming spaxels are analyzed so that this metallicity calibrator can be used. Based on the tight empirical relation between metallicity and local stellar surface mass density at fixed stellar mass, ALM regions are identified by those spaxels with observed metallicity lower than the expected value by more than $0.14$\,dex. We find few spaxels whose metallicities are higher than expected by this amount. By investigating the global properties of galaxies having ALM regions and their local properties, we present the following findings:

\begin{enumerate}

\item ALM regions are found in about 25\% of the MaNGA star-forming galaxies. The low-metallicity spaxels comprise about 10\% of the total number of star-forming spaxels over this same mass range.

\item The incidence rate of ALM regions decreases systematically with increasing stellar mass, from about 30\% at 10$^9$ M$_{\odot}$ to nearly zero at 10$^{11}$ M$_{\odot}$. The incidence rate is also much higher for bluer NUV$-r$ colors (and therefore higher global specific SFRs), reaching about 70\% for galaxies bluer than NUV$-r$ = 1.5. Roughly 50\% of the strongly disturbed galaxies (Asymmetry $>$ 0.2) have ALM regions, compared to only 25\% of the less-disturbed galaxies. Nevertheless, the majority of ALM galaxies (82\%) are not strongly disturbed. We find no strong trends with Concentration (a proxy for Hubble type).

\item On local kpc scales, the ALM regions have higher sSFRs and younger stellar populations. Specifically, the ALM incidence rate rises steeply as a function of both D$_n 4000$, sSFR, and EW(\Ha), reaching over 50\% for the youngest and most-actively-star-forming regions.

%\item The locations and sizes of ALM regions are related to galaxy interaction properties. In mergers, the ALM spaxels cover a larger portion of the star-forming gas ($\sim$40 percent), and are found with equal likelihood across the inner and outer regions. In galaxies that are in close-pairs or are isolated galaxies, the ALM regions are more common in the outer region ($>1R_e$) than in the inner regions. This trend is stronger at higher stellar mass ($>10^{9.6}$\Msun).

\item The locations of ALM regions are strongly related to galaxy interaction. ALM regions can be found at all radii (including the centers) of mergers, while they tend to be located in the outer disk in close pairs and isolated galaxies. 

\item Interesting morphologies of ALM regions are found. Some close pairs have an azimuthally asymmetric metallicity distribution, with lower metallicities on the sides closer to the companions. Also, low-metallicity outer rings are seen in the more massive galaxies ($>10^{\sim9.6}$\Msun).

\item We have proposed that these ALM regions trace the sites of accretion of gas with a metallicity significantly less than that of the pre-existing ISM. This could be mass transfer or inflow in mergers or strong interactions, but their existence in non-interacting galaxies suggests that accretion of gas from the CGM, IGM, and/or from gas-rich dwarf satellites can also be important.

\item Several independent lines of evidence suggest that the duration of the ALM regions is a few hundred\,Myr. Interpreting the incidence rate of 25 percent as a duty cycle, this would correspond one accretion event per \,Gyr. The implied time-averaged accretion rate of this metal-poor gas is $\sim1$\Msunyr, similar to the SFR of late-type galaxies over the stellar mass range $\sim10^9$ to 10$^{10}$ \Msun. Therefore, fed by bursty accretion events, local galaxies on the star-forming sequence can continue their star formation without running out of gas.

\end{enumerate}

\acknowledgments

HCH would like to acknowledge the helpful discussion with Christy Tremonti, Michael Dopita, and Fr\'ed\'eric Vogt. The authors are grateful to the anonymous referee for constructive suggestions.

Funding for the Sloan Digital Sky Survey IV has been provided by the Alfred P. Sloan Foundation, the U.S. Department of Energy Office of Science, and the Participating Institutions. SDSS-IV acknowledges
support and resources from the Center for High-Performance Computing at
the University of Utah. The SDSS web site is www.sdss.org.

SDSS-IV is managed by the Astrophysical Research Consortium for the 
Participating Institutions of the SDSS Collaboration including the 
Brazilian Participation Group, the Carnegie Institution for Science, 
Carnegie Mellon University, the Chilean Participation Group, the French Participation Group, Harvard-Smithsonian Center for Astrophysics, 
Instituto de Astrof\'isica de Canarias, The Johns Hopkins University, 
Kavli Institute for the Physics and Mathematics of the Universe (IPMU) / 
University of Tokyo, the Korean Participation Group, Lawrence Berkeley National Laboratory, 
Leibniz Institut f\"ur Astrophysik Potsdam (AIP),  
Max-Planck-Institut f\"ur Astronomie (MPIA Heidelberg), 
Max-Planck-Institut f\"ur Astrophysik (MPA Garching), 
Max-Planck-Institut f\"ur Extraterrestrische Physik (MPE), 
National Astronomical Observatories of China, New Mexico State University, 
New York University, University of Notre Dame, 
Observat\'ario Nacional / MCTI, The Ohio State University, 
Pennsylvania State University, Shanghai Astronomical Observatory, 
United Kingdom Participation Group,
Universidad Nacional Aut\'onoma de M\'exico, University of Arizona, 
University of Colorado Boulder, University of Oxford, University of Portsmouth, 
University of Utah, University of Virginia, University of Washington, University of Wisconsin, 
Vanderbilt University, and Yale University.

\bibliography{metallicity}{}
%\bibliography{quasars_radio}{}
\bibliographystyle{aasjournal}
%\bibliography{master_0612,additional}

\appendix

We have used a specific metallicity indicator in this paper. To show that our results are insensitive to this choice, we present here some of our most crucial results using different metallicity indicators. In doing so, we tried to select a diversity of metallicity calibrators: some use \oiii, some use \sii; some based on empirical methods anchored to the direct method, while some based on photoionization model. Our main conclusions do not change using these different metallicity calibrators. 

We first summarize the metallicity calibrators we use.
\section{Summary of metallicity calibrators}

\subsection{O3N2 (PP04)}
This is a hybrid calibrator which uses the direct method for low metallicity (\Ometallicity $\lesssim 8.7$) \Hii\ regions and photoionization models for high-metallicity (\Ometallicity $\gtrsim 8.7$) \Hii\ regions. The calibrator we use is Equation 3 of \cite{Pettini2004}, 
\begin{equation}
\label{eq:O3N2-PP04}
12+\log({\rm O/H}) = 8.73 - 0.32 \times {\rm O3N2},\ {\rm where}\ {\rm O3N2} = \log \frac{[{\rm O_{III}} ]\lambda5007 / {\rm H_\beta}}{[{\rm N_{II}} ]\lambda6584 / {\rm H_\alpha}}.
\end{equation}
It is valid for $-1< {\rm O3N2} < 1.9$. Since the wavelength of \Ha\ is close to \nii\ and \Hb\ is close to \oiii, O3N2 calibrators are insensitive to the dust extinction.

\subsection{O3N2 (Marino+13)}
It is calibrated using the direct method for the \Hii\ regions in the CALIFA survey. The calibrator is Equation 2 of \cite{Marino2013},

\begin{equation}
12+\log({\rm O/H}) = 8.533 - 0.214 \times {\rm O3N2},
\end{equation}
with O3N2 defined in Eq.~\ref{eq:O3N2-PP04}. The calibrator is valid for $-1.1< {\rm O3N2} < 1.7$. Again, it is insensitive to dust extinction. 

\subsection{Calibrators from PG16}
This is calibrated using the counterpart method anchored to the direct method. \cite{Pilyugin2016} provide two types of calibrators. One uses \oii\ and the other one uses \sii, which are called `R calibration' and `S calibration' respectively in \cite{Pilyugin2016}. Here we follow their notations for the flux ratios:

\begin{equation}
\begin{array}{l}
R_2 = [{\rm O_{II}} ]\lambda3727, 9/ {\rm H_\beta},  \\
R_3 = [{\rm O_{III}} ]\lambda4959, 5007/ {\rm H_\beta}, \\
N_2 = [{\rm N_{II}} ]\lambda6548, 84/ {\rm H_\beta}, \\
S_2 = [{\rm S_{II}} ]\lambda6717, 31/ {\rm H_\beta}.
\end{array}
\end{equation}

The R and S calibration consist of two equations for two metallicity ranges. The upper branch ($\log N_2 \ge -0.6$) of the R calibration, their Equation 4, is
\begin{equation}
\label{eq:R-upper}
12+\log({\rm O/H}) = 8.589 + 0.022 \log (R_3/R_2) + 0.399 \log N_2 + (-0.137 + 0.164 \log(R_3/R_2) + 0.589 \log N_2) \times \log R_2,
\end{equation}
and the R calibration for the lower branch ($\log N_2 < -0.6$) is 
\begin{equation}
\label{eq:R-lower}
12+\log({\rm O/H}) = 7.932 + 0.944 \log (R_3/R_2) + 0.695 \log N_2 + (0.970 - 0.291 \log(R_3/R_2) - 0.019 \log N_2) \times \log R_2.
\end{equation}
The calibration explicitly takes the excitation parameter $P=R_3/(R_2+R_3)$ into account, whose spirit is similar to the \oiii/\oii\ ratios we use to derive the metallicity from the model grid in Sec.~\ref{sec:Z-calibrator}. Because the dependence on the excitation parameter at the high metallicity end is weak, the upper-branch calibration (Eq.~\ref{eq:R-upper}) can be reduced to a function of $N_2$ and $R_2$ only (see their Equation 8). The reduced form only uses \oii, \nii, and \Hb, making it similar to O2N2-based calibrators, although it does not explicitly depend on the \oii/\nii\ flux ratio. Only 0.4\% of our SF spaxels are in the lower branch, but we still choose to use the full form (Eq.~\ref{eq:R-upper} and ~\ref{eq:R-lower}) to derive the metallicity.

The S calibration for the upper branch is, Equation 6 in \cite{Pilyugin2016},
\begin{equation}
\label{eq:S-upper}
12+\log({\rm O/H}) = 8.424 + 0.030 \log (R_3/S_2) + 0.751 \log N_2 + (-0.349 + 0.182 \log(R_3/S_2) + 0.508 \log N_2) \times \log S_2,
\end{equation}
and the S calibration for the lower branch ($\log N_2 < -0.6$) is 
\begin{equation}
\label{eq:S-lower}
12+\log({\rm O/H}) = 8.072 + 0.789 \log (R_3/S_2) + 0.726 \log N_2 + (1.069 - 0.170 \log(R_3/S_2) + 0.022 \log N_2) \times \log S_2.
\end{equation}
Similar to the R calibration, the reduced form for the upper branch is also available (see their Equation 9).

\subsection{R23 (Tremonti+04)}
This is calibrated by using Bayesian statistics to fit several emission lines (\Ha, \Hb, \oiii, \oii, \nii, \sii) to the theoretical models. Based on this, \cite{Tremonti2004} also provide the calibration for R23 in their Equation 1:
\begin{equation}
\label{eq:R23-cali}
12+\log({\rm O/H}) = 9.185 - 0.313 x - 0.264 x^2 - 0.321 x^3,
\end{equation}
where $x= \log($R23) = $\log($(\oii3927,9 + \oiii4959 + \oiii5007)/\Hb). This R23 calibration is valid for the upper branch of the double-valued R23-metallicity relation. \nii/\oii\ ratios can be used to separate the upper- and lower-branch for R23 \citep{Kewley2008}. Only 0.002\% of our SF spaxel sample are in the lower branch ($\log($\nii/\oii$)<-1.2$), so we apply Equation~\ref{eq:R23-cali} for the entire sample without special treatment for the lower-branch spaxels.

\section{Results using different metallicity calibrators}

We follow the same procedure described in Sec.~\ref{sec:def-lowZ} for each metallicity calibrator. All calibrators use the extinction-corrected line fluxes. With the metallicity derived from each calibrator, we reproduce Fig.~\ref{fig:new-MZrelation},, Fig.~\ref{fig:mass-dep}, and Fig.~\ref{fig:new-hist_dZ}. $\sigma_{\rm Z}$ is set to the standard deviation of the best-fit Gaussian profile in each alternative version of Fig.~\ref{fig:new-hist_dZ}. Typically, the residual (the black line in Fig.~\ref{fig:new-hist_dZ}) intersects with the fitted Gaussian at around $-2\sigma_{\rm Z}$. Some exceptions are O3N2 (PP04) and O3N2 (Marino+13) that their intersections are $\sim -1.7 \sigma_{\rm Z}$. Following Sec.~\ref{sec:def-lowZ}, we define spaxels deviating more than $-2\sigma_{\rm Z}$ as ALM spaxels.

\begin{figure*}[htbp] %  figure placement: here, top, bottom, or page
   \centering
   \includegraphics[height=2.5in]{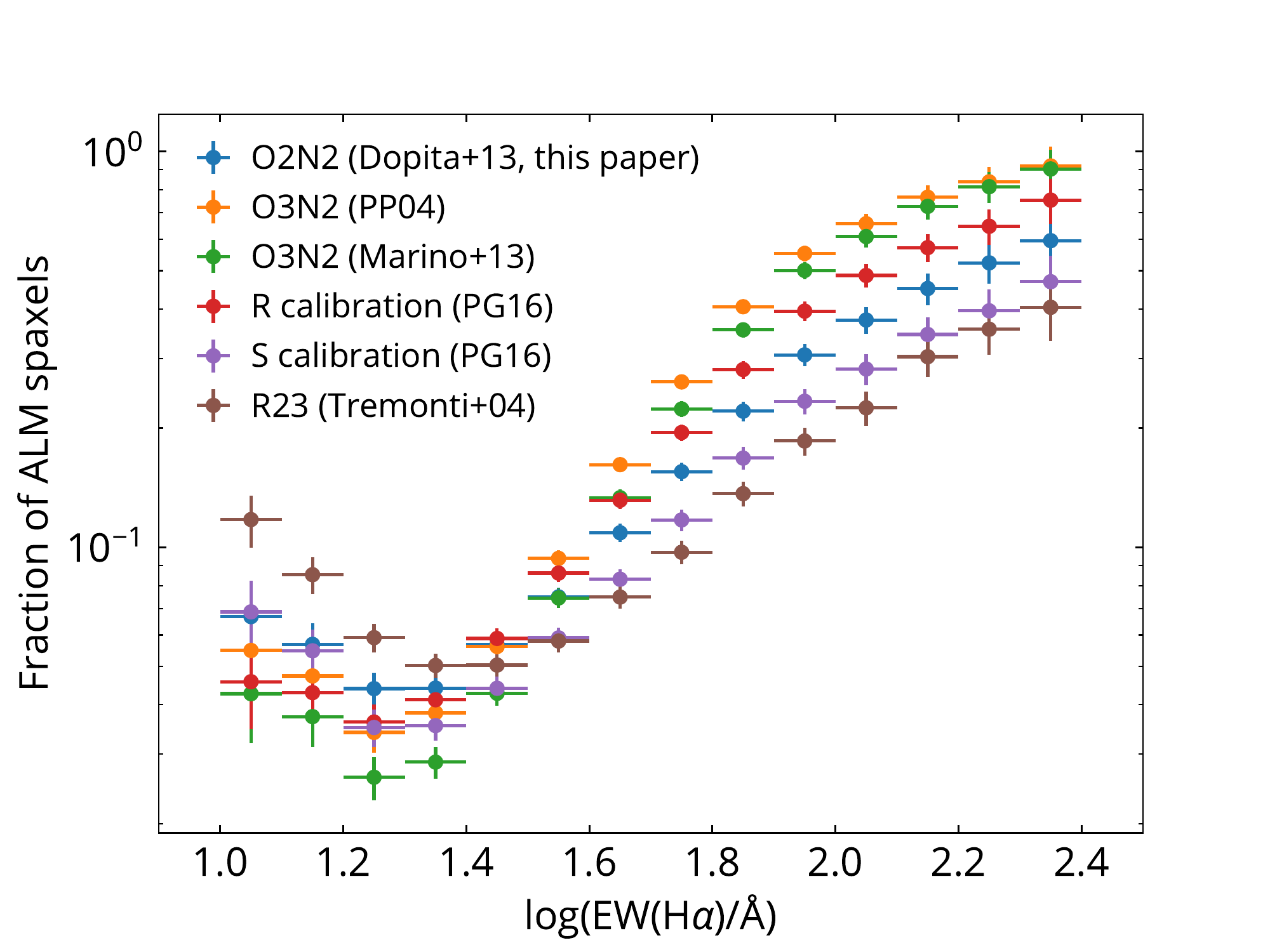}
   \includegraphics[height=2.5in]{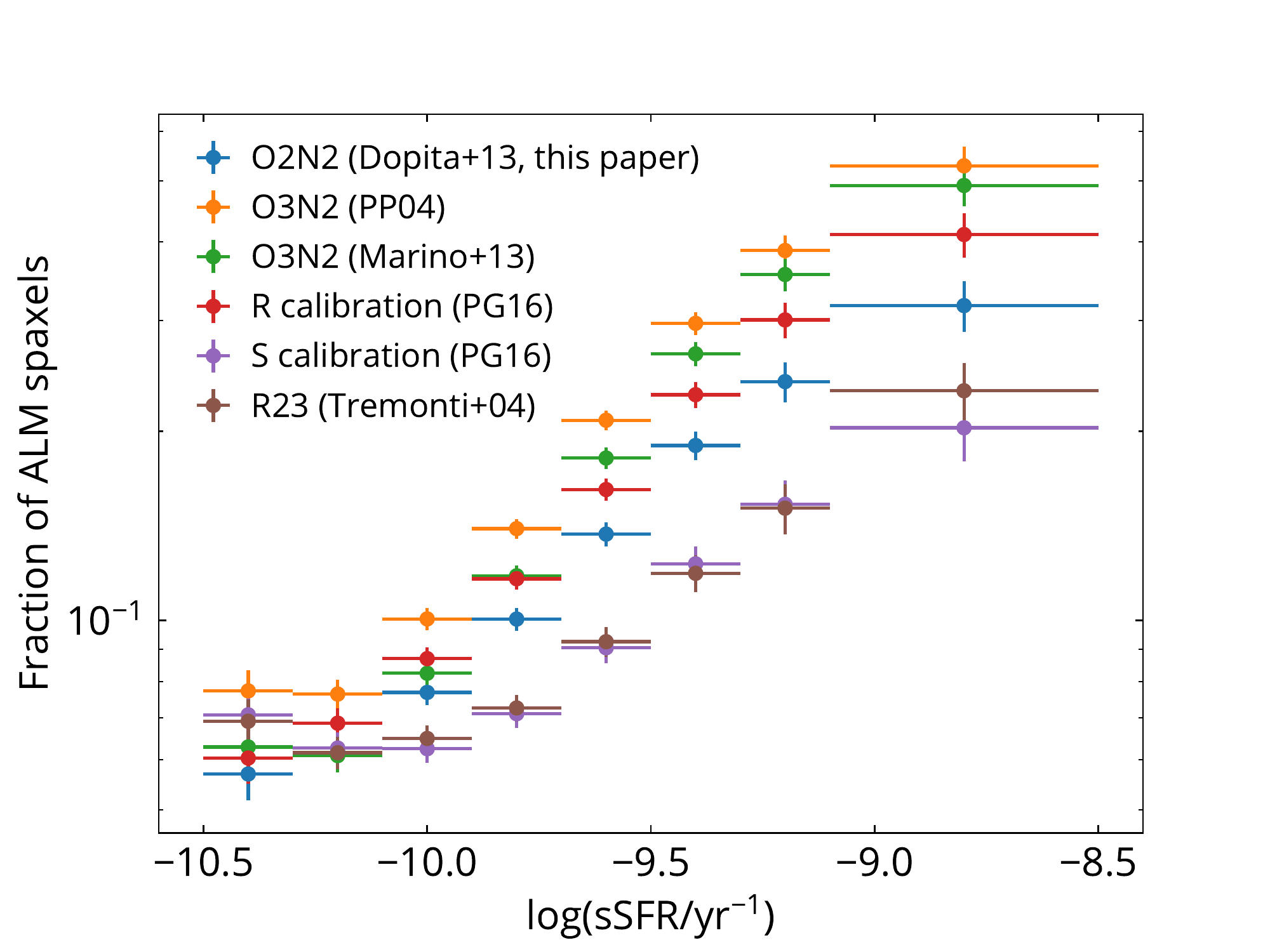}   

\caption{Reproduced Fig.~\ref{fig:new-local-prop} using different metallicity calibrators. The trend remains the same, only differing in the amplitude. The O2N2 metallicity calibrator we used throughout the paper is representative for all calibrators tested here.}
\label{fig:calibrator-test}
\end{figure*}

Fig.~\ref{fig:calibrator-test} presents our key results using these metallicity calibrators. The overall trend is the same, in that the fraction of ALM spaxels rises significantly with increasing sSFR and EW(\Ha). At low sSFR and EW(\Ha), all metallicity calibrators give similar ALM fractions, with 5\% at EW(\Ha)$=20$\,\AA\ and $<10$\% at sSFR$<10^{-10}$\,yr$^{-1}$. The rising amplitudes differ from calibrator to calibrator. ALM fractions from two O3N2 calibrators (PP04 and Marino+13) rise the most steeply towards the higher EW(\Ha) and sSFR. The ALM fraction can even reach $90$\% in the highest bin of EW(\Ha). The R calibration (PG16) and the O2N2-based calibrator we use in this paper also show significant increasing ALM fractions towards the higher end of EW(\Ha) and sSFR, but not as strong as the two O3N2 calibrators. R23 (Tremonti+04) also shows the similar trend. Interesting, calibrators involving \sii\ give a weaker trend. The S calibration has a similar amplitude to the R23 calibrator. 

We emphasize that we select a variety of metallicity calibrators. The O3N2 calibrators are insensitive to extinction correction, supporting that extinction plays a minor role in our results. Also, some use \oiii\ (two O3N2 and R23) while some do not, and some are calibrated through the direct method (two O3N2 and the two from PG16) while others are based on the photoionization models. Some explicitly take the ionization parameter (or excitation parameter) into account, like the R and S calibration from PG16 (empirical method) and the O2N2-based calibrator we use in this paper (photoionization model). The O2N2-based calibrator has a moderate trend, making it a good representative to present the results throughout the paper. Therefore, our results and conclusions in this paper do not change with different metallicity calibrators. 

\end{document}